# Electronic Origins of Elastic Behavior in Rocksalt, Zinc-Blende and Wurtzite 3d Transition-Metal Nitrides

J. Cañas[1,2] and O. Ambacher[1]

[1] *Institute for Sustainable Systems Engineering (INATECH), University Freiburg, Emmy-Noether-Str. 2, D-79110 Freiburg, Germany*

[2] *Univ. Grenoble Alpes, CNRS, Grenoble INP, Institut Néel, 38000 Grenoble, France*

Corresponding author: Jesus Cañas Fernandez
email: jesus.canas@neel.cnrs.fr
ORCID: 0000-0003-0202-6987

**ABSTRACT**

We present a comprehensive ab-initio study of the elastic properties of the IV-period transition metal nitrides, highlighting the correlation between their mechanical behavior and electronic structure. By analyzing the electronic density of states, we reveal how the occupation of bonding or antibonding orbitals determines the evolution of elastic properties along the period. Furthermore, we examine trends across different crystal structures, specifically comparing rocksalt, zincblende and wurtzite phases, to uncover how variations in crystal symmetry and atomic coordination influence their elastic properties.

## I. INTRODUCTION

Transition metal nitrides (TMNs) in the rocksalt phase, such as TiN, exhibit exceptional hardness, high melting points, and low chemical reactivity, making them highly attractive for wear-resistant coatings on tools and dies. [1] Their outstanding mechanical durability also makes them valuable for automotive components. Extensive research has been devoted over the last decades to screening materials and elucidating their mechanical properties for this purpose. [2-7]

In contrast, wurtzite nitrides such as GaN combine robust mechanical properties with piezoelectricity. [8] These features make them well-suited for devices that interconvert electrical and mechanical vibrations, including bulk acoustic wave (BAW) devices. In such applications, the stiffness coefficient directly influences the piezoelectric coupling coefficient, rendering it a critical design parameter. [9,10] Accordingly, hard wurtzite nitrides with strong piezoelectric responses are essential for improving electromechanical conversion efficiency in BAW resonators.
Recent studies have explored combining these two families of nitrides to enhance piezoelectric coefficients. The piezoelectric response of wurtzite nitrides can be significantly improved by alloying with rocksalt-phase transition metal nitrides, such as AlN or GaN alloys with ScN. [11] However, near the phase transition, the softening of the elastic properties partially compensates for the gains in BAW device efficiency achieved through enhanced piezoelectricity. Understanding the trends in elastic coefficients and the influence of crystal lattice structure on the overall elastic properties of nitrides is therefore crucial, and it is the matter of the present paper to elucidate the mechanisms governing the mechanical properties of these phases.

In this contribution, we perform an ab initio investigation of the elastic properties of IV-period TMNs in their rocksalt, zincblende and wurtzite phases. We analyze their electronic density of states (DOS) to establish the relationship between elastic behavior and electronic structure, focusing on the occupation of bonding and antibonding orbitals. We compare trends across nitrides with different coordinations, tetrahedral versus octahedral, and different crystal symmetries, cubic and hexagonal, to elucidate the electronic mechanisms governing their elastic properties.

## II. AB-INITIO CALCULATIONS

We have performed the electronic structure calculations using the PWscf code of the Quantum Espresso (QE) software package. [12] The unit cell consisted of 8 atoms for all the materials considered. We used the generalized gradient approximation (GGA) Perdew-Burke-Ernzerhof (PBE) functional for exchange correlation. [13] The interactions of valence electrons with the atomic nuclei and core electrons are described by pseudopotentials taken from the open-source Standard Solid State Pseudopotentials (SSSP PBE Efficiency v1.3.0) library. [14] All calculations were performed within a non-spin-polarized framework, assuming a spin-degenerate electronic configuration, in order to provide a consistent description of the complete series and to analyze the general trends in bonding and electronic structure. The electron wave functions and density are represented using plane waves. The corresponding kinetic energy and charge density cutoffs were chosen according to the recommended values of the SSSP pseudopotential library, using the values corresponding to the most demanding pseudopotential in each system. For instance, for Fe-containing systems, the kinetic energy cutoffs for the wave functions and the charge density were set to 90 Ry and 1080 Ry, respectively. Brillouin zones were mapped using a Monkhorst-Pack mesh of $20 \times 20 \times 20$ k points. The convergence threshold was set to $10^{-8}$ Ry for the total energy and to $10^{-6}$ Ry/Bohr for the forces on atoms. The elastic coefficients were calculated using the Thermo_pw code. [15] A set of independent strains was generated, each including 10 different small deformation values, with ionic positions relaxed for each case. The stress in the strained structures was obtained using self-consistent calculations. The resulting strain–stress relationships were fitted with a quartic polynomial, and the elastic coefficient was obtained as the derivative at zero strain.

The calculations show good agreement with the lattice parameters (generally well below 1%) and elastic coefficients (generally within 10%) reported from GGA simulations in the literature for most compounds and phases. [1, 2, 16-24] The elastic constants are also in very good agreement with HSE06 calculations for the particular case of GaN. [25] As expected, our calculated elastic constants are systematically smaller than those obtained using the LDA. This trend is consistent with the well-known tendency of the LDA to underestimate equilibrium lattice parameters by approximately 2% compared to GGA, leading to stiffer predicted elastic properties. [2, 18, 26]. Furthermore, there is reasonably good agreement with the experimentally determined elastic coefficients for w-GaN, rs-TiN, rs-VN, and rs-ScN. [27, 28, 29]

The CrN, MnN, and FeN compounds represent an exception to the overall accuracy observed throughout the series, notably in the $C_{44}$ elastic coefficient. This deviation is related to the fact that the present calculations were performed without considering spin polarization, whereas previous studies have demonstrated that magnetic ordering, particularly in the rocksalt phases, plays a key role in the stabilization and elastic properties of these compounds. [4, 24, 30, 31] Nevertheless, the use of a consistent non-spin-polarized framework across the complete series allows us to identify and analyze the general trends associated with the electronic structure evolution. The present approach is therefore intended primarily for the qualitative understanding of the relationship between bonding characteristics and elastic trends, rather than for providing highly accurate quantitative predictions for magnetic systems.  A table comparing the calculated lattice parameters and elastic constants with previously reported calculations, as well as with the available experimental data, is provided in the Supplementary Information.

## III. ROCKSALT IV-PERIOD NITRIDES

Rocksalt is a cubic crystal structure in which each atom has six first-nearest neighbors. This structure is typical of transition metal nitrides such as ScN and TiN, as its symmetry stabilizes the spatial distribution of d-electrons. In addition to the first-nearest neighbors, there are twelve second-nearest neighbors that also play a significant role in the bonding characteristics of the compound.

To gain insight into the chemical bonding of transition metal nitrides, it is instructive to examine the specific case of ScN. Figure 1 presents the total density of states (DOS) and the projected (p)DOS onto the atomic orbitals of rocksalt-ScN (rs-ScN). The DOS reveals a semicore state arising from the N(2s) orbital, followed by a complex valence band primarily composed of N(2p) and Sc(3d) states. The conduction band (CB) is mainly derived from Sc(3d) orbitals.

Our calculations predict rs-ScN to exhibit semimetallic behavior, with the Fermi level located precisely at the point where the valence and conduction bands intersect. However, this is a well-known artifact associated with the delocalization of d-electrons in standard density functional theory (DFT) methods. Experimentally, rs-ScN exhibits an indirect band gap of 1.1 eV. [32] Previous works have employed more accurate approaches based on hybrid functionals to better reproduce this band gap. [23] Nevertheless, despite the known limitations of PBE in describing band gaps, the calculated electronic structure provides a consistent qualitative basis for interpreting the bonding characteristics and their relationship to the elastic properties.

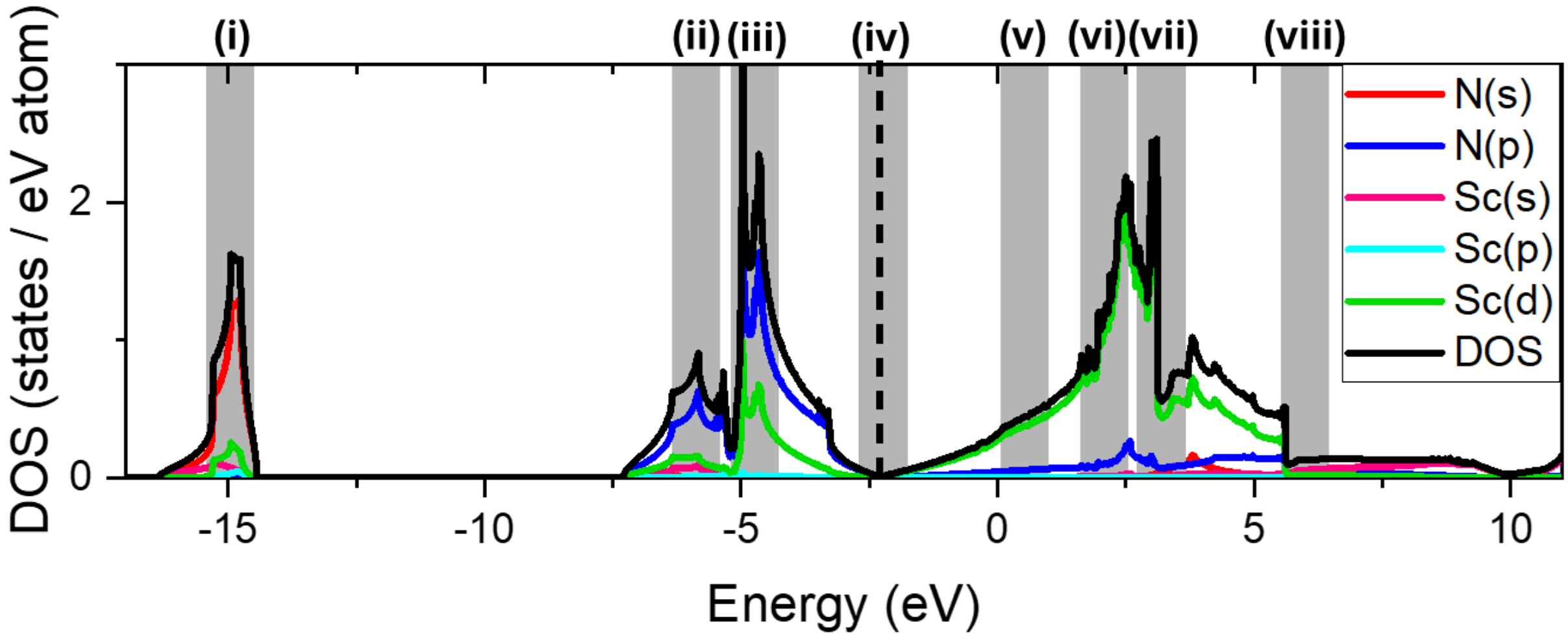

***FIG. 1.*** *Total and projected density of states (DOS and pDOS) for rocksalt ScN, showing contributions from Sc and N atomic orbitals. Various energy ranges are highlighted with gray shading to indicate the energy windows corresponding to the integrated local density of states (ILDOS) maps shown in Fig. 2.*

To gain a deeper understanding of the density of states, integrated local density of states (ILDOS) maps of the electronic density, corresponding to the energy windows highlighted in Figure 1, are presented in Figure 2. This spatial representation allows for the identification of the character and bonding nature of the electronic states within distinct energy ranges.

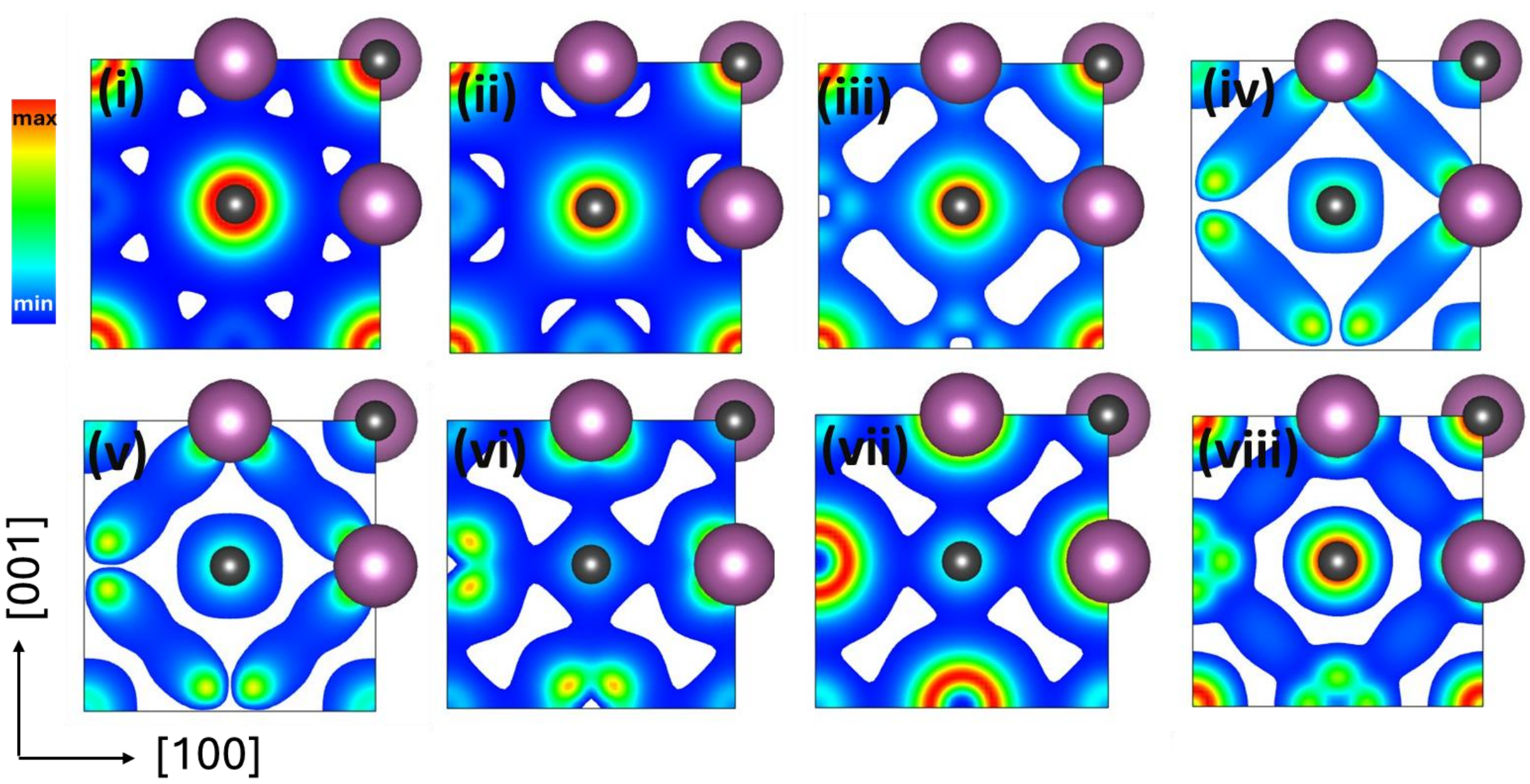

***FIG. 2.*** *Integrated local density of states (ILDOS) maps for rocksalt ScN, corresponding to the gray-shaded energy regions in Fig. 1. Each panel visualizes the spatial distribution of electronic states within a specific energy window, illustrating the character of the states. An additional representation using three-dimensional isosurfaces is provided in the Supplementary Information.*

Region (i) shows a semicore N(2s) level centered around –15 eV, which plays a negligible role in bonding due to its highly localized character and minimal interplay with Sc orbitals. Region (ii) represents the lower part of the valence band, ranging from –7.5 to –5 eV. It displays constructive overlap between N(2p), Sc(3d) and Sc(4s) states, which highlight the bonding character of the interaction. Region (iii) captures the upper part of the valence band, from –5 to –2.5 eV. It exhibits a strongly directional constructive interaction between N(2p) orbitals and the Sc(3d) states pointing directly towards the nitrogen atoms ($e_g$ symmetry: $dz^2$ and $dx^2$-$y^2$). The occupied states in regions (ii) and (iii) contribute to the formation of σ-like bonds between Sc and N atoms. These bonding interactions are crucial for the structural stability and mechanical properties of rs-ScN.

Above the Fermi level, the conduction band begins, predominantly composed of unoccupied Sc(3d) states. Regions (iv) and (v) reveal constructive overlap among Sc(3d) orbitals, which are pointing diagonally between nitrogen atoms ($t_{2g}$ symmetry: dxy, dxz, dyz). The states in regions (iv) and (v) exhibit a σ-like bond character between second-nearest Sc neighbors (denoted $\sigma_{2nd}$). Although these states are unoccupied and therefore do not contribute directly to the elastic properties of rs-ScN, their bonding character reveals the unconventional nature of its electronic band gap. Since both the valence and conduction bands in rs-ScN exhibit bonding character, a nontrivial dependence of the band gap on pressure and temperature can be expected.

Appearing in the 1.5 to 4 eV range, regions (vi) and (vii) correspond to antibonding interactions ($\sigma^*_{2nd}$) between second-nearest neighbor Sc(3d) orbitals with $t_{2g}$ symmetry. Due to the relatively weak coupling between second-neighbor orbitals, the bonding–antibonding splitting is small, and these states merge into a broad, continuous band.

At even higher energies, antibonding σ-like states (σ*) are observed in region (viii), resulting from the interaction between N(2p) orbitals and Sc(3d, 4s) states. The decoupled contributions from $e_g$ and $t_{2g}$ d-orbital symmetries in the rocksalt structure simplify the interpretation of these states. However, in region (vii), a small component of constructive bonding between first-nearest neighbors can still be identified.

Based on this detailed ILDOS analysis of the spatial distribution of electronic states in ScN, we can extend our understanding to the DOS across the entire fourth period of rocksalt nitride compounds. Figure 3 presents the DOS for these nitrides, ranging from potassium nitride (KN) to germanium nitride (GeN), with the nitrogen 2s semicore level used as a common energy reference.

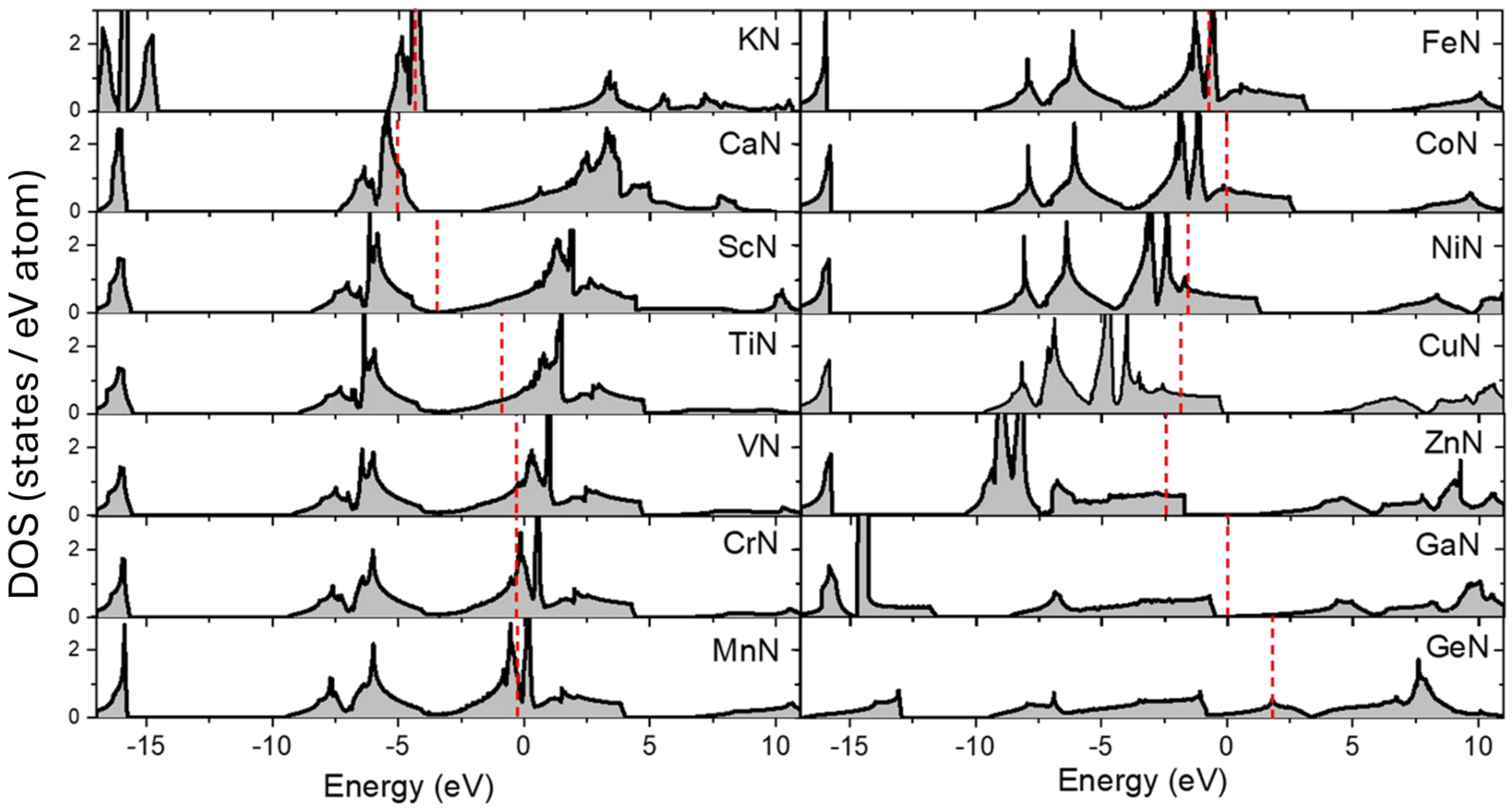

***FIG. 3.*** *Total density of states (DOS) for rocksalt nitrides of IV-period elements from K to Ge. The plot illustrates the evolution of the electronic structure across the period. The energy scale has been aligned such that the maximum of the N(2s) semicore state is fixed. The pDOS for each of the compounds presented can be found in the Supplementary Information.*

The strong similarity in the DOS profiles from KN to CuN is noteworthy, not only in terms of the energy level arrangement but also in the spatial distribution of electronic states within the crystal structures. This recurring pattern allows for a qualitative interpretation of the elastic properties of these materials. One can adopt a simplified view in which the overall electronic structure remains relatively consistent across the series, while the Fermi level progressively rises with increasing atomic number (Z) of the metal, reflecting the greater number of electrons contributed by the metal atoms.

The elastic coefficients $C_{11}$, $C_{12}$ and $C_{44}$ are plotted as functions of the atomic number Z in figures 4(a), 4(b), and 4(c), respectively. We observe that both $C_{11}$ and $C_{12}$ exhibit an initial increase followed by a decrease, with their respective maxima occurring in VN and CoN. In contrast, $C_{44}$ shows a more complex trend, including negative values for KN, MnN, and FeN, which violate the Born stability criteria. Additionally, the bulk modulus is presented as a function of the relaxed lattice parameter in Figure 4(d). The lattice parameter decreases and bulk modulus increases showing a negative correlation from KN to CrN. MnN and FeN show the same lattice parameter with similar bulk modulus. From CoN to ZnN the lattice parameter increases again, and the bulk modulus decreases until GaN breaks the trend.

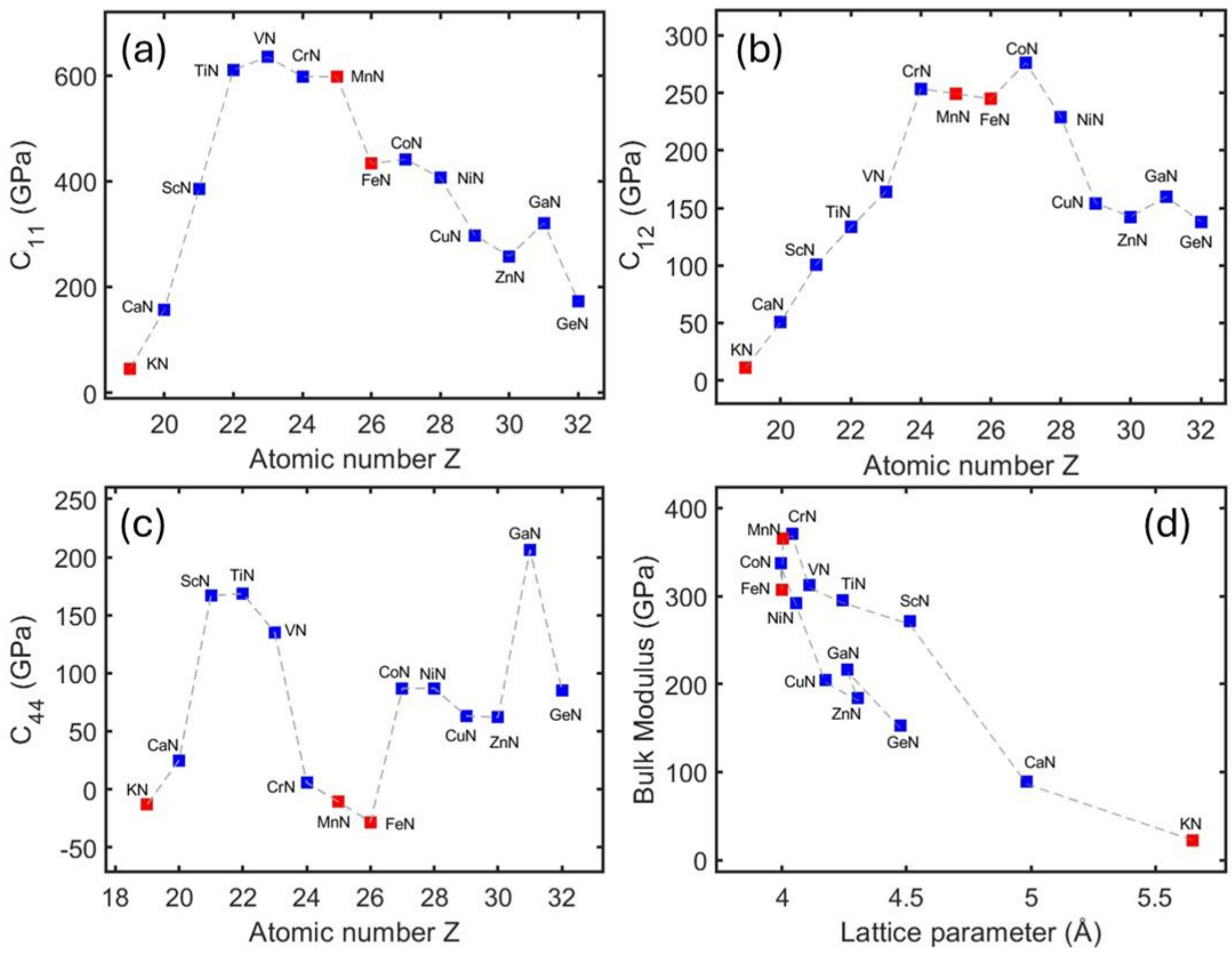

***FIG. 4.*** *Elastic coefficients of rocksalt IV-period nitrides. Panels (a), (b), and (c) show $C_{11}$, $C_{12}$, $C_{44}$, respectively, plotted as a function of atomic number Z. Panel (d) shows the bulk modulus as a function of the equilibrium lattice parameter. The elastically stable compounds are displayed in blue, and the unstable ones in red.*

The evolution of the lattice parameter and elastic coefficients can be understood from the perspective of the DOS. KN exhibits a small bulk modulus and presents an unstable behavior due to the absence of d-states participation in bonding. CaN shows an increased bulk modulus and overall elastic stability, arising from the promotion of Ca s-electrons into d-states, which allow $\sigma$ bonds to strengthen. To further refine this analysis, we examine the partial orbital occupations for each TMN rocksalt crystal (Löwdin charges), as summarized in table 1. ScN continues the trend of lattice contraction and stiffening, as its d-electron shell primarily consist of states associated with the $\sigma$ bonding interaction between Sc(d) and N(p) states in region (iii), consistent with the observed increase in $e_g$ symmetry d-electron occupation. With increasing atomic number *Z*, the $t_{2g}$ symmetry d-states are rapidly populated, strengthening the $\sigma_{2nd}$ bonds up to CrN, while $e_g$ symmetry d-electrons are filled more gradually but still provide a small contribution to $\sigma$ bonds. Consequently, the crystal undergoes further stiffening accompanied by a continued reduction in lattice parameter.

From CrN onward, d-states responsible for $\sigma^*_{2nd}$ antibonds begin to fill, and three elements deviate from the previously observed trend, exhibiting more complex behavior. MnN, FeN, and CoN display very similar lattice parameters. This can be attributed to the moderate increase in $e_g$ electron occupation, which strengthen $\sigma$ bonds, coupled with a strong increase in $t_{2g}$ electrons populating $\sigma^*_{2nd}$ antibonds, effectively balancing their overall impact. The bulk moduli of these compounds is relatively similar, with minor differences arising from the interplay between first- and second-neighbor interactions. Furthermore, when the lattice parameter becomes sufficiently small, *d–d* interactions can destabilize the crystal against shear, as observed in FeN and MnN. Further increasing *Z* slightly raises the bulk modulus for CoN, likely because the partial filling of $\sigma^*_{2nd}$ antibonding states restabilizes the crystal against shear by weakening second neighbour interactions. Beyond CoN, further increases in *Z* reduce the stiffness and expand the lattice parameter up to ZnN, due to the progressive filling of σ* antibonds, as observed in region (viii) of the Figure 2.

Regarding the overall trends of the elastic coefficients, the behavior of $C_{11}$ surprisingly correlates with the filling of $\sigma_{2nd}$ and $\sigma^*_{2nd}$ states via the occupation of $t_{2g}$ orbitals, suggesting that second-neighbor interactions play a significant role in determining this elastic coefficient. In contrast, the trend in $C_{12}$ correlates more strongly with the occupation of σ and σ* states through orbitals with $e_g$ symmetry, making it primarily dependent on first-neighbor interactions. The behavior of $C_{44}$ is more complex, reflecting an interplay between $t_{2g}$ electron population and lattice parameter: its values turn negative for the maximum occupation of $\sigma_{2nd}$ and the smallest lattice parameter. Directional σ N(p)–Me(d) bonding states are expected to provide resistance to shear deformation due to their strong directional character. In contrast, second-neighbor d-d bonding states are more sensitive to shear-induced changes in metal–metal distances and orbital overlap. Under shear deformation, the strengthening of d-d interactions can lower the energy of these states, potentially leading to a negative contribution to the shear modulus when they are occupied. [5,6] Consequently, a sufficiently small lattice parameter combined with maximal occupation of $\sigma_{2nd}$ bonds destabilizes MnN and FeN against shear and renders CrN nearly unstable. Nonetheless, spin-polarized calculations indicate that CrN, MnN, and FeN are magnetically stabilized against shear. [24, 30]

Beyond this simple picture, variations in the density of states with increasing Z can also contribute to the elastic properties. A relatively minor effect is the increased splitting of the σ bonding and σ* antibonding states (between regions (ii) and (vii)) as the lattice parameter decreases, driven by an enhanced transfer integral. This splitting reaches a maximum for CoN and FeN. A more significant effect occurs at higher Z, where the d-states shift to lower energies due to the stronger nuclear attraction. Although this trend persists across the period, it becomes particularly relevant from ZnN onward, as the main d-state contributions decouple from regions (iv)–(vii) and remain concentrated in regions (ii) and (iii). In ZnN, the d-shell is fully occupied, and d-state participation in bonding is strongly reduced.

For GaN, the d-states appear as semicore levels with modest interaction with the N(2s) states due to energy overlap. However, because the d-shell is filled, the bonding pattern changes substantially, and the simple picture of a continuous filling of similar states no longer applies. In rs-GaN, the bonding is instead dominated by the strong participation of Ga(p) states. This shift explains the overall increase in the elastic coefficients and bulk modulus of GaN compared with the decreasing trend observed in the preceding TMNs. Further raising the Fermi level in GeN leads to the filling of antibonding states, which decreases the stiffness and increases the lattice parameter. It is also worth noting that the band-gap nature in rs-ScN and rs-GaN is different: in the latter, it originates from the splitting of first-neighbor σ bonding and antibonding states.

*__TABLE 1.__ Partial electronic occupations of atomic orbitals (in elemental charge units) for each element in the rocksalt IV-period transition metal nitrides. An asterisk (*) denotes compounds that are elastically unstable. The populations may exceed the nominal occupation limits of individual orbital as multiple radial shells are considered.*

| Rocksalt | Sc | Ti | V | Cr | Mn* | Fe* | Co | Ni | Cu | Zn |
|---|---|---|---|---|---|---|---|---|---|---|
| **N** | 5.90 | 6.03 | 5.82 | 5.91 | 5.18 | 5.28 | 5.14 | 5.15 | 5.46 | 5.40 |
| **s** | 1.58 | 1.66 | 1.57 | 1.64 | 1.31 | 1.35 | 1.33 | 1.44 | 1.52 | 1.35 |
| **p** | 4.32 | 4.38 | 4.25 | 4.27 | 3.87 | 3.93 | 3.80 | 3.71 | 3.93 | 4.05 |
| **Me** | 10.05 | 10.85 | 12.12 | 12.97 | 14.79 | 15.69 | 16.83 | 17.81 | 18.53 | 19.57 |
| **s** | 2.36 | 2.42 | 2.37 | 2.42 | 2.21 | 2.21 | 2.30 | 2.39 | 2.37 | 2.11 |
| **p** | 5.99 | 5.99 | 5.99 | 5.99 | 6.93 | 6.90 | 7.02 | 7.04 | 6.87 | 6.70 |
| **d** | 1.70 | 2.44 | 3.76 | 4.56 | 5.66 | 6.59 | 7.51 | 8.38 | 9.29 | 9.88 |
| **d-$e_g$** | 0.46 | 0.47 | 0.63 | 0.72 | 0.86 | 0.91 | 0.97 | 1.29 | 1.67 | 1.87 |
| **d-$t_{2g}$** | 0.26 | 0.50 | 0.83 | 1.04 | 1.31 | 1.59 | 1.85 | 1.93 | 1.98 | 1.89 |

## II. ZINCBLENDE IV-PERIOD NITRIDES

Zincblende is a cubic crystal structure in which each atom is coordinated by four first-nearest neighbors. This structure is uncommon in conventional nitrides, where the ionic character of fourfold coordination typically favors the wurtzite phase energetically. Nevertheless, studying the zincblende structure allows the role of coordination in determining the elastic properties of nitrides to be assessed, without the additional complexity arising from the structural non-idealities characteristic of wurtzite. Unlike the rocksalt structure, where the second-nearest neighbors are positioned along the diagonals in between the first-nearest neighbors, in zincblende the second-nearest neighbors are partially screened by the first neighbors, thereby making first-neighbor interactions the dominant factor in governing the elastic properties.

Figure 5 shows the total DOS and the pDOS onto the atomic orbitals of zincblende ScN (zb-ScN). The DOS exhibits a structure similar to that of the rocksalt phase for energies below the Fermi level, with a semicore state arising from N(2s) and a complex valence band primarily formed by interactions between N(2p) and Sc(3d) states. In contrast, the change in coordination strongly alters the Sc(3d)–Sc(3d) second-neighbor interactions, leading to the opening of a band gap in zb-ScN and modifying the conduction-band structure through hybrid Sc(3d)–Sc(3d) interactions with mixed $t_{2g}$ and $e_g$ character.

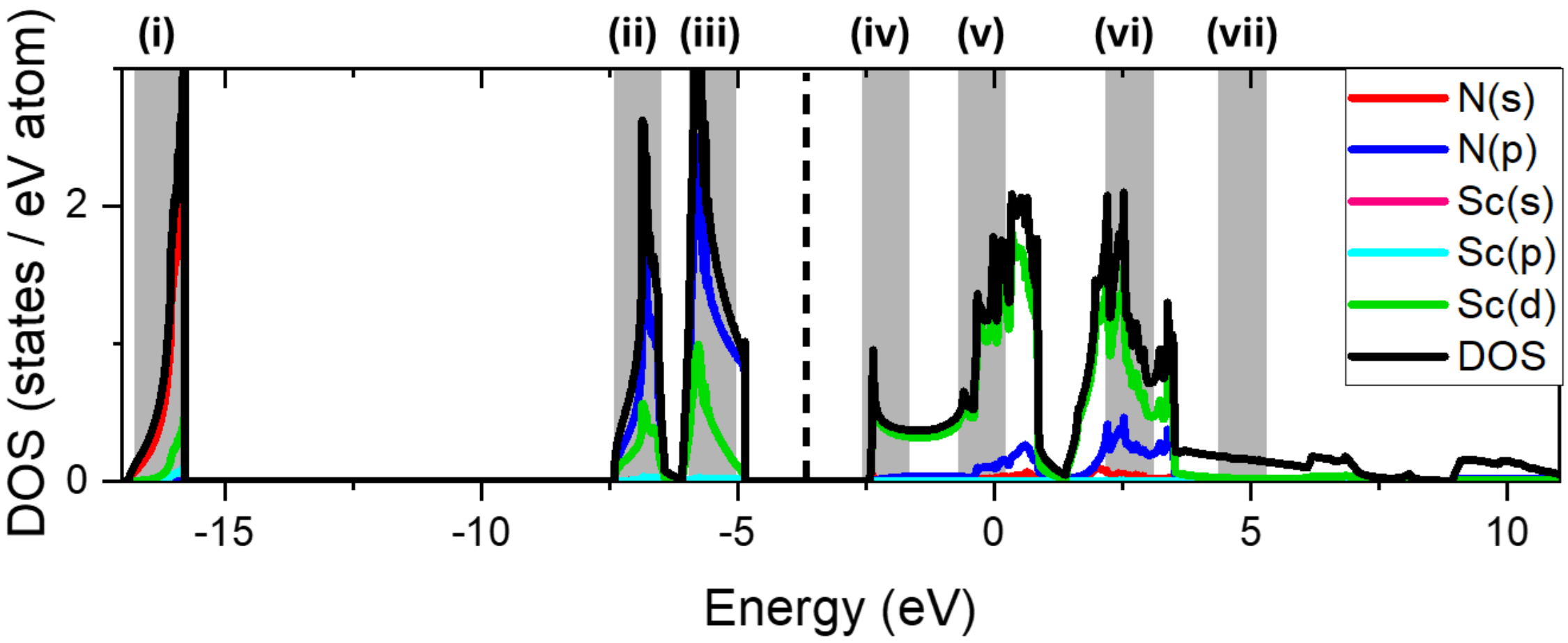

***FIG. 5.*** *Total and projected density of states (DOS and pDOS) for zincblende ScN, showing contributions from Sc and N atomic orbitals. Various energy ranges are highlighted with gray shading to indicate the energy windows corresponding to the integrated local density of states (ILDOS) maps shown in Fig. 6.*

Further insight into the electronic density configuration can be obtained by examining sections of the ILDOS on the (110) plane of zb-ScN shown in Figure 6, corresponding to the energy windows highlighted in Figure 5. This spatial representation enables the identification of the character and bonding nature of the electronic states across distinct energy ranges, as summarized below.

Region (i) corresponds to a semicore N(2s) level centered around –16 eV. This state is highly localized and exhibits negligible interaction with Sc orbitals, thereby playing no significant role in bonding, similar to what is observed in the rocksalt phase.

The valence band consists of two regions: (ii) from –7.5 to –6.25 eV and (iii) from –6.25 to –5 eV, which display very similar features. In both cases, constructive overlap occurs between N(2p) orbitals and Sc(3d) states with hybrid $e_g$ and $t_{2g}$ character, confirming their bonding nature. These occupied states contribute to the formation of σ-like Sc–N bonds, which are essential for the structural stability and mechanical properties of zb-ScN. However, due to the nontrivial symmetry matching between the atomic coordination and the *d* and *p* orbitals, charge-density lobes appear misaligned with the direct Sc–N bonds. This imperfect symmetry matching helps explain the higher energy of the zincblende phase compared with the rocksalt phase in TMNs. In fact, the roles d-electrons with $e_g$ and $t_{2g}$ symmetries are much more coupled in zincblende, as both contribute simultaneously to the first-neighbor N(p)–Sc(d) interactions and to the second-neighbor Sc(d)–Sc(d) interactions.

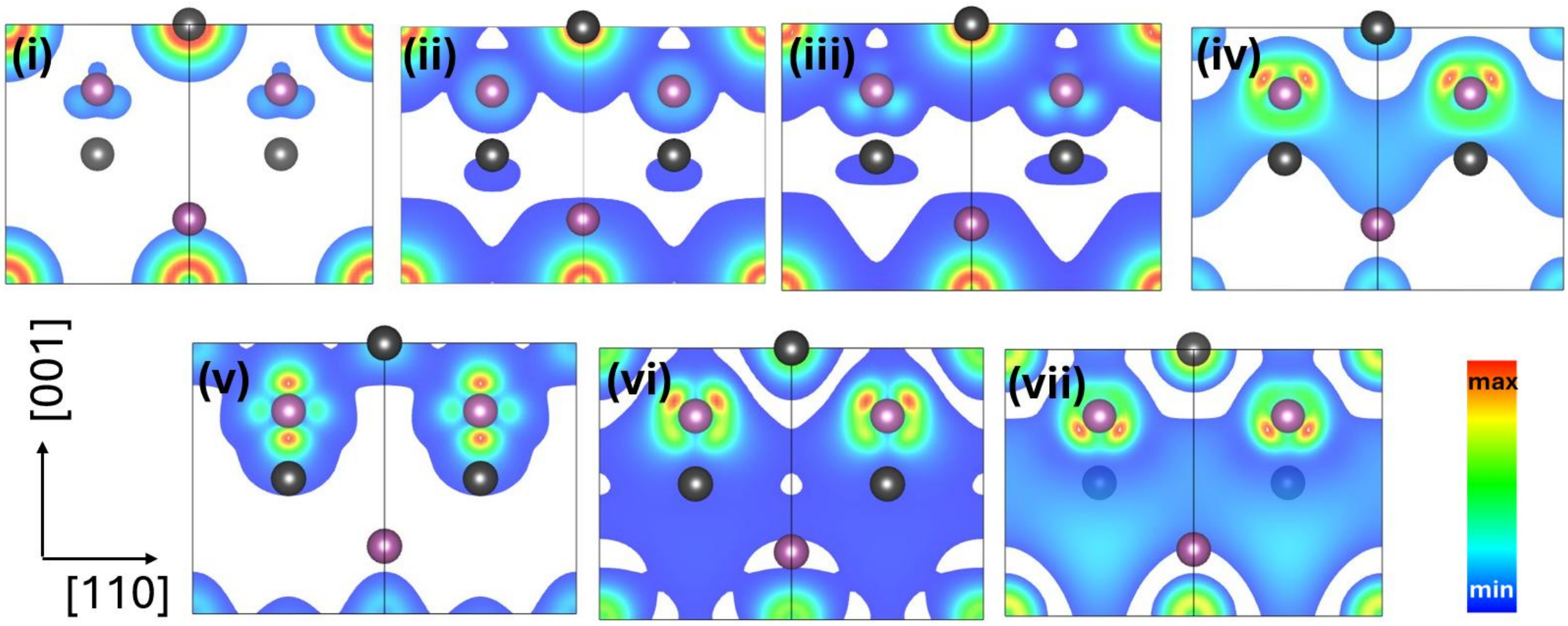

***FIG. 6.*** *Integrated local density of states (ILDOS) maps for zincblende ScN, corresponding to the gray-shaded energy regions in Fig. 5. Each panel visualizes the spatial distribution of electronic states within a specific energy window, illustrating the character of the states. An additional representation using three-dimensional isosurfaces is provided in the Supplementary Information.*

The conduction band is primarily composed of unoccupied Sc(3d) states, with additional contributions from N(2p) orbitals at higher energies. The CB minimum in region (iv) reveals highly delocalized overlap of Sc(3d) orbitals (with slight $t_{2g}$ character) from second-neighbor interactions, projecting charge-density lobes into the interstitial regions of the ScN lattice. At higher energies (–1 to 0 eV), region (v) is characterized by Sc(3d) orbitals with slight $e_g$ character, also projecting into interstitial sites and producing weak lateral overlap with N(2p) orbitals.
Regions (vi) and (vii), appearing at 2.5–3.5 eV and 4.5–5.5 eV, likely correspond to antibonding interactions, predominantly of Sc(3d) orbitals with $t_{2g}$ symmetry, and N orbitals with *s* and *p* character.

Based on the detailed ILDOS analysis of the spatial distribution of electronic states in zb-ScN, we can extend our understanding to the DOS across the entire IV-period of rocksalt nitride compounds. Figure 7 shows the DOS for these nitrides, from potassium nitride (KN) to germanium nitride (GeN), using the nitrogen 2s semicore level as a common reference.
As previously observed for rocksalt, the DOS profiles from KN to CuN display strong similarities, not only in the arrangement of energy levels but also in the spatial distribution of electronic states within the crystal structures. This recurring pattern enables a qualitative interpretation of the elastic properties of these materials. A simplified perspective can thus be adopted, in which the overall electronic structure remains relatively consistent across the series, while the Fermi level rises progressively with increasing atomic number of the metal, reflecting the greater electron contribution from the metal atoms. Nevertheless, subtle effects must also be considered. The σ-bond splitting increases slightly with decreasing lattice parameters, but a more significant effect is the progressive lowering of the *d*-states at higher *Z*, driven by the stronger nuclear attraction.

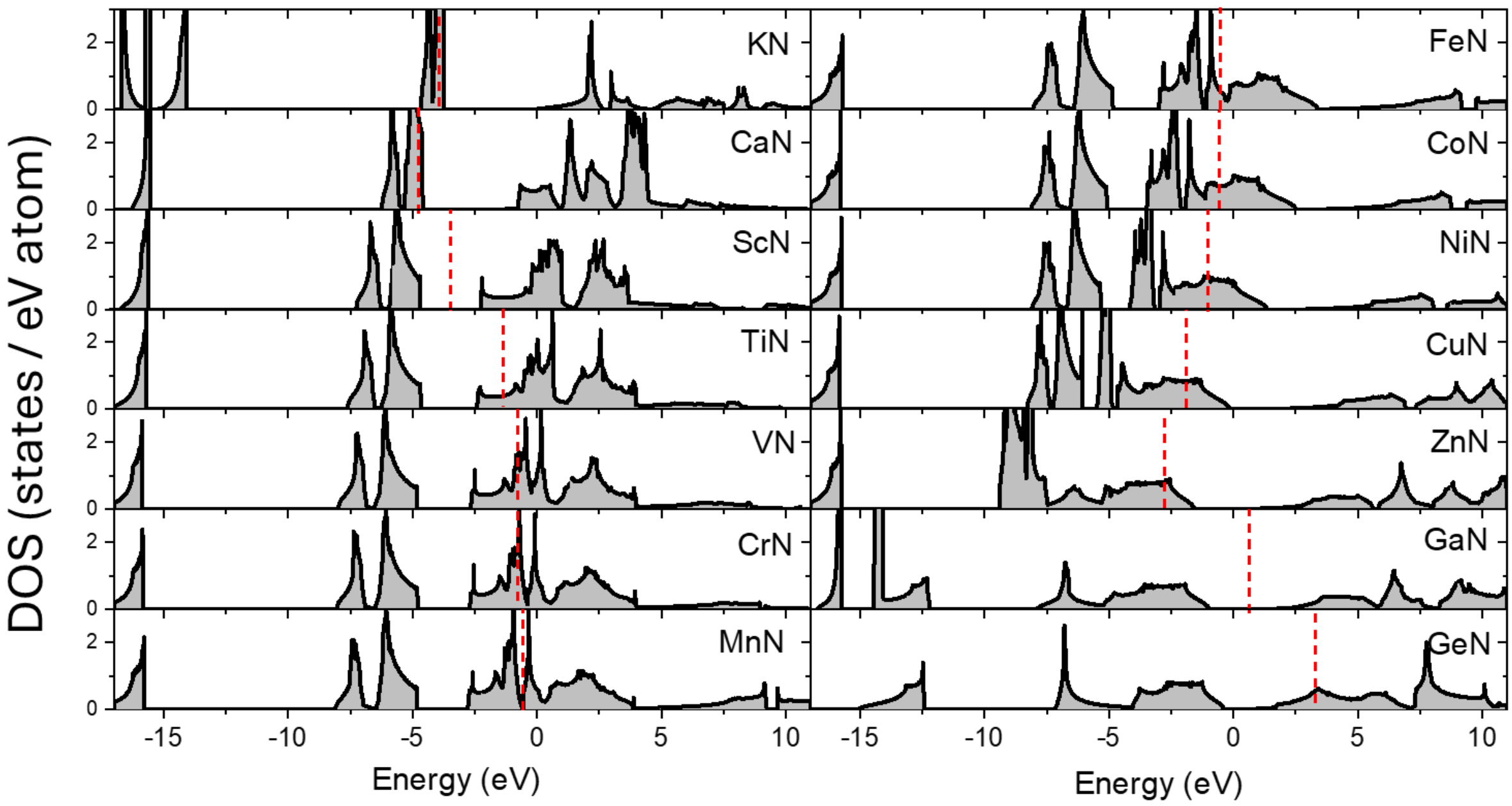

***FIG. 7.*** *Total density of states (DOS) for zincblende nitrides of IV-period elements from K to Ge. The plot illustrates the evolution of the electronic structure across the period. The energy scale has been aligned such that the maximum of the N(2s) semicore state is fixed. The pDOS for each of the compounds presented can be found in the Supplementary Information.*

The elastic coefficients $C_{11}$, $C_{12}$ and $C_{44}$ are plotted as functions of the atomic number Z in figures 8(a), 8(b), and 8(c), respectively. Both $C_{11}$ and $C_{12}$ exhibit an initial increase followed by a decrease, with their maxima occurring at similar Z for MnN. In contrast, $C_{44}$ shows a more complex evolution, including negative values for KN, CrN, and GeN, which violate the Born stability criteria. The bulk modulus, plotted as a function of the relaxed lattice parameter in Figure 8(d), shows a negative correlation up to MnN: as the lattice parameter decreases, the bulk modulus increases. MnN and FeN exhibit nearly identical lattice parameters and similar bulk moduli. From FeN to ZnN, the lattice parameter increases again while the bulk modulus decreases, until GaN breaks this trend. Overall, zincblende compounds are less stiff than their rocksalt counterparts, primarily due to their lower coordination number, which reduces the number of strong nearest-neighbor interactions per atom.

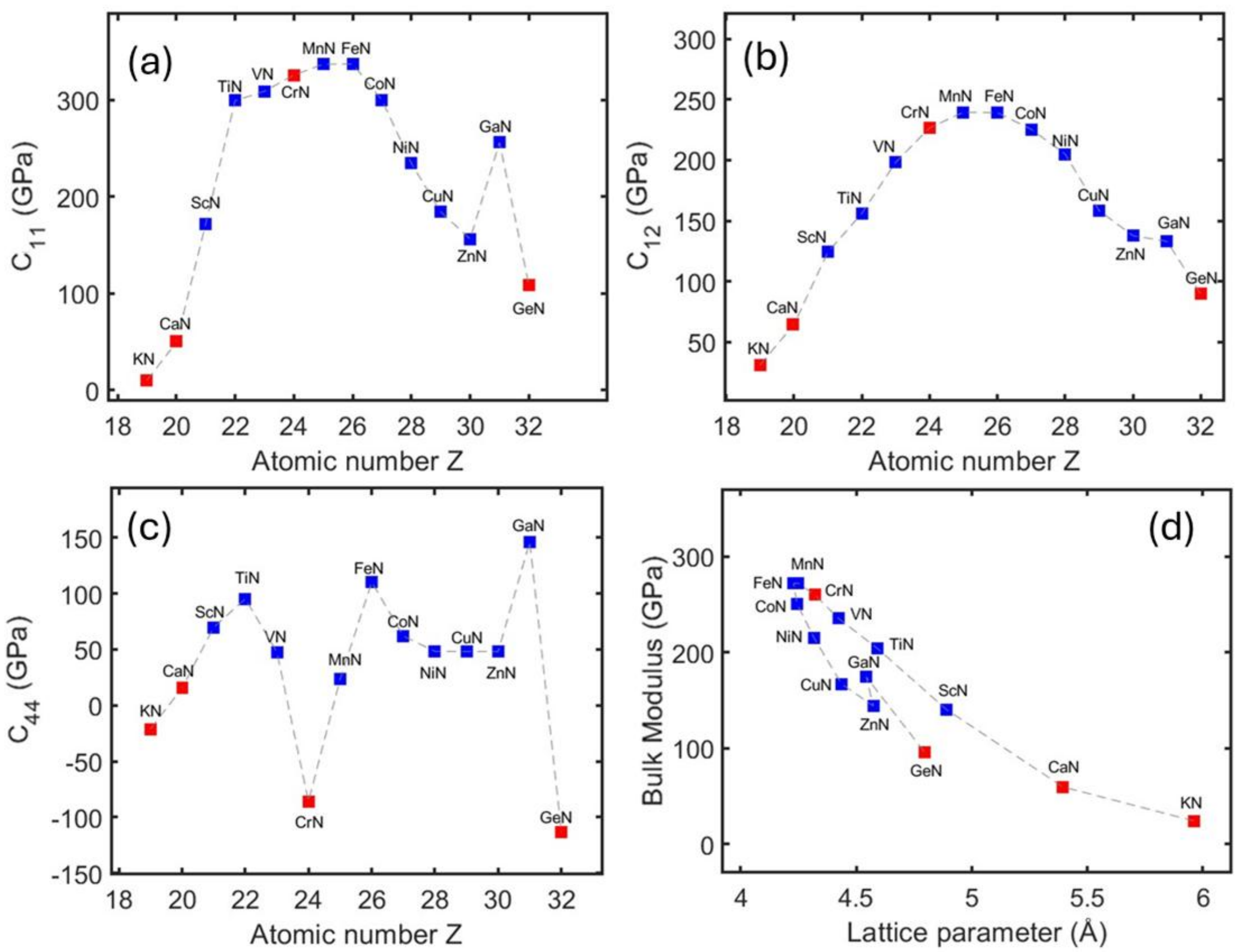

***FIG. 8.*** *Elastic coefficients of zincblende IV-period nitrides. Panels (a), (b), and (c) show $C_{11}$, $C_{12}$, $C_{44}$, respectively, plotted as a function of atomic number Z. Panel (d) shows the bulk modulus as a function of the equilibrium lattice parameter. The elastically stable compounds are displayed in blue, and the unstable ones in red.*

The evolution of the lattice parameter and elastic coefficients in zincblende nitrides can be interpreted from the DOS. KN exhibits a very small bulk modulus and mechanical instability, reflecting the absence of d-state participation in bonding. In CaN, the promotion of Ca s-electrons into d-states enables the formation of σ bonds, leading to elastic stabilization and an increased bulk modulus. Further insights can be obtained from the orbital occupations in TMNs (Löwdin charges), summarized in Table 2.

ScN continues the trend of stiffening and lattice contraction as the d-electron shell fills mainly states associated with the $\sigma$ bonding interaction between Sc(d) and N(p) states in region (iii) of Figure 6. As Z increases, the $t_{2g}$ states are predominantly filled in TiN, further stiffening the crystal and reducing the lattice parameter due to delocalized, metallic-like bonding between second-neighbor d orbitals. Further stiffening comes from the mild constructive overlap between N(p) and d-states with $e_g$ character up to FeN, as seen in region (v) of Figure 6. From there onward, $t_{2g}$ electrons continue to fill up to ZnN, occupying antibonding states first in region (vi) and then in region (vii) of Figure 7, which explains the progressive decrease in stiffness and increase in lattice parameter with higher Z.

***TABLE 2.*** *Partial electronic occupations of atomic orbitals (in elemental charge units) for each element in the zincblende IV-period transition metal nitrides. An asterisk (*) denotes compounds that are elastically unstable. The populations may exceed the nominal occupation limits of individual orbital as multiple radial shells are considered.*

| **Zincblende** | **Sc** | **Ti** | **V** | **Cr*** | **Mn** | **Fe** | **Co** | **Ni** | **Cu** | **Zn** |
|---|---|---|---|---|---|---|---|---|---|---|
| **N** | 5.93 | 6.01 | 5.77 | 5.82 | 5.11 | 5.22 | 5.09 | 5.10 | 5.45 | 5.55 |
| **s** | 1.60 | 1.66 | 1.59 | 1.63 | 1.31 | 1.36 | 1.35 | 1.41 | 1.49 | 1.51 |
| **p** | 4.34 | 4.35 | 4.18 | 4.19 | 3.80 | 3.86 | 3.74 | 3.69 | 3.96 | 4.04 |
| **Me** | 10.01 | 10.81 | 12.14 | 13.02 | 1486 | 15.75 | 16.89 | 17.87 | 18.53 | 19.44 |
| **s** | 2.41 | 2.46 | 2.42 | 2.46 | 2.26 | 2.27 | 2.36 | 2.44 | 2.41 | 2.53 |
| **p** | 5.98 | 5.98 | 5.98 | 5.99 | 6.92 | 6.90 | 7.05 | 7.09 | 6.89 | 7.05 |
| **d** | 1.63 | 2.37 | 3.73 | 4.58 | 5.68 | 6.58 | 7.48 | 8.33 | 9.23 | 9.85 |
| **d-$e_g$** | 0.24 | 0.29 | 0.67 | 1.00 | 1.43 | 1.66 | 1.74 | 1.84 | 1.95 | 1.99 |
| **d-$t_{2g}$** | 0.38 | 0.60 | 0.80 | 0.86 | 0.94 | 1.09 | 1.33 | 1.55 | 1.78 | 1.95 |

Regarding the overall trends of each elastic coefficient, contrary to the rocksalt case, the behaviors of $C_{11}$ cand $C_{12}$ are similar across the full row. Both increase up to MnN and FeN, reaching comparable values, and then decrease until the influence of the p-electrons from GaN becomes significant. This difference can be attributed to the reduced impact of second-neighbor bonding in zincblende and the higher symmetry of rocksalt, which leads to a clearer separation between $t_{2g}$ and $e_g$ derived contributions, affecting $C_{11}$ cand $C_{12}$ differently. In zincblende, by contrast, these contributions are more strongly mixed, resulting in more homogeneous trends for $C_{11}$ cand $C_{12}$.

$C_{44}$ exhibits a complex behavior in the zincblende series, becoming negative not only for the extreme compounds (KN and GeN) but also for the central compound CrN. The reduction of $C_{44}$ around CrN can be attributed to the position of the Fermi level within a region where filling of p–d bonding states (region v) destabilizes the crystal against shear. In this energy range, the d-$e_g$ orbitals can overlap more directionally with N(p) orbitals under shear distortion, potentially lowering their energy and thus rendering the structure unstable against shear. This interpretation is based on the calculated equilibrium electronic structure and the orbital character and symmetry of the relevant states; explicit calculations of the electronic structure under shear strain would be required to directly confirm this mechanism. Nonetheless, because second-neighbor d–d interactions are much weaker in zincblende than in rocksalt, the non-magnetic phases of MnN and FeN are comparatively more stable in the zincblende structure [4,24].

Finally, in zb-GaN, the significant overlap of Ga s and p contributions, together with the tetrahedral coordination of the zinc-blende structure, are consistent with substantial $sp^3$ hybridization. This more directional bonding character results in an overall increase in the elastic coefficients, interrupting the decreasing trend observed for the preceding TMNs. In GeN, the further rise of the Fermi level leads to the filling of antibonding states, which reduces stiffness and expands the lattice parameter.

## III. WURTZITE IV-PERIOD NITRIDES

Wurtzite is a hexagonal crystal structure in which each atom is tetrahedrally coordinated by four first-nearest neighbors. Unlike zincblende, the wurtzite phase is the thermodynamically stable form for conventional group-III nitrides such as GaN, as its hexagonal symmetry better accommodates the partially ionic character of the bonds.

The electronic properties of wurtzite differ from those of zincblende. For example, AlN exhibits a direct bandgap in the wurtzite phase, but not in zincblende, due to differences in the Brillouin zone structures [33]. Nevertheless, the charge density distribution in these compounds is very similar, and consequently the elastic properties of ideal wurtzite, particularly the bulk modulus, closely resemble those of zincblende, since they are governed by the same type of bonds and the same coordination environment [34].

Nonetheless, wurtzite crystals are well known to exhibit non-ideal c/a ratios or internal u parameters. Large deviations in these parameters can significantly impact both the structure and the elastic properties, as they introduce asymmetry into the tetrahedrally coordinated bonds. In an extreme case, when $u = 0.5$, the distorted wurtzite structure transforms into a layered hexagonal (lh) phase, featuring three equivalent in-plane bonds and two weaker out-of-plane bonds, thereby creating a new coordination environment.

Therefore, the wurtzite system can be analyzed in comparison with zincblende when the u parameter is close to its ideal value. In contrast, in the layered hexagonal limit, the vertical lattice parameter c approaches values more similar to those of rocksalt crystals, leading to a stiffening of the crystal.

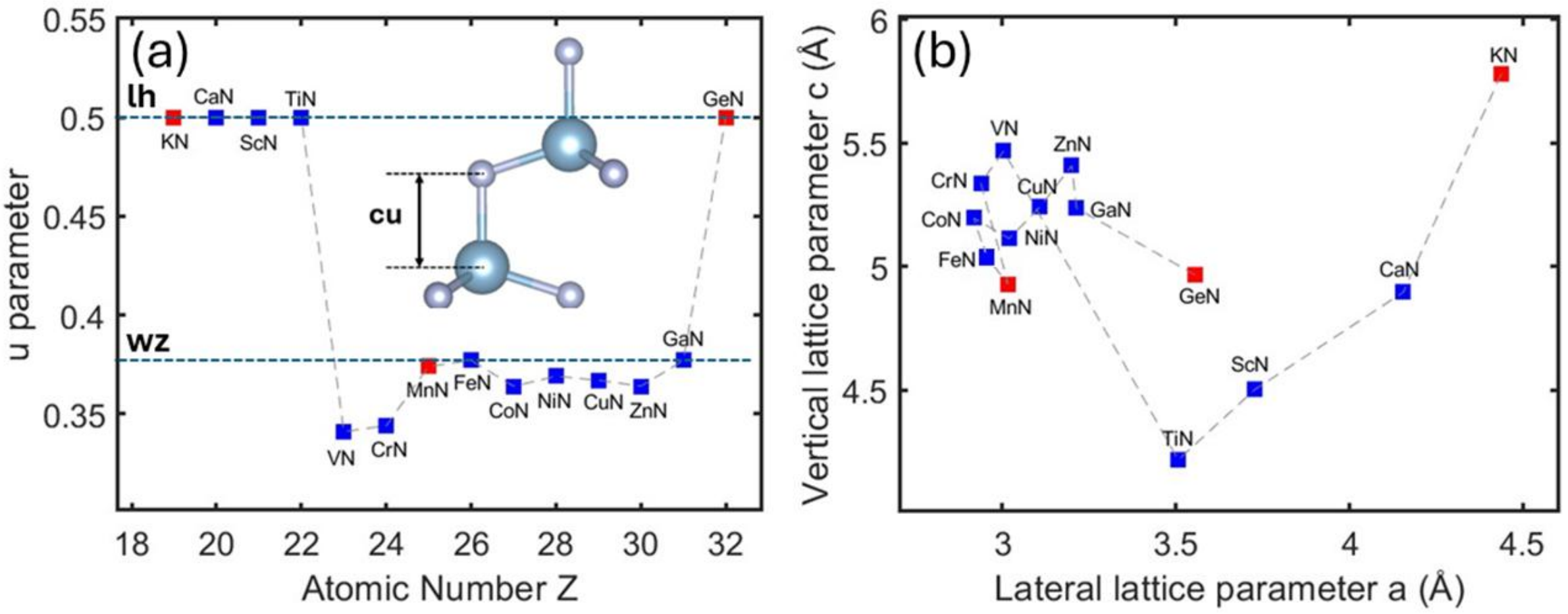

***FIG. 9.*** *Structural parameters of hexagonal IV-period nitrides. (a) Equilibrium u parameter versus atomic number and (b) Vertical lattice parameter c versus lateral lattice parameter a.*

The structural parameters of hexagonal IV-period nitrides are represented in Figure 9 (a) and (b). The first compounds up to TiN adopt layered hexagonal structures with u = 0.5. Regarding the lattice parameters, a clear decreasing trend with Z is observed for these elements, with their vertical lattice parameter resembling that of their rocksalt phase.

The remaining transition-metal nitrides generally exhibit u parameters below the ideal wurtzite value, stabilizing the structure by shortening the vertical bond. The lateral lattice parameter for compounds from VN to NiN slightly varies, while the vertical lattice parameter changes according to the different u values. From NiN to ZnN, the compounds show a clear upward trend in both lattice parameters, following the same tendency as their zincblende counterparts. GaN exhibits a nearly ideal wurtzite structure, whereas GeN reverts to a layered configuration due to destabilization induced by its p electrons.

Figure 10 shows the total DOS and the pDOS onto the atomic orbitals of layered hexagonal ScN. The DOS again exhibits a structure similar to the other crystalline phases for energies below the Fermi level, with a semicore N(2s) state and a complex valence band primarily formed by interactions between N(2p) and Sc(3d) states. In lh-ScN, the lower portion of the valence band is more rocksalt-like than zincblende-like, owing to the participation of Sc(s) electrons.

Nevertheless, the fivefold coordination significantly modifies the Sc(3d)–Sc(3d) second-neighbor interactions present in rocksalt, opening a band gap in lh-ScN that is smaller than in the zincblende phase because of the relatively stronger Sc(3d)–Sc(3d) overlap. The conduction band structure is also more similar to rocksalt; however, the layered hexagonal symmetry does not allow a clear separation between $t_{2g}$ and $e_g$ states. Instead, the d-states are distinguished along the vertical axis ($z^2$), in plane ($x^2$-$y^2$, xy) and diagonal axis (xz, yz).

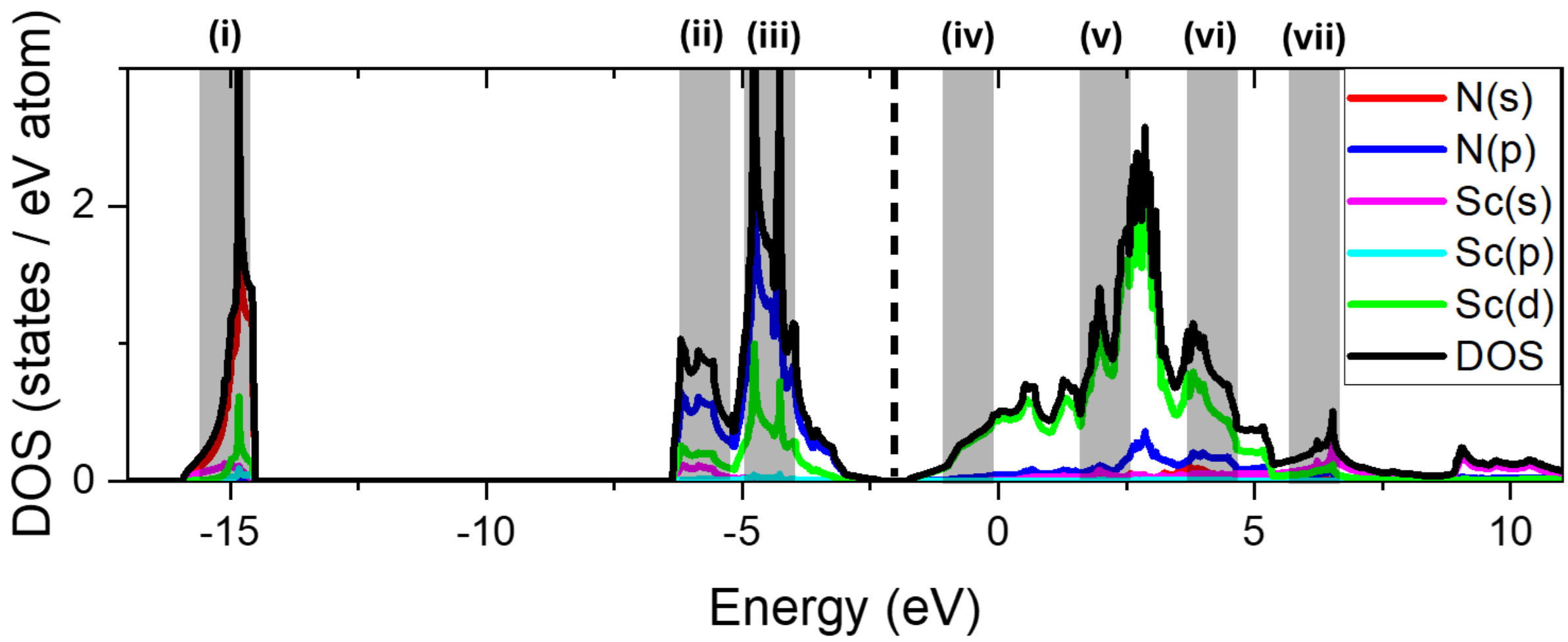

***FIG. 10.*** *Total and projected density of states (DOS and pDOS) for layered hexagonal ScN, showing contributions from Sc and N atomic orbitals. Various energy ranges are highlighted with gray shading to indicate the energy windows corresponding to the integrated local density of states (ILDOS) maps shown in Fig. 11.*

A detailed analysis of the electronic density configuration can be achieved by representing sections of the ILDOS on the a-plane of layered hexagonal ScN, corresponding to the energy windows highlighted in Figure 10 and presented in Figure 11. This spatial representation allows for identification of the character and bonding nature of the electronic states within distinct energy ranges, as described below.

Region (i) corresponds to a semicore N(2s) level centered around –15 eV, which plays a negligible role in bonding due to its highly localized character and minimal interaction with Sc orbitals, similar to its behavior in rocksalt and zincblende phases.

The valence band consists of two regions: (ii) from –6 eV to –5 eV and (iii) from –5 eV to –3 eV. Region (ii) exhibits constructive overlap of Sc(s) and Sc(d) orbitals, both in-plane and along the diagonals, with N(p) orbitals as well as with other Sc(s) and Sc(d) orbitals. Region (iii) is dominated by bonding interactions between vertical and in-plane Sc(d) orbitals and N(p) orbitals. The occupied states in these regions contribute to the formation of σ-like bonds between Sc and N atoms along both vertical and in-plane directions. These bonding interactions determine the structural stability and mechanical properties of lh-ScN.

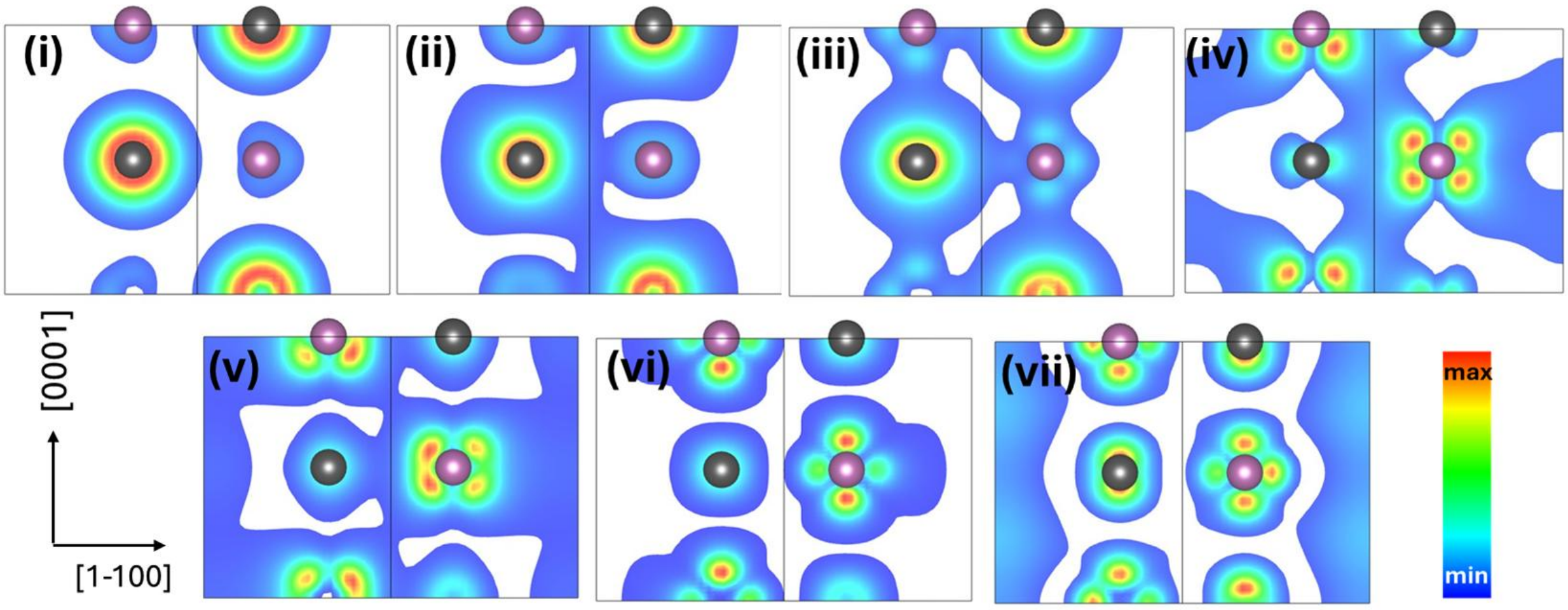

*FIG. 11. Integrated local density of states (ILDOS) maps for layered hexagonal ScN, corresponding to the gray-shaded energy regions in Fig. 10. Each panel visualizes the spatial distribution of electronic states within a specific energy window, illustrating the character of the states. An additional representation using three-dimensional isosurfaces is provided in the Supplementary Information.*

The conduction band is predominantly composed of unoccupied Sc(3d) states, with additional N(2p) contributions at higher energies, similar to other ScN phases. The CB minimum in region (iv) reveals a delocalized constructive overlap between diagonal Sc(3d) orbitals of second neighbors, delocalized within the diagonal interstitial positions of the ScN lattice. At higher energies, region (v) shows the transition from the constructive overlap observed in region (iv) to destructive interactions among the same diagonal d-states. At even higher energies, the destructive overlap between Sc(d) and N(p) states corresponds to sigma antibonding interactions ($\sigma^*$) between N(p) orbitals and Sc(3d) orbitals, with predominantly vertical character in region (vi) and in-plane character in region (vii).

Figure 12 presents the DOS for these nitrides, ranging from potassium nitride (KN) to germanium nitride (GeN), using the nitrogen 2s semicore level as a common reference. As previously observed, the DOS profiles from KN to CuN show similarities in the arrangement of energy levels; however, variations in the u parameter along the period lead to corresponding changes in the spatial distribution of electronic states. By combining the ILDOS analysis of layered hexagonal ScN with the strong resemblance between the DOS of ideal wurtzite and zincblende, we can extend this understanding to the DOS across the IV- period of hexagonal nitride compounds. This allows for a qualitative interpretation of the elastic properties of these materials, with a rising Fermi level as discussed earlier, while taking into account differences in crystal structure and the minor effects of Z and lattice parameters.

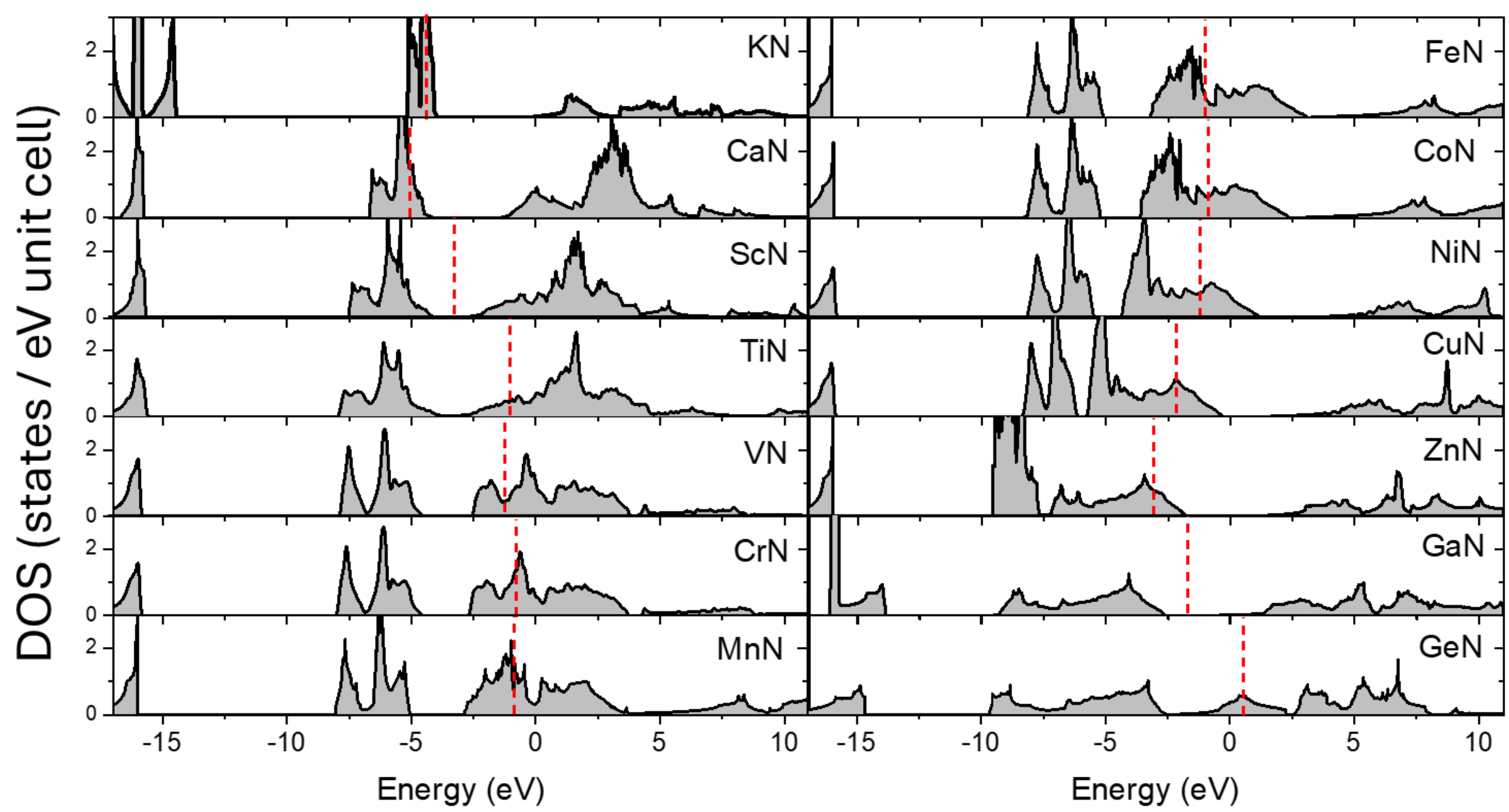

***FIG. 12.*** *Total density of states (DOS) for hexagonal nitrides of IV-period elements from K to Ge. The plot illustrates the evolution of the electronic structure across the period. The energy scale has been aligned such that the maximum of the N(2s) semicore state is fixed. The pDOS for each of the compounds presented can be found in the Supplementary Information.*

The five independent elastic coefficients of the wurtzite structure, $C_{11}$, $C_{12}$, $C_{13}$ , $C_{33}$, and $C_{44}$, are plotted as functions of the atomic number Z in figures 13(a-e), respectively.

$C_{11}$ and $C_{12}$ display similar trends: they increase for the first four layered-hexagonal compounds, reflecting stabilization from the addition of d-electrons. For the remaining TMNs, the values remain roughly constant with a slight negative trend, except for MnN, which is highly unstable in this configuration. GaN stabilizes the crystal, particularly increasing $C_{11}$, whereas lh-GeN becomes unstable due to an increase in $C_{12}$.

$C_{13}$ increases up to CrN before decreasing for the heavier elements, while $C_{33}$ rises up to TiN and then decreases, with the trend interrupted only by GaN. $C_{44}$ exhibits again a very complex behavior, showing instability or near-instability for several compounds, including KN, CrN, MnN, and CuN, with maxima observed for lh-TiN and wz-GaN.

The bulk modulus is shown as a function of the relaxed lateral lattice parameter in Figure 13(f). A clear upward trend is observed for the first layered-hexagonal compounds up to TiN, as the addition of d-electrons strengthens the crystal. Their bulk moduli are approximately 10% larger than their zincblende counterparts but remain lower than those of the rocksalt phases. For the subsequent transition-metal nitrides up to CoN, the bulk modulus remains fairly constant, with the exception of the unstable MnN. Beyond this, a decreasing trend is observed across the remaining transition metals, followed by the stiffening of GaN and the softer GeN. Compounds approaching ideal wurtzite geometry exhibit bulk moduli similar to zincblende; in the case of GaN, the difference is less than 1%, and for most other elements it remains below 4%, with the exception of ZnN (7%).

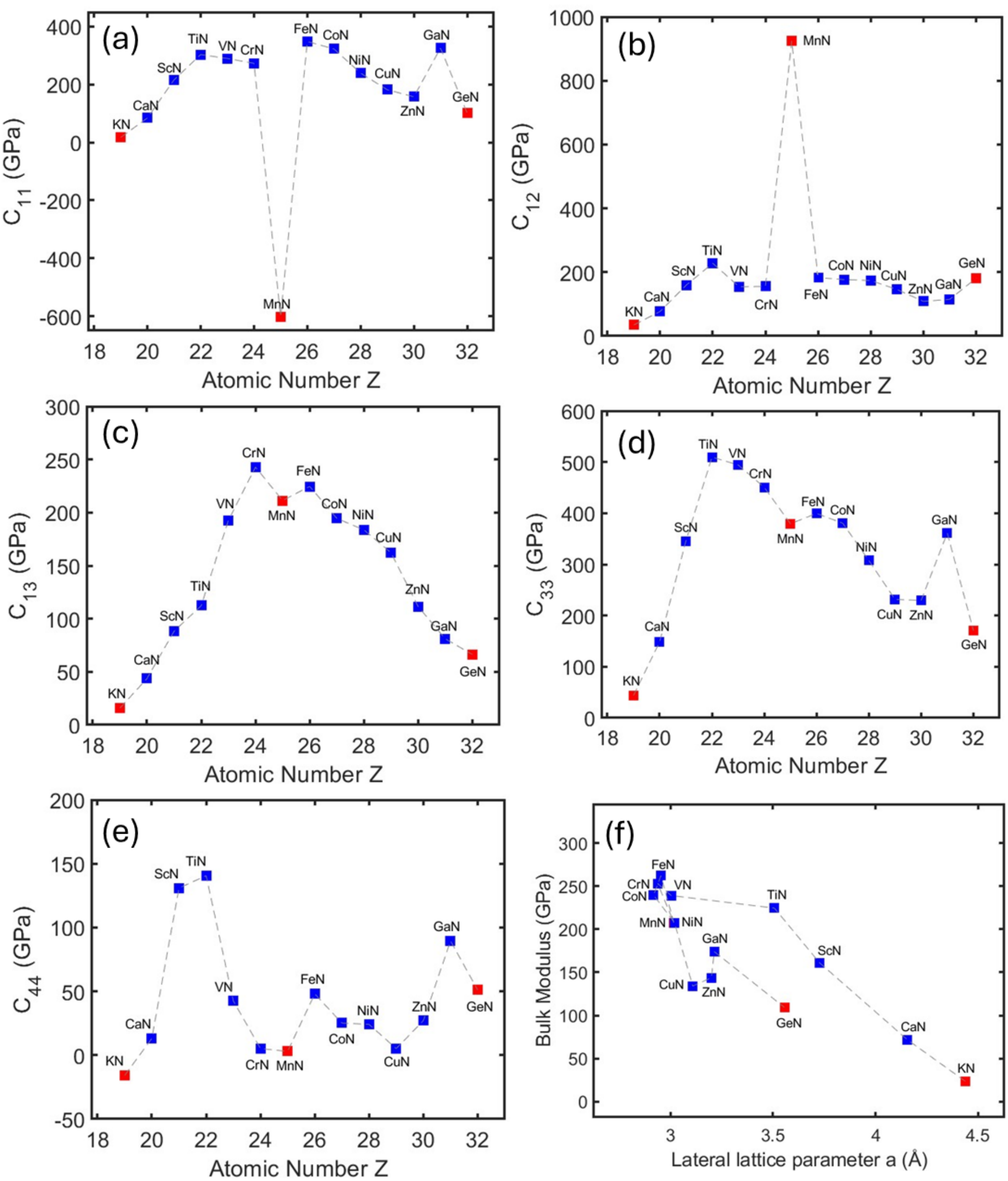

***FIG. 13.*** *Elastic coefficients of hexagonal IV-period nitrides. Panels (a), (b), (c), (d) and (e) show $C_{11}$, $C_{12}$, $C_{13}$, $C_{33}$ and $C_{44}$, respectively, plotted as a function of atomic number Z. Panel (f) shows the bulk modulus as a function of the equilibrium lateral lattice parameter a. The elastically stable compounds are displayed in blue, and the unstable ones in red.*

The evolution of the lattice parameter and elastic coefficients can be interpreted from the DOS. As in previous cases, KN exhibits a small bulk modulus and unstable behavior due to the absence of d-state participation in bonding. In CaN, this contribution increases through the promotion of Ca s-electrons into d-states, allowing constructive overlap with N(p) orbitals, as seen in region (ii), and giving rise to σ-like bonds.

To refine this analysis, the partial orbital occupations of the transition-metal nitrides (Löwdin charges) are summarized in Table 3. ScN continues the trend of lattice contraction and stiffening as its d-electron shell predominantly fills vertical (($z^2$) states associated with N(p)–Sc(d) bonding. As Z increases, lateral ($x^2$-$y^2$, xy) and diagonal (xz, yz) d-states are progressively occupied in TiN, further stiffening the crystal and reducing the lattice parameter. This effect is largely driven by second-neighbor d–d interactions, as illustrated in region (iv).

From TiN onward, the structure corresponds to wurtzite lattices, so the ILDOS paradigm shown in Figure 11 for lh-ScN is no longer strictly applicable. In this regime, the zincblende framework provides a more relevant comparison. The wurtzite compounds exhibit a behavior similar to zincblende: the addition of d-electrons continues to stiffen the crystal up to FeN. An exception is MnN, which remains unstable. Unlike zincblende, no significant lateral lattice parameter contraction is observed due to the slight variations in the u parameter. The stiffening arises from the gradual filling of mildly constructive overlaps in vertical and lateral directions between N(p) and d-states with predominant diagonal and in-plane characters. This mechanism is analogous to the stiffening observed in zincblende caused by the constructive overlap between N(p) orbitals and $e_g$ d-states.

Further increases in the Fermi level populate antibonding states between N(p) and metal d-orbitals, causing lattice expansion and crystal softening. In the case of GaN, p-electrons dominate, leading to $sp^3$-type bonding with a reduced role of d-states, which stiffens the lattice, whereas in GeN, the filling of antibonding states destabilizes and softens the crystal.

With respect to the elastic coefficients, all values increase for the initial layered-hexagonal compounds as the d-shell becomes progressively filled. From there, as Z increases, $C_{11}$ follows the same trend as the bulk modulus. $C_{12}$ exhibits relatively small variations among the different wurtzite compounds; however, these variations generally reflect the bulk modulus trend, weighted by the u parameter, with larger u values corresponding to higher $C_{12}$. $C_{13}$ shows a trend similar to the bulk modulus but reaches a maximum earlier, at CrN, likely due to a lower u parameter, which stiffens lateral bonds in VN and CrN.

$C_{33}$ displays distinct behavior, with lh-compounds exhibiting the highest values. This is likely because vertical compression in wurtzite can be accommodated by modifying u, whereas in lh structures, vertical bonds are directly squeezed, leading to higher resistance against deformation.

From VN onward, $C_{33}$ decreases with increasing d-electron count, but GaN shows a pronounced increase as p-electrons restiffen the u-parameter related flexibility.
Finally, $C_{44}$ exhibits a complex trend similar to zincblende, with clear minima for CrN, MnN, and CuN, reflecting the occupation of d-states that destabilize the crystal under shear.

***TABLE 3.*** *Partial electronic occupations of atomic orbitals (in elemental charge units) for each element in the hexagonal IV-period transition metal nitrides. An asterisk (*) denotes compounds that are elastically unstable. The populations may exceed the nominal occupation limits of individual orbital as multiple radial shells are considered.*

| **Hexagonal** | **Sc** | **Ti** | **V** | **Cr** | **Mn*** | **Fe** | **Co** | **Ni** | **Cu** | **Zn** |
|---|---|---|---|---|---|---|---|---|---|---|
| **N** | 5.93 | 6.00 | 5.75 | 5.81 | 5.13 | 5.23 | 5.08 | 5.10 | 5.46 | 5.55 |
| **s** | 1.59 | 1.65 | 1.59 | 1.64 | 1.32 | 1.36 | 1.35 | 1.41 | 1.50 | 1.51 |
| **p** | 4.34 | 4.35 | 4.16 | 4.17 | 3.81 | 3.87 | 3.73 | 3.69 | 3.96 | 4.04 |
| **Me** | 10.02 | 10.86 | 12.14 | 13.03 | 14.78 | 15.74 | 16.89 | 17.86 | 18.53 | 19.43 |
| **s** | 2.37 | 2.42 | 2.42 | 2.46 | 2.2 | 2.27 | 2.36 | 2.44 | 2.41 | 2.53 |
| **p** | 5.98 | 5.98 | 5.98 | 5.99 | 6.92 | 6.90 | 7.05 | 7.08 | 6.88 | 7.05 |
| **d** | 1.66 | 2.45 | 3.74 | 4.57 | 5.66 | 6.57 | 7.48 | 8.33 | 9.23 | 9.85 |
| **d $z^2$** | 0.46 | 0.49 | 0.77 | 0.81 | 0.89 | 1.06 | 1.39 | 1.88 | 1.79 | 1.96 |
| **d xz, yz** | 0.26 | 0.49 | 0.70 | 0.94 | 1.26 | 1.45 | 1.59 | 1.74 | 1.88 | 1.97 |
| **d $x^2$-$y^2$, xy** | 0.34 | 0.49 | 0.79 | 0.94 | 1.12 | 1.30 | 1.46 | 1.64 | 1.84 | 1.97 |

## IV. CONCLUSION

In conclusion, we have provided a comprehensive analysis of the elastic properties of IV-period transition metal nitrides, demonstrating the close relationship between their mechanical behavior and electronic structure. Our study shows that the occupation of bonding and antibonding orbitals plays a key role in determining the evolution of elastic coefficients across the period. By comparing rocksalt, zincblende, and wurtzite phases, we have highlighted how differences in crystal symmetry and atomic coordination influence the elastic response, offering a unified perspective on the interplay between electronic structure and mechanical properties in TMNs.

## SUPPLEMENTARY MATERIAL

The Supplementary Information contains the projected densities of states for all compounds and crystal structures considered in this work, three-dimensional ILDOS isosurface representations corresponding to the selected energy windows discussed in the main text for the three ScN phases, and a complete compilation of the calculated elastic properties together with available experimental data and previously reported theoretical results for comparison.

## ACKNOWLEDGEMENT

This project has received funding from the European Union under the Marie Skłodowska-Curie grant agreement No 101146464 and from the German Science Foundation (DFGproject: "Polrock" 530081697).

## AUTHOR DECLARATION

### Conflict of Interest

The authors have no conflicts to disclose.

### Data Availability

The data supporting this study's findings are available from the corresponding author upon reasonable request.

**Supplementary information**

# Electronic Origins of Elastic Behavior in Rocksalt, Zinc-Blende and Wurtzite 3d Transition-Metal Nitrides

J. Cañas[1,2] and O. Ambacher[1]

[1]*Institute for Sustainable Systems Engineering (INATECH), University Freiburg, Emmy-Noether-Str. 2, D-79110 Freiburg, Germany*

[2]*Univ. Grenoble Alpes, CNRS, Grenoble INP, Institut Néel, 38000 Grenoble, France*

## Comparison of the Calculated Elastic Coefficients with Calculated and Experimental Literature Values

A comprehensive comparison of the structural and elastic properties calculated in this work with those reported in the literature is presented in Table 1. The calculated elastic constants are in very good agreement with previous GGA calculations, with deviations generally well below 10% across the materials considered in this work. Some of the larger discrepancies are observed for the wurtzite compounds, which can largely be attributed to differences in the internal structural parameter (u), whose value is not always reported or consistently optimized in the literature. As expected, our calculated elastic constants are systematically smaller than those obtained using the LDA. This trend is consistent with the well-known tendency of the LDA to underestimate equilibrium lattice parameters by approximately 2%, leading to stiffer predicted elastic properties.

The comparison with available experimental data is also satisfactory. Experimental measurements remain scarce for many of the compounds investigated because several are metastable, difficult to synthesize, or unstable under ambient conditions. Nevertheless, for compounds with reliable experimental data, the agreement is generally good. For example, the calculated elastic constants of wurtzite GaN and rocksalt TiN, VN and ScN agree with the most reliable experimental measurements to within approximately 10–20% for the most accurately determined elastic constants. It should be noted that the shear elastic constants are generally more difficult to measure experimentally and therefore exhibit a larger scatter in the reported values.

The CrN, MnN, and FeN compounds represent an exception to the overall accuracy observed throughout the series, notably in the $C_{44}$ elastic coefficient. This deviation is related to the fact that the present calculations were performed without considering spin polarization, whereas previous studies have demonstrated that magnetic ordering, particularly in the rocksalt phases, plays a key role in the stabilization and elastic properties of these compounds.

| Lattice | MeN | a Å | c Å | u | Author | Year | Method | C11 GPa | C12 GPa | C44 GPa | C33 GPa | C44 GPa | C66 GPa | BM (Voigt) GPa | DOI |
|---|---|---|---|---|---|---|---|---|---|---|---|---|---|---|---|
| rs | KN | 5.614 | | | Cañas et al. | 2026 | GGA | 45.9 | 11.0 | -13.2 | x | x | x | 22.6 | |
| rs | CaN | 4.987 | | | Cañas et al. | 2026 | GGA | 156.8 | 50.7 | 24.9 | x | x | x | 86.1 | |
| rs | ScN | 4.504 | | | Pupyrev et al. | 2024 | EXP | 333.3 | 112.2 | 95.5 | | | | 185.9 | https://doi.org/10.1063/5.0237166 |
| rs | ScN | 4.504 | | | Pupyrev et al. | 2024 | EXP | 341.6 | 130.7 | 106.3 | | | | 201.0 | https://doi.org/10.1063/5.0237166 |
| rs | ScN | 4.463 | | | Liu et al. | 2014 | LDA | 470.1 | 99.4 | 164.3 | | | | 223.0 | http://dx.doi.org/10.1088/0953-8984/26/2/025404 |
| rs | ScN | 4.543 | | | Liu et al. | 2014 | GGA | 399.3 | 95.9 | 157.6 | | | | 197.0 | https://doi.org/10.1142/S0217984915500098 |
| rs | ScN | 4.516 | | | Holec et al. | 2012 | GGA | 390.0 | 105.0 | 166.0 | | | | 200.0 | https://doi.org/10.1103/PhysRevB.85.064101 |
| rs | ScN | 4.444 | | | Brik et al. | 2012 | LDA | 418.7 | 101.8 | 173.5 | | | | 207.4 | https://doi.org/10.1016/j.commatsci.2011.08.008 |
| rs | ScN | 4.516 | | | Brik et al. | 2012 | GGA | 354.1 | 100.2 | 170.0 | | | | 184.8 | https://doi.org/10.1016/j.commatsci.2011.08.008 |
| rs | ScN | 4.501 | | | Ekuma et al. | 2012 | GGA | 453.0 | 99.0 | 185.0 | | | | 217.0 | https://doi.org/10.1063/1.4751260 |
| rs | ScN | 4.510 | | | Friak et al. | 2018 | GGA | 388.0 | 106.0 | 166.0 | | | | 200.0 | https://doi.org/10.3390/nano8121049 |
| rs | ScN | 4.480 | | | Shoaib et al. | 2013 | GGA | 384.6 | 94.1 | 156.1 | | | | 190.9 | https://doi.org/10.1016/j.commatsci.2013.06.015 |
| rs | ScN | 4.510 | | | Zeghoum et al. | 2013 | GGA | 385.0 | 104.4 | 167.8 | | | | 197.9 | https://doi.org/10.1134/S1990793125700022 |
| rs | ScN | 4.514 | | | Cañas et al. | 2026 | GGA | 384.3 | 100.4 | 166.8 | x | x | x | 195.0 | |
| rs | TiN | 4.241 | | | Liu et al. | | EXP | | | | | | | | http://dx.doi.org/10.1088/0953-8984/26/2/025404 |
| rs | TiN | | | | Kim et al. | 1992 | EXP | 625.0 | 165.0 | 163.0 | | | | 318.3 | https://doi.org/10.1063/1.351651 |
| rs | TiN | | | | Meng al. | 1995 | EXP | 507.0 | 96.0 | 163.0 | | | | 233.0 | https://doi.org/10.1063/1.351651 |
| rs | TiN | 4.253 | | | Holec et al. | 2012 | GGA | 560.0 | 135.0 | 163.0 | | | | 276.7 | https://doi.org/10.1103/PhysRevB.85.064101 |
| rs | TiN | 4.184 | | | Liu et al. | 2014 | LDA | 712.3 | 123.2 | 171.1 | | | | 319.6 | http://dx.doi.org/10.1088/0953-8984/26/2/025404 |
| rs | TiN | 4.258 | | | Liu et al. | 2014 | GGA | 603.0 | 118.7 | 159.6 | | | | 280.1 | http://dx.doi.org/10.1088/0953-8984/26/2/025404 |
| rs | TiN | 4.185 | | | Brik et al. | 2012 | LDA | 648.9 | 129.0 | 193.9 | | | | 302.3 | https://doi.org/10.1016/j.commatsci.2011.08.008 |
| rs | TiN | 4.250 | | | Brik et al. | 2012 | GGA | 537.7 | 117.7 | 175.4 | | | | 257.7 | https://doi.org/10.1016/j.commatsci.2011.08.008 |
| rs | TiN | 4.260 | | | Fulcher et al. | 2012 | GGA | 531.0 | 118.0 | 166.0 | | | | 255.7 | https://doi.org/10.1103/PhysRevB.85.184106 |
| rs | TiN | 4.256 | | | Wang et al. | 2010 | GGA | 610.0 | 137.0 | 158.0 | | | | 294.7 | https://doi.org/10.1016/j.commatsci.2010.03.014 |
| rs | TiN | 4.250 | | | Wang et al. | 2017 | GGA | 589.0 | 125.0 | 166.0 | | | | 279.7 | https://doi.org/10.1016/j.actamat.2017.01.017 |
| rs | TiN | 4.270 | | | Lazar et al. | 2007 | GGA | 604.0 | 136.0 | 162.0 | | | | 292.0 | https://doi.org/10.1103/PhysRevB.76.174112 |
| rs | TiN | 4.246 | | | Yang et al. | 2009 | GGA | 579.0 | 129.0 | 180.0 | | | | 279.0 | https://doi.org/10.1016/j.jallcom.2009.06.023 |
| rs | TiN | 4.255 | | | Mota et al. | 2015 | GGA | 596.8 | 112.2 | 153.7 | | | | 273.7 | https://doi.org/10.1016/j.phpro.2015.05.077 |
| rs | TiN | 4.246 | | | Nagao et al. | 2006 | GGA | 585.0 | 137.0 | 165.0 | | | | 286.6 | https://doi.org/10.1103/PhysRevB.73.144113 |
| rs | TiN | | | | Balasubramanian et al. | 2018 | GGA | 590.0 | 169.0 | 164.0 | | | | 309.0 | https://doi.org/10.1016/j.actamat.2018.04.033 |
| rs | TiN | 4.244 | | | Cañas et al. | 2026 | GGA | 610.5 | 133.3 | 168.5 | x | x | x | 292.3 | |
| rs | VN | | | | Kim et al. | 1992 | EXP | 533.0 | 135.0 | 133.0 | | | | 267.7 | https://doi.org/10.1063/1.351651 |
| rs | VN | 4.128 | | | Lazar et al. | 2007 | GGA | 636.0 | 162.0 | 126.0 | | | | 320.0 | https://doi.org/10.1103/PhysRevB.76.174112 |
| rs | VN | 4.120 | | | Wang et al. | 2017 | GGA | 628.7 | 144.6 | 147.4 | | | | 306.0 | https://doi.org/10.1016/j.actamat.2017.01.017 |
| rs | VN | 4.056 | | | Brik et al. | 2012 | LDA | 763.2 | 148.7 | 158.3 | | | | 353.5 | https://doi.org/10.1016/j.commatsci.2011.08.008 |
| rs | VN | 4.119 | | | Brik et al. | 2012 | GGA | 628.7 | 144.6 | 147.4 | | | | 306.0 | https://doi.org/10.1016/j.commatsci.2011.08.008 |
| rs | VN | 4.127 | | | Holec et al. | 2012 | LDA | 660.0 | 174.0 | 118.0 | | | | 336.0 | https://doi.org/10.1016/j.commatsci.2011.08.008 |
| rs | VN | 4.057 | | | Liu et al. | 2014 | GGA | 751.1 | 178.9 | 126.6 | | | | 369.6 | https://doi.org/10.1016/j.commatsci.2011.08.008 |
| rs | VN | 4.133 | | | Liu et al. | 2014 | GGA | 620.5 | 166.8 | 116.5 | | | | 318.0 | https://doi.org/10.1103/PhysRevB.85.064101 |
| rs | VN | | | | Balasubramanian et al. | 2018 | GGA | 623.0 | 251.0 | 122.0 | | | | 375.0 | https://doi.org/10.1016/j.actamat.2018.04.033 |
| rs | VN | 4.114 | | | Cañas et al. | 2026 | GGA | 634.6 | 146.2 | 135.1 | x | x | x | 309.0 | |
| rs | CrN | 4.165 | | | Almer et al. | 2003 | EXP | 540.0 | 27.0 | 88.0 | | | | 198.0 | https://doi.org/10.1063/1.1582351 |

| | | | | | | | | | | | | | |
|---|---|---|---|---|---|---|---|---|---|---|---|---|---|
| rs | CrN | 3.987 | Liu et al. | 2014 | LDA | 702.8 | 227.1 | 9.3 | | | | 385.7 | http://dx.doi.org/10.1088/0953-8984/26/2/025404 |
| rs | CrN | 4.064 | Liu et al. | 2014 | GGA | 569.2 | 209.0 | 4.6 | | | | 329.1 | http://dx.doi.org/10.1088/0953-8984/26/2/025404 |
| rs | CrN | 4.063 | Brik et al. | 2012 | LDA | 518.3 | 220.2 | 9.5 | | | | 319.6 | https://doi.org/10.1016/j.commatsci.2011.08.008 |
| rs | CrN | 4.032 | Brik et al. | 2012 | GGA | 502.8 | 214.2 | 4.1 | | | | 310.4 | https://doi.org/10.1016/j.commatsci.2011.08.008 |
| rs | CrN | 4.131 | Ming et al. | 2015 | GGA | 502.4 | 213.8 | 10.4 | | | | 310.0 | https://doi.org/10.1142/S0217984915500098 |
| rs | CrN | | Zhou et al. | 2014 | GGA | (538.0) | (88.0) | (143.0) | | | | 238.0 | https://doi.org/10.1103/PhysRevB.90.184102 |
| rs | CrN | | Balasubramanian et al. | 2018 | GGA | 582.0 | 270.0 | 8 (126) | | | | 373.0 | https://doi.org/10.1016/j.actamat.2018.04.033 |
| rs | CrN | 4.045 | Cañas et al. | 2026 | GGA | 597.2 | 253.7 | 5.6 | x | x | x | 368.2 | |
| rs | MnN | 3.945 | Liu et al. | 2014 | LDA | 682.1 | 241.6 | -13.3 | | | | 388.4 | http://dx.doi.org/10.1088/0953-8984/26/2/025404 |
| rs | MnN | 4.025 | Liu et al. | 2014 | GGA | 550.0 | 217.8 | -8.5 | | | | 328.5 | http://dx.doi.org/10.1088/0953-8984/26/2/025404 |
| rs | MnN | | Balasubramanian et al. | 2018 | GGA | | | -9 (122) | | | | | https://doi.org/10.1016/j.actamat.2018.04.033 |
| rs | MnN | 4.005 | Cañas et al. | 2026 | GGA | 597.3 | 249.3 | -10.9 | x | x | x | 365.3 | |
| rs | FeN | 3.927 | Liu et al. | 2014 | LDA | 543.1 | 299.3 | -44.5 | | | | 380.6 | http://dx.doi.org/10.1088/0953-8984/26/2/025404 |
| rs | FeN | 4.010 | Liu et al. | 2014 | GGA | 428.7 | 263.4 | -13.3 | | | | 318.5 | http://dx.doi.org/10.1088/0953-8984/26/2/025404 |
| rs | FeN | | Balasubramanian et al. | 2018 | GGA | | | -44 (69) | | | | | https://doi.org/10.1016/j.actamat.2018.04.033 |
| rs | FeN | 3.999 | Cañas et al. | 2026 | GGA | 432.9 | 244.8 | -28.2 | x | x | x | 307.5 | |
| rs | CoN | 3.927 | Liu et al. | 2014 | LDA | 518.0 | 278.9 | 56.2 | | | | 358.6 | http://dx.doi.org/10.1088/0953-8984/26/2/025404 |
| rs | CoN | 4.015 | Liu et al. | 2014 | GGA | 417.9 | 237.2 | 89.1 | | | | 297.4 | http://dx.doi.org/10.1088/0953-8984/26/2/025404 |
| rs | CoN | | Balasubramanian et al. | 2018 | GGA | 510.0 | 230.0 | 55.0 | | | | 320.0 | https://doi.org/10.1016/j.actamat.2018.04.033 |
| rs | CoN | 3.996 | Cañas et al. | 2026 | GGA | 441.2 | 276.3 | 86.9 | x | x | x | 331.3 | |
| rs | NiN | 3.981 | Liu et al. | 2014 | LDA | 485.1 | 230.8 | 89.1 | | | | 315.6 | http://dx.doi.org/10.1088/0953-8984/26/2/025404 |
| rs | NiN | 4.076 | Liu et al. | 2014 | GGA | 383.0 | 194.1 | 86.3 | | | | 257.1 | http://dx.doi.org/10.1088/0953-8984/26/2/025404 |
| rs | NiN | | Balasubramanian et al. | 2018 | GGA | 368.0 | 214.0 | 80.0 | | | | 265.0 | https://doi.org/10.1016/j.actamat.2018.04.033 |
| rs | NiN | 4.059 | Cañas et al. | 2026 | GGA | 406.5 | 229.2 | 86.9 | x | x | x | 288.8 | |
| rs | CuN | 4.084 | Liu et al. | 2014 | LDA | 395.8 | 186.5 | 64.7 | | | | 256.3 | http://dx.doi.org/10.1088/0953-8984/26/2/025404 |
| rs | CuN | 4.188 | Liu et al. | 2014 | GGA | 308.9 | 155.6 | 60.5 | | | | 206.7 | http://dx.doi.org/10.1088/0953-8984/26/2/025404 |
| rs | CuN | | Balasubramanian et al. | 2018 | GGA | 305.0 | 176.0 | 61.0 | | | | 219.0 | https://doi.org/10.1016/j.actamat.2018.04.033 |
| rs | CuN | 4.173 | Cañas et al. | 2026 | GGA | 297.5 | 154.1 | 63.1 | x | x | x | 201.9 | |
| rs | ZnN | 4.203 | Liu et al. | 2014 | LDA | 328.1 | 151.2 | 72.9 | | | | 210.2 | http://dx.doi.org/10.1088/0953-8984/26/2/025404 |
| rs | ZnN | 4.313 | Liu et al. | 2014 | GGA | 249.4 | 127.2 | 61.7 | | | | 167.9 | http://dx.doi.org/10.1088/0953-8984/26/2/025404 |
| rs | ZnN | | Balasubramanian et al. | 2018 | GGA | 245.0 | 134.0 | 59.0 | | | | 171.0 | https://doi.org/10.1016/j.actamat.2018.04.033 |
| rs | ZnN | 4.305 | Cañas et al. | 2026 | GGA | 257.6 | 142.4 | 62.0 | x | x | x | 180.8 | |
| rs | GaN | 4.276 | Soykan et al. | 2014 | GGA | 343.2 | 157.4 | 202.2 | | | | 219.3 | https://doi.org/10.19113/sdufbed.38099 |
| rs | GaN | 4.265 | Cañas et al. | 2026 | GGA | 321.0 | 159.8 | 206.3 | x | x | x | 213.5 | |
| rs | GeN | 4.485 | Cañas et al. | 2026 | GGA | 172.1 | 138.1 | 85.3 | x | x | x | 149.4 | |
| zb | KN | 5.960 | Cañas et al. | 2026 | GGA | 10.1 | 31.0 | -21.1 | x | x | x | 24.0 | |
| zb | CaN | 5.395 | Cañas et al. | 2026 | GGA | 49.2 | 64.5 | 15.5 | x | x | x | 59.4 | |
| zb | ScN | 4.840 | Liu et al. | 2014 | LDA | 186.9 | 140.9 | 73.8 | | | | 156.2 | http://dx.doi.org/10.1088/0953-8984/26/2/025404 |
| zb | ScN | 4.925 | Liu et al. | 2014 | GGA | 172.4 | 124.9 | 69.8 | | | | 140.7 | http://dx.doi.org/10.1088/0953-8984/26/2/025404 |
| zb | ScN | | Ghebouli et al. | 2021 | GGA | 171.6 | 124.3 | 70.8 | | | | 140.1 | https://doi.org/10.1016/j.jmrt.2021.07.073 |
| zb | ScN | 4.800 | Ghebouli et al. | 2021 | LDA | 187.6 | 143.2 | 72.2 | | | | 157.8 | https://doi.org/10.1016/j.jmrt.2021.07.073 |
| zb | ScN | 4.888 | Ghebouli et al. | 2021 | GGA | 173.4 | 129.7 | 74.7 | | | | 144.3 | https://doi.org/10.1016/j.jmrt.2021.07.073 |
| zb | ScN | 4.891 | Cañas et al. | 2026 | GGA | 171.4 | 124.2 | 69.8 | x | x | x | 139.9 | |
| zb | TiN | 4.529 | Liu et al. | 2014 | LDA | 322.2 | 176.4 | 103.1 | | | | 225.0 | http://dx.doi.org/10.1088/0953-8984/26/2/025404 |

| | | | | | | | | | | | | | | | |
|---|---|---|---|---|---|---|---|---|---|---|---|---|---|---|---|
| zb | TiN | 4.609 | | | Liu et al. | 2014 | GGA | 292.2 | 154.1 | 95.3 | | | | 200.1 | http://dx.doi.org/10.1088/0953-8984/26/2/025404 |
| zb | TiN | 4.613 | | | Mota et al. | 2015 | GGA | 305.7 | 174.7 | 84.0 | | | | 218.4 | https://doi.org/10.1016/j.phpro.2015.05.077 |
| zb | TiN | 4.592 | | | Cañas et al. | 2026 | GGA | 299.2 | 156.0 | 95.2 | x | x | x | 203.7 | |
| zb | VN | 4.368 | | | Liu et al. | 2014 | LDA | 346.8 | 229.4 | 43.2 | | | | 268.5 | http://dx.doi.org/10.1088/0953-8984/26/2/025404 |
| zb | VN | 4.446 | | | Liu et al. | 2014 | GGA | 309.4 | 196.5 | 42.4 | | | | 234.1 | http://dx.doi.org/10.1088/0953-8984/26/2/025404 |
| zb | VN | 4.424 | | | Cañas et al. | 2026 | GGA | 308.9 | 198.3 | 47.6 | x | x | x | 235.1 | |
| zb | CrN | 4.262 | | | Liu et al. | 2014 | LDA | 361.7 | 258.7 | -77.9 | | | | 293.0 | http://dx.doi.org/10.1088/0953-8984/26/2/025404 |
| zb | CrN | 4.342 | | | Liu et al. | 2014 | GGA | 320.5 | 221.8 | -54.5 | | | | 254.7 | http://dx.doi.org/10.1088/0953-8984/26/2/025404 |
| zb | CrN | 4.323 | | | Ming et al. | 2015 | GGA | (267.8) | (175.9) | (92.5) | | | | 206.5 | https://doi.org/10.1142/S0217984915500098 |
| zb | CrN | 4.321 | | | Cañas et al. | 2026 | GGA | 325.7 | 226.7 | -85.8 | x | x | x | 259.7 | |
| zb | MnN | 4.188 | | | Liu et al. | 2014 | LDA | 372.7 | 279.7 | 24.0 | | | | 310.7 | http://dx.doi.org/10.1088/0953-8984/26/2/025404 |
| zb | MnN | 4.269 | | | Liu et al. | 2014 | GGA | 331.4 | 235.5 | 42.9 | | | | 267.5 | http://dx.doi.org/10.1088/0953-8984/26/2/025404 |
| zb | MnN | 4.248 | | | Cañas et al. | 2026 | GGA | 336.8 | 239.6 | 23.5 | x | x | x | 271.9 | |
| zb | FeN | 4.160 | | | Liu et al. | 2014 | LDA | 379.2 | 282.8 | 114.2 | | | | 314.9 | http://dx.doi.org/10.1088/0953-8984/26/2/025404 |
| zb | FeN | 4.243 | | | Liu et al. | 2014 | GGA | 334.6 | 234.6 | 110.6 | | | | 267.9 | http://dx.doi.org/10.1088/0953-8984/26/2/025404 |
| zb | FeN | 4.229 | | | Cañas et al. | 2026 | GGA | 337.0 | 239.2 | 110.2 | x | x | x | 271.8 | |
| zb | CoN | 4.177 | | | Liu et al. | 2014 | LDA | 347.4 | 266.0 | 71.2 | | | | 293.1 | http://dx.doi.org/10.1088/0953-8984/26/2/025404 |
| zb | CoN | 4.265 | | | Liu et al. | 2014 | GGA | 297.6 | 219.9 | 65.3 | | | | 245.8 | http://dx.doi.org/10.1088/0953-8984/26/2/025404 |
| zb | CoN | 4.244 | | | Cañas et al. | 2026 | GGA | 299.7 | 225.5 | 61.9 | x | x | x | 250.2 | |
| zb | NiN | 4.241 | | | Liu et al. | 2014 | LDA | 278.1 | 246.2 | 49.9 | | | | 256.8 | http://dx.doi.org/10.1088/0953-8984/26/2/025404 |
| zb | NiN | 4.336 | | | Liu et al. | 2014 | GGA | 232.9 | 200.8 | 43.5 | | | | 211.5 | http://dx.doi.org/10.1088/0953-8984/26/2/025404 |
| zb | NiN | 4.318 | | | Cañas et al. | 2026 | GGA | 234.3 | 205.1 | 48.3 | x | x | x | 214.8 | |
| zb | CuN | 4.344 | | | Liu et al. | 2014 | LDA | 222.4 | 199.9 | 41.1 | | | | 207.4 | http://dx.doi.org/10.1088/0953-8984/26/2/025404 |
| zb | CuN | 4.452 | | | Liu et al. | 2014 | GGA | 183.0 | 160.1 | 37.7 | | | | 167.7 | http://dx.doi.org/10.1088/0953-8984/26/2/025404 |
| zb | CuN | 4.436 | | | Cañas et al. | 2026 | GGA | 184.1 | 158.4 | 48.1 | x | x | x | 166.9 | |
| zb | ZnN | 4.472 | | | Liu et al. | 2014 | LDA | 178.0 | 159.4 | 47.4 | | | | 165.6 | http://dx.doi.org/10.1088/0953-8984/26/2/025404 |
| zb | ZnN | 4.588 | | | Liu et al. | 2014 | GGA | 144.7 | 126.6 | 42.6 | | | | 132.6 | http://dx.doi.org/10.1088/0953-8984/26/2/025404 |
| zb | ZnN | 4.575 | | | Cañas et al. | 2026 | GGA | 156.2 | 137.6 | 48.0 | x | x | x | 143.8 | |
| zb | GaN | | | | Shimada et al. | 1998 | LDA | 285.0 | 161.0 | 149.0 | | | | 202.3 | https://doi.org/10.1063/1.368739 |
| zb | GaN | 4.460 | | | Kim et al. | 1996 | LDA | 296.0 | 154.0 | 206.0 | | | | 201.3 | https://doi.org/10.1103/PhysRevB.53.16310 |
| zb | GaN | | | | Wright et al. | 1997 | LDA | 293.0 | 159.0 | 155.0 | | | | 203.7 | https://doi.org/10.1063/1.366114 |
| zb | GaN | | | | Lepkowski et al. | 2015 | LDA | 287.0 | 158.0 | 159.0 | | | | 201.0 | https://doi.org/10.1063/1.4914416 |
| zb | GaN | 4.489 | | | Caro et al. | 2012 | HSE | 288.6 | 154.1 | 166.0 | | | | 198.9 | https://doi.org/10.1103/PhysRevB.86.014117 |
| zb | GaN | 4.543 | | | Cañas et al. | 2026 | GGA | 255.9 | 133.1 | 146.1 | x | x | x | 174.0 | |
| zb | GeN | 4.795 | | | Cañas et al. | 2026 | GGA | 108.5 | 89.5 | -113 | x | x | x | 95.9 | |
| wz | KN | 4.438 | 5.783 | 0.500 | Cañas et al. | 2026 | GGA | 17.3 | 34.7 | 15.8 | 43.9 | -15.9 | -8.7 | 23.4 | |
| wz | CaN | 4.155 | 4.897 | 0.500 | Cañas et al. | 2026 | GGA | 84.6 | 76.7 | 43.7 | 148.7 | 13.2 | 4.0 | 71.8 | |
| wz | ScN | 3.727 | 4.503 | 0.500 | Cañas et al. | 2026 | GGA | 216.3 | 159.9 | 88.2 | 345.0 | 130.9 | 28.2 | 161.2 | |
| wz | TiN | 3.290 | 5.270 | 0.375 | Wang et al. | 2010 | GGA | 265.0 | 183.0 | 97.0 | 201.0 | 58.0 | 41.0 | 165.0 | https://doi.org/10.1016/j.commatsci.2010.03.014 |
| wz | TiN | 3.510 | 4.237 | 0.500 | Kim et al. | 2022 | GGA | 310.0 | 260.0 | 110.0 | 510.0 | 145.0 | 25.0 | 232.2 | https://doi.org/10.1111/jace.18722 |
| wz | TiN | 3.507 | 4.219 | 0.500 | Cañas et al. | 2026 | GGA | 303.4 | 227.9 | 112.3 | 509.7 | 140.6 | 37.7 | 224.6 | |
| wz | VN | 3.004 | 5.468 | 0.341 | Cañas et al. | 2026 | GGA | 289.8 | 154.4 | 192.5 | 494.3 | 42.6 | 67.7 | 239.2 | |
| wz | CrN | 2.881 | 5.549 | 0.375 | Ming et al. | 2015 | GGA | 199.0 | 178.5 | 178.5 | 199.0 | 76.1 | 10.3 | 185.3 | https://doi.org/10.1142/S0217984915500098 |
| wz | CrN | 2.938 | 5.339 | 0.344 | Cañas et al. | 2026 | GGA | 273.1 | 155.5 | 242.9 | 450.5 | 5.2 | 58.9 | 253.2 | |
| wz | MnN | 3.017 | 4.930 | 0.374 | Cañas et al. | 2026 | GGA | -602.3 | 924.7 | 211.3 | 379.4 | 3.0 | -763.5 | 207.7 | |
| wz | FeN | 2.954 | 5.042 | 0.377 | Cañas et al. | 2026 | GGA | 348.5 | 182.9 | 224.6 | 399.5 | 48.0 | 82.8 | 262.3 | |
| wz | CoN | 2.918 | 5.197 | 0.364 | Cañas et al. | 2026 | GGA | 322.8 | 176.6 | 194.5 | 380.4 | 25.2 | 73.1 | 239.7 | |
| wz | NiN | 3.020 | 5.114 | 0.369 | Cañas et al. | 2026 | GGA | 239.2 | 174.1 | 183.6 | 307.6 | 24.1 | 30.6 | 207.6 | |
| wz | CuN | 3.109 | 5.243 | 0.367 | Cañas et al. | 2026 | GGA | 182.1 | 145.6 | 162.2 | 231.4 | 5.2 | 18.2 | 170.6 | |
| wz | ZnN | 3.200 | 5.410 | 0.364 | Cañas et al. | 2026 | GGA | 157.5 | 108.6 | 111.4 | 229.6 | 27.1 | 24.5 | 134.1 | |
| wz | GaN | | | | Yamaguchi et al | 1997 | EXP | 365.0 | 135.0 | 114.0 | 381.0 | 109.0 | 115.0 | 204.1 | https://doi.org/10.1088/0953-8984/9/1/025 |
| wz | GaN | 3.190 | 5.190 | | Schwarz et al. | 1997 | EXP | 377.0 | 160.0 | 114.0 | 209.0 | 81.4 | 109.0 | 193.2 | https://doi.org/10.1063/1.118503 |
| wz | GaN | | | | Polian et al. | 1996 | EXP | 390.0 | 145.0 | 106.0 | 398.0 | 105.0 | 123.0 | 210.2 | https://doi.org/10.1063/1.361236 |
| wz | GaN | 3.162 | 5.142 | 0.377 | Kim et al. | 1996 | LDA | 396.0 | 144.0 | 100.0 | 392.0 | 91.0 | 126.0 | 208.0 | https://doi.org/10.1103/PhysRevB.53.16310 |
| wz | GaN | | | | Wright et al. | 1997 | LDA | 367.0 | 135.0 | 103.0 | 405.0 | 95.0 | 116.0 | 202.3 | https://doi.org/10.1063/1.366114 |
| wz | GaN | 3.232 | 5.268 | 0.376 | Duan et al. | 2008 | LDA | 334.0 | 132.0 | 99.0 | 372.0 | 86.0 | 101.0 | 188.9 | https://doi.org/10.1063/1.2831486 |
| wz | GaN | | | | Lepowski et al. | 2015 | LDA | 365.0 | 138.0 | 98.0 | 402.0 | 96.0 | 113.5 | 200.0 | https://doi.org/10.1063/1.4914416 |
| wz | GaN | 3.210 | 5.237 | 0.376 | Shimada et al. | 1998 | LDA | 350.0 | 140.0 | 104.0 | 376.0 | 101.0 | 115.0 | 196.9 | https://doi.org/10.1063/1.368739 |
| wz | GaN | 3.233 | 5.228 | 0.377 | Usman et al. | 2011 | GGA | 329.0 | 109.0 | 80.0 | 357.0 | 91.0 | 110.0 | 172.6 | https://doi.org/10.1021/jp207141k |
| wz | GaN | 3.225 | 5.257 | | Fan et al. | 2016 | GGA | 321.0 | 107.0 | 84.0 | 334.0 | 88.0 | 107.0 | 169.6 | https://doi.org/10.1155/2016/8612892 |
| wz | GaN | 3.180 | 5.185 | 0.377 | Caro et al. | 2012 | HSE | 368.6 | 131.6 | 95.7 | 406.2 | 101.7 | 118.5 | 198.8 | https://doi.org/10.1103/PhysRevB.86.014117 |
| wz | GaN | 3.214 | 5.237 | 0.377 | Cañas et al. | 2026 | GGA | 325.7 | 114.9 | 81.0 | 361.2 | 89.5 | 105.4 | 174.1 | |
| wz | GeN | 3.556 | 4.967 | 0.500 | Cañas et al. | 2026 | GGA | 101.3 | 180.9 | 66.4 | 153.4 | 51.1 | -39.8 | 109.3 | |

Table 1: Comparison of the calculated lattice parameters and elastic coefficients of the rocksalt, zincblende, and hexagonal phases of fourth-period binary nitrides with calculated and experimental values reported in the literature. Red compound names indicate unstable compounds, while values in parentheses correspond to spin-polarized calculations.

## Projected Density of States for Rocksalt 4th Period Nitrides

The projected densities of states (pDOS) for the rocksalt binary metal nitrides are presented in Figures 1–14, illustrating the evolution of the electronic states across the series and the progressive shift of the Fermi level with increasing atomic number. This representation not only allows the general trends across the series to be identified but also provides insight into the electronic structure of each individual compound. For the transition-metal nitrides, the pDOS distributions remain qualitatively similar across the series, with the main changes arising from the progressive filling of the electronic states as the Fermi level shifts upward. This evolution provides a consistent framework for understanding the observed trends in the elastic properties in terms of the filling of bonding and antibonding states. In contrast, GaN and GeN exhibit distinctly different pDOS distributions due to their reduced d-orbital contributions, accompanied by changes in the ILDOS distribution and, consequently, in the nature and spatial distribution of the chemical bonding. These differences highlight the distinct bonding mechanisms in these compounds compared with the transition-metal nitrides.

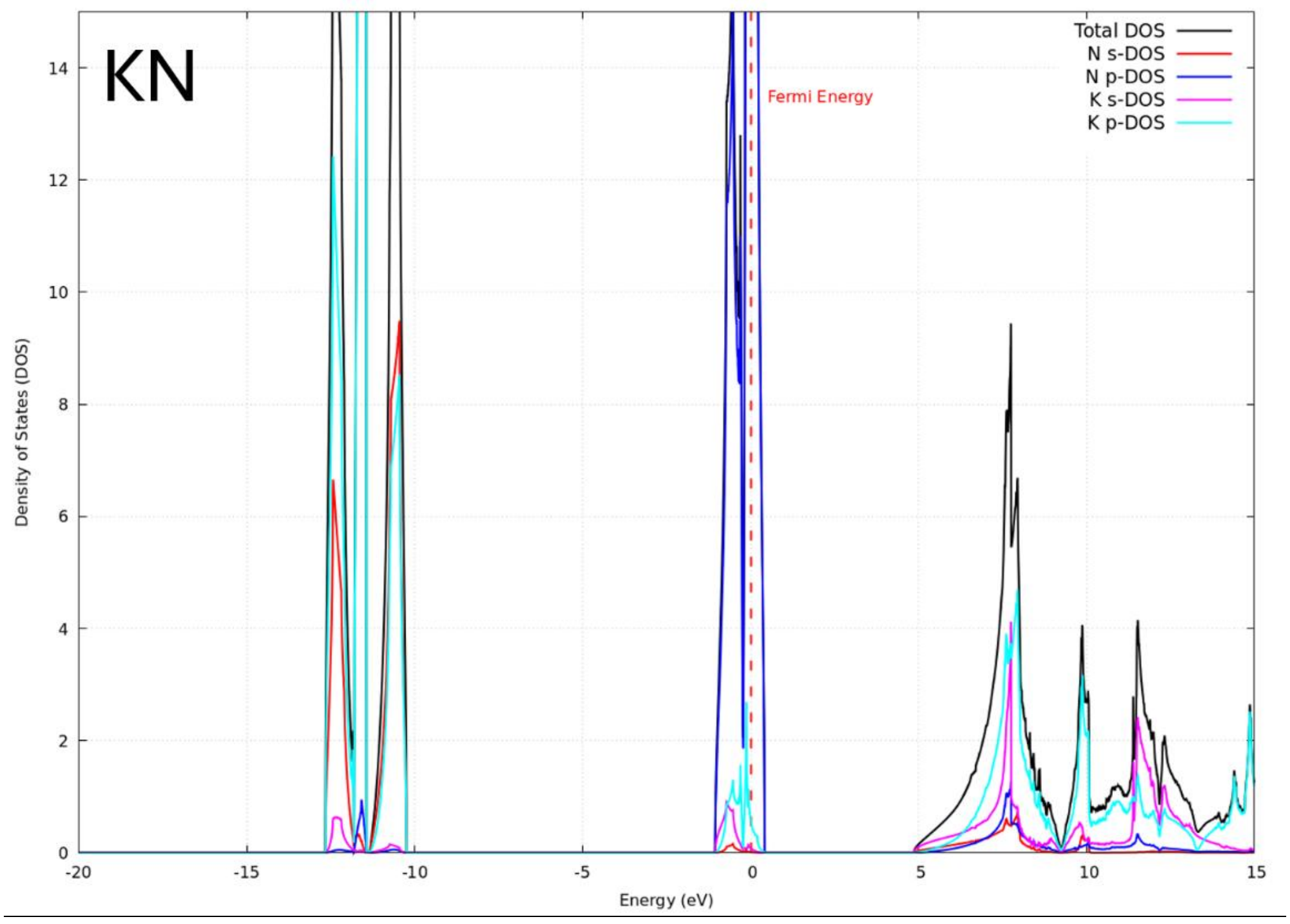

Figure 1: Projected local density of states for rocksalt KN.

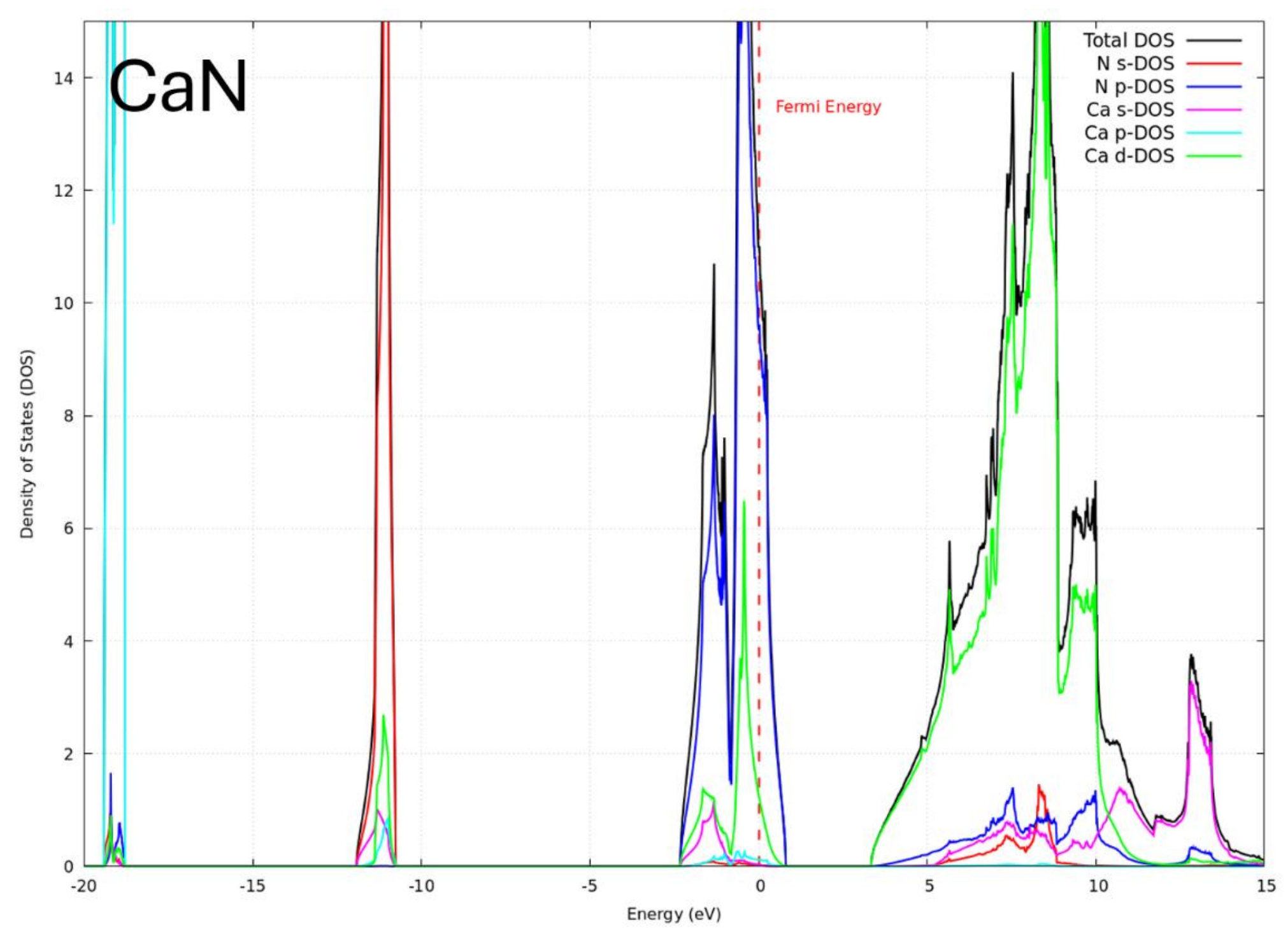

Figure 2: Projected local density of states for rocksalt CaN.

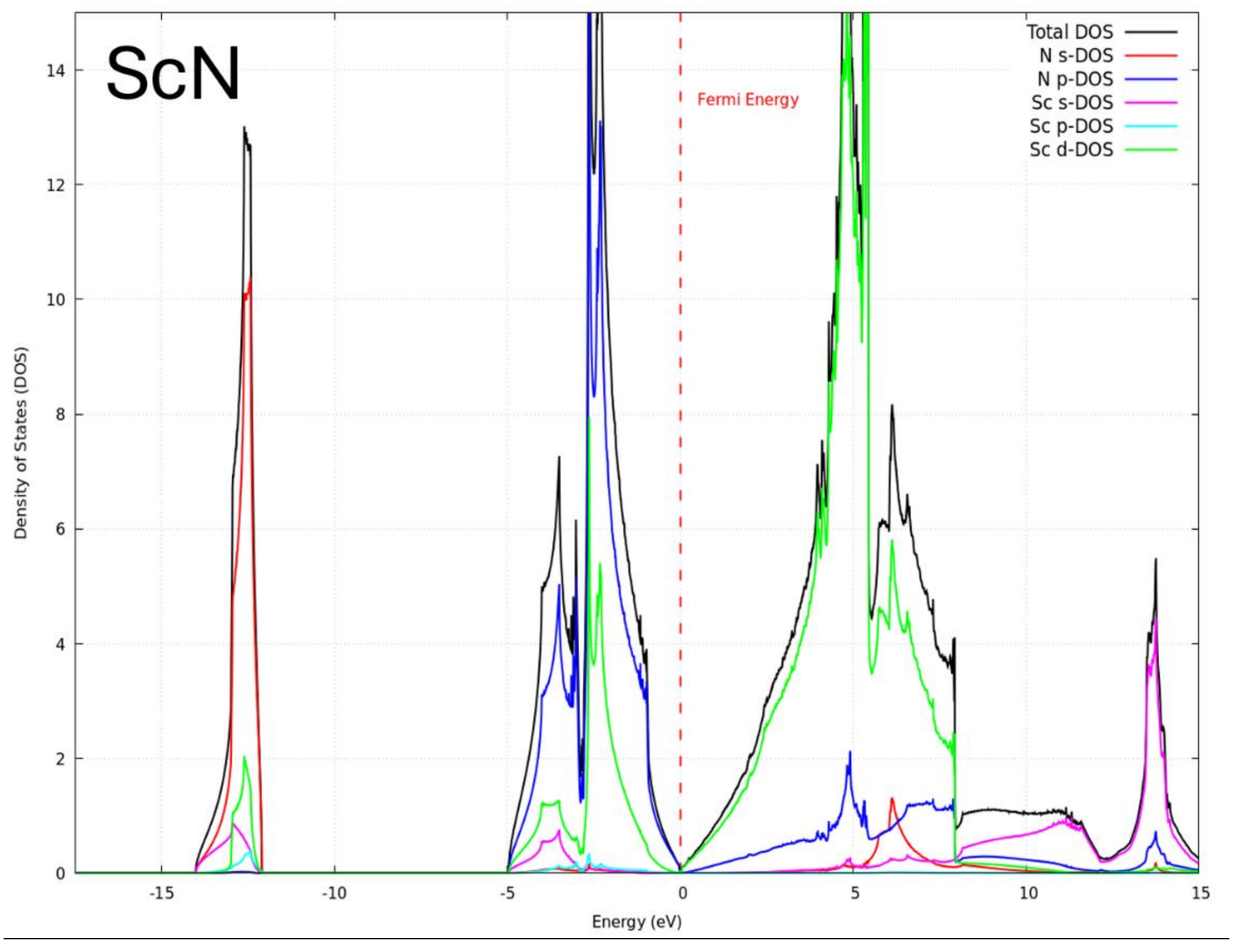

Figure 3: Projected local density of states for rocksalt ScN.

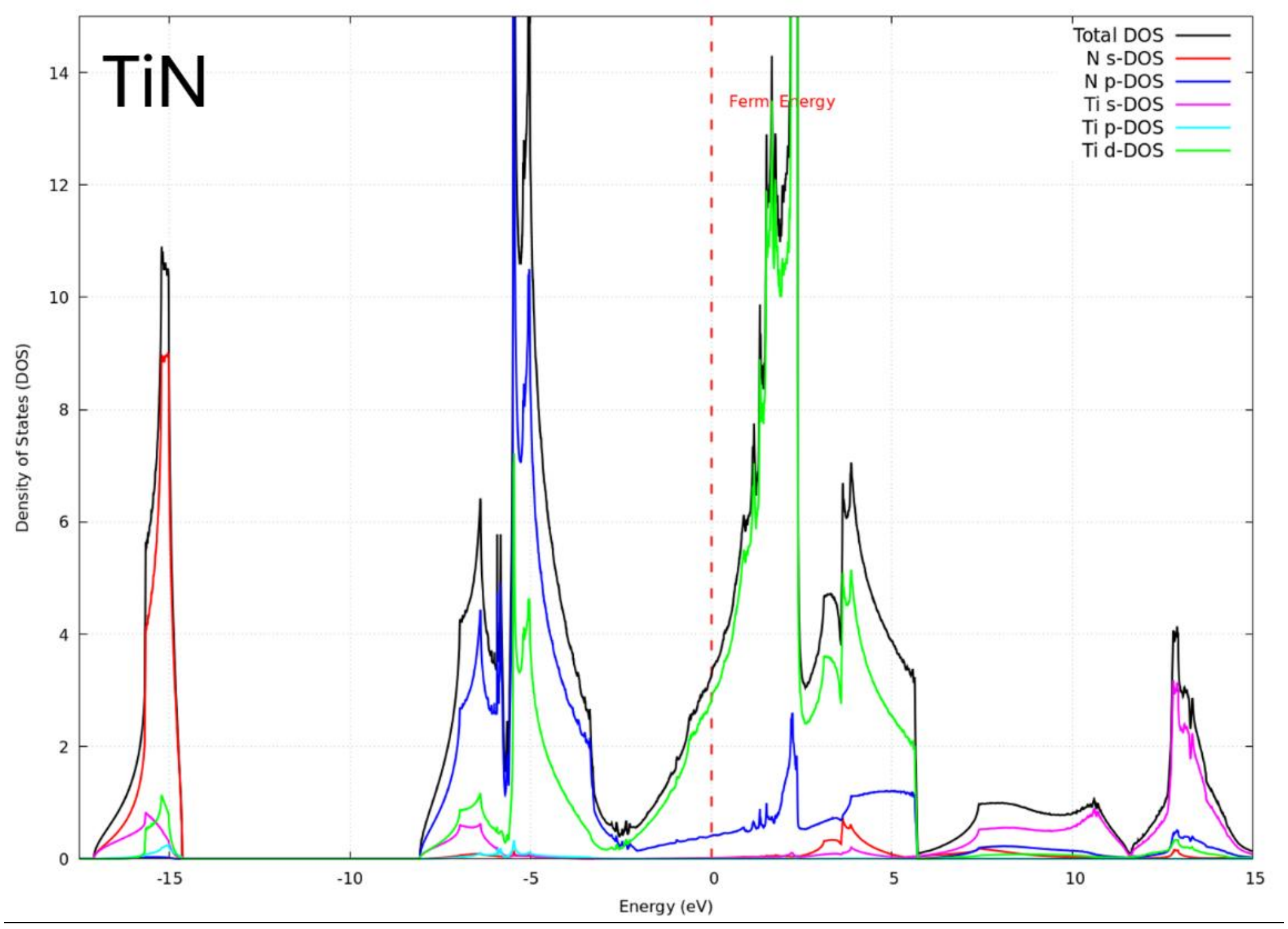

Figure 4: Projected local density of states for rocksalt TiN.

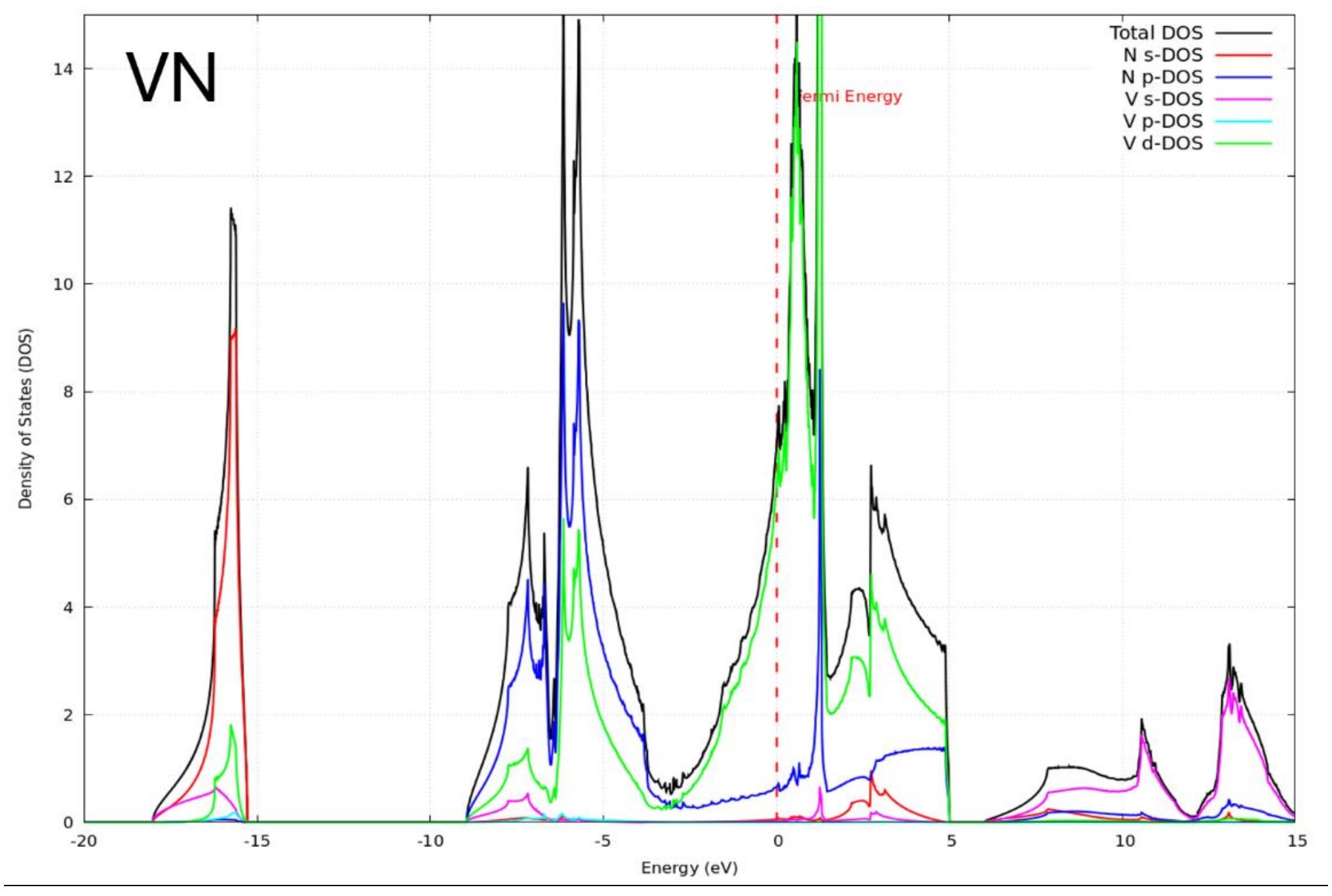

Figure 5: Projected local density of states for rocksalt VN.

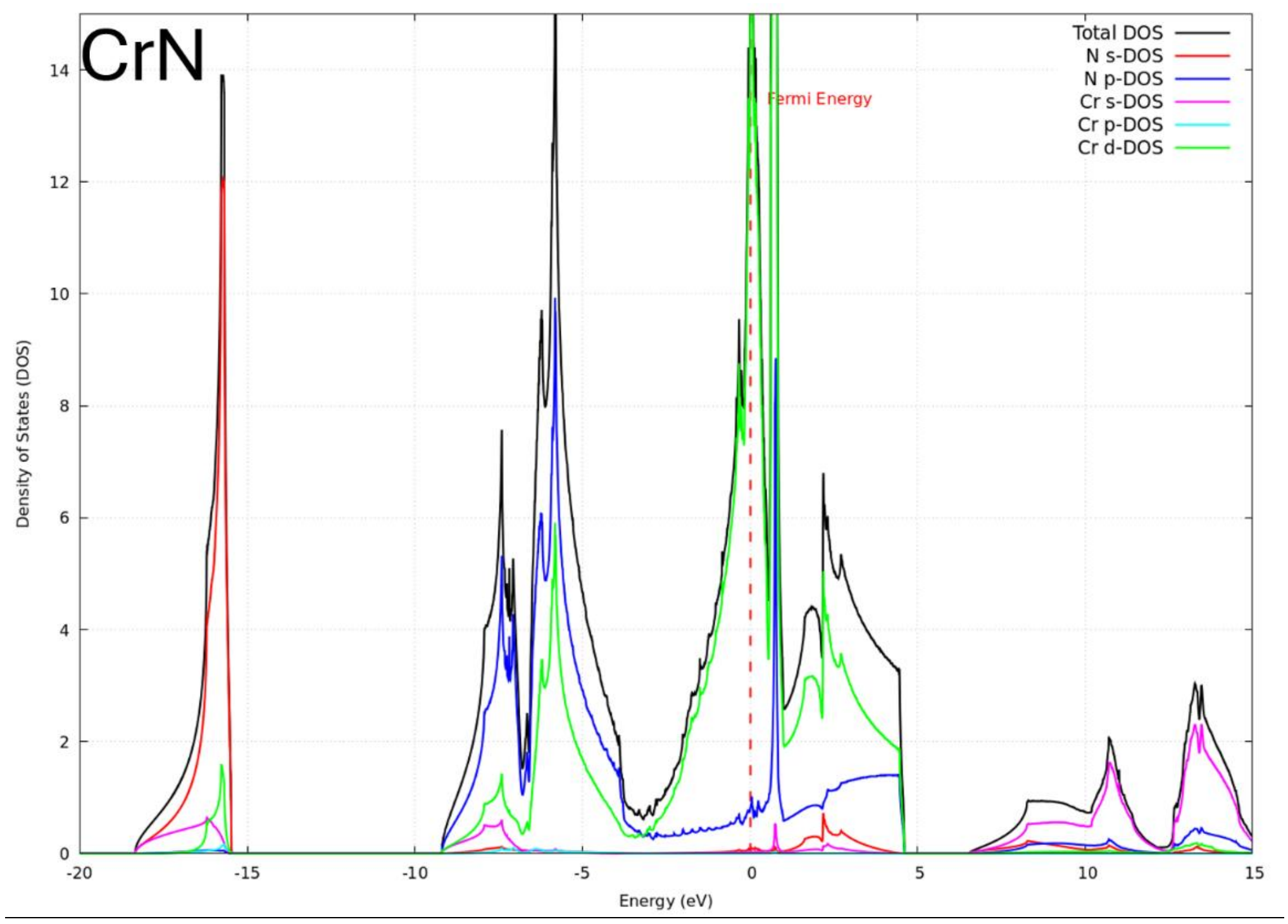

Figure 6: Projected local density of states for rocksalt CrN.

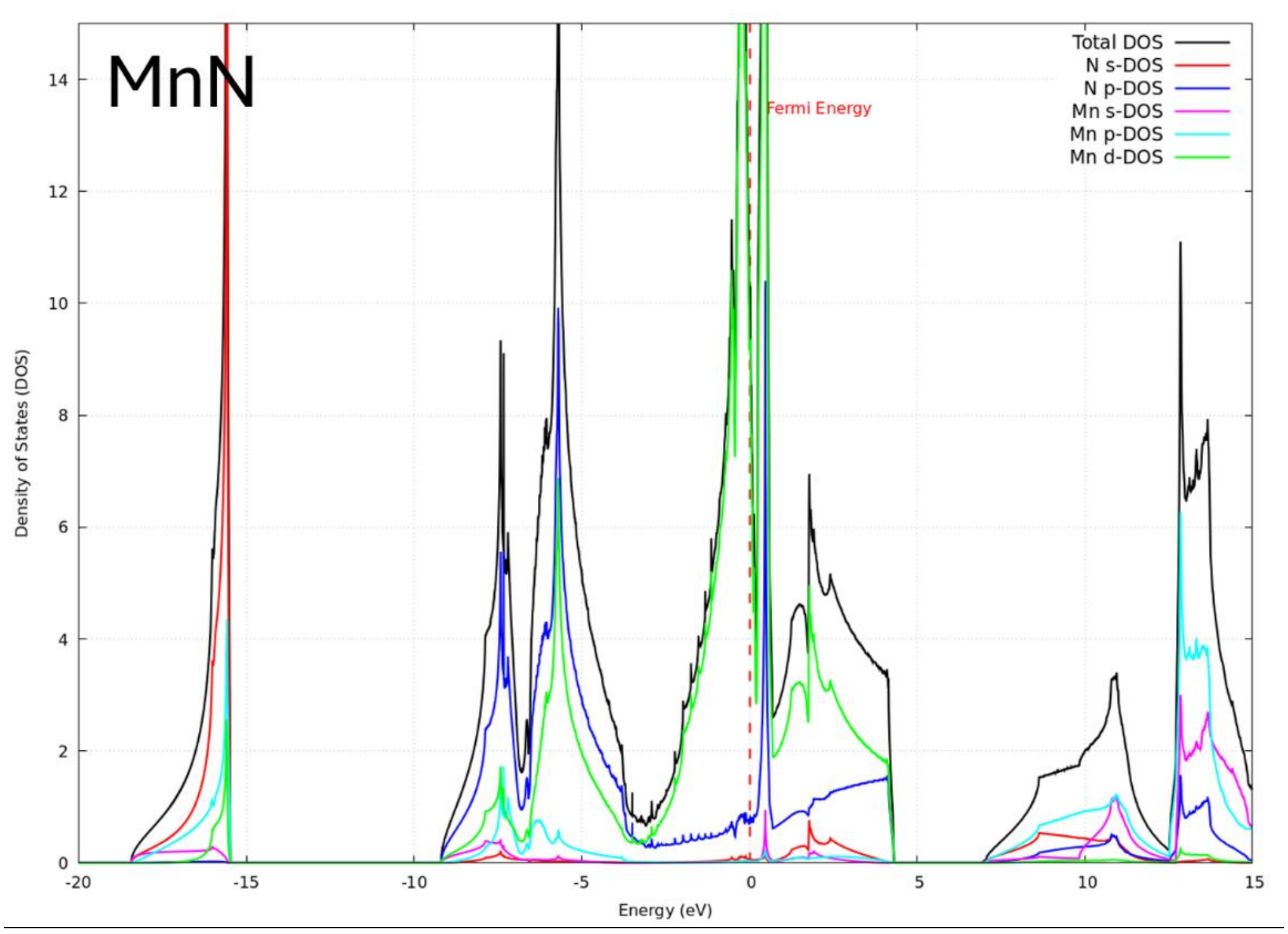

Figure 7: Projected local density of states for rocksalt MnN.

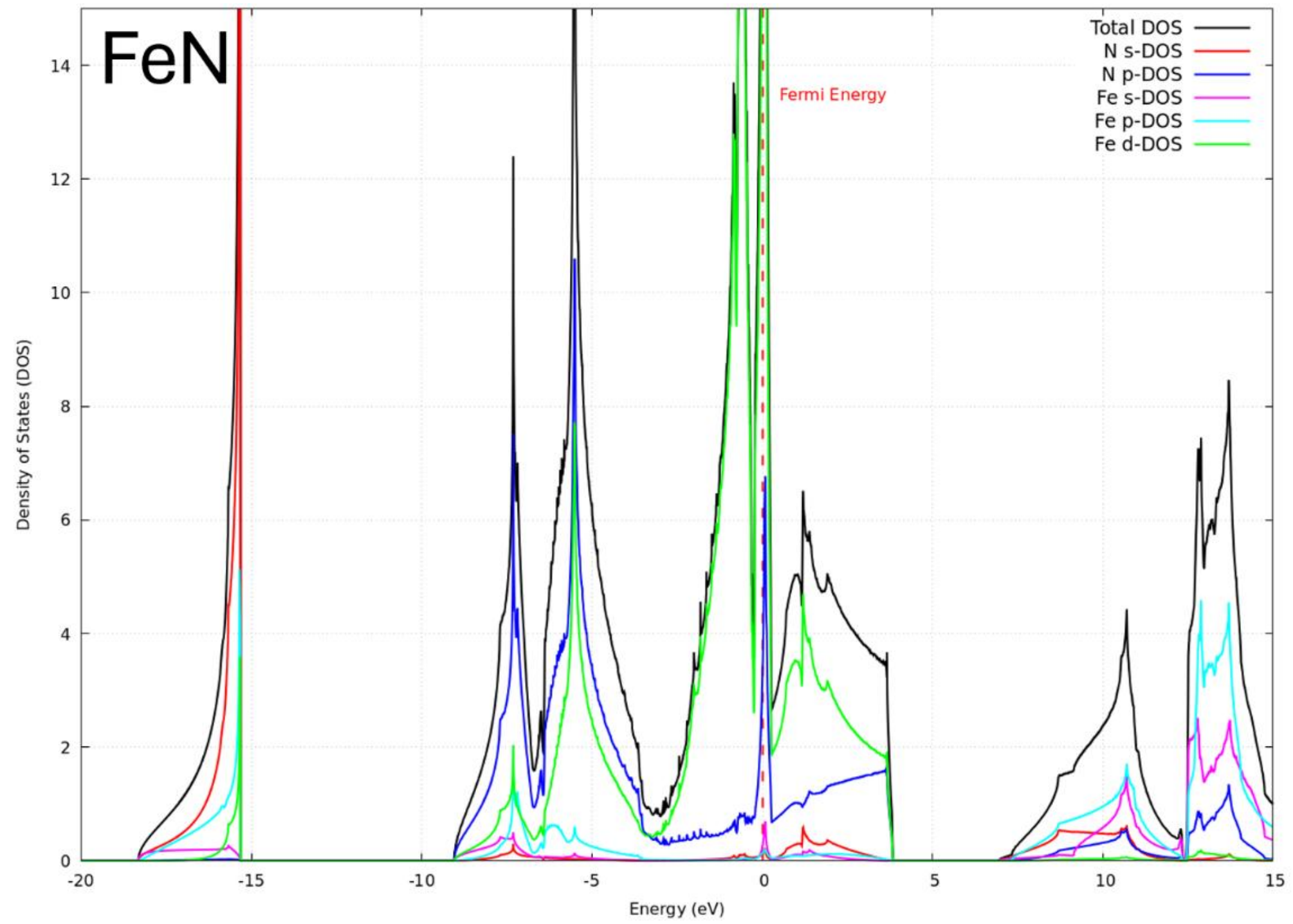

Figure 8: Projected local density of states for rocksalt FeN.

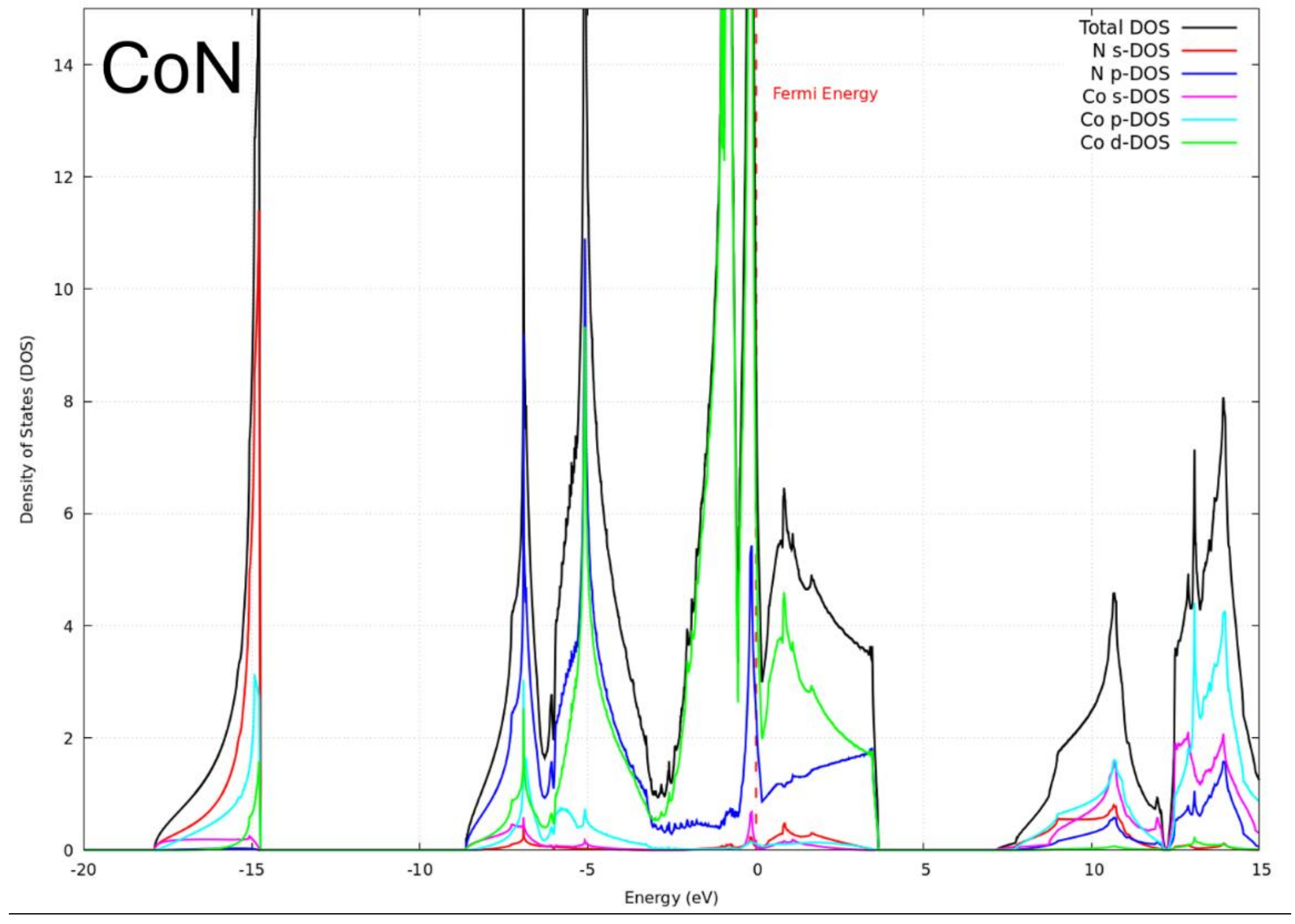

Figure 9: Projected local density of states for rocksalt CoN.

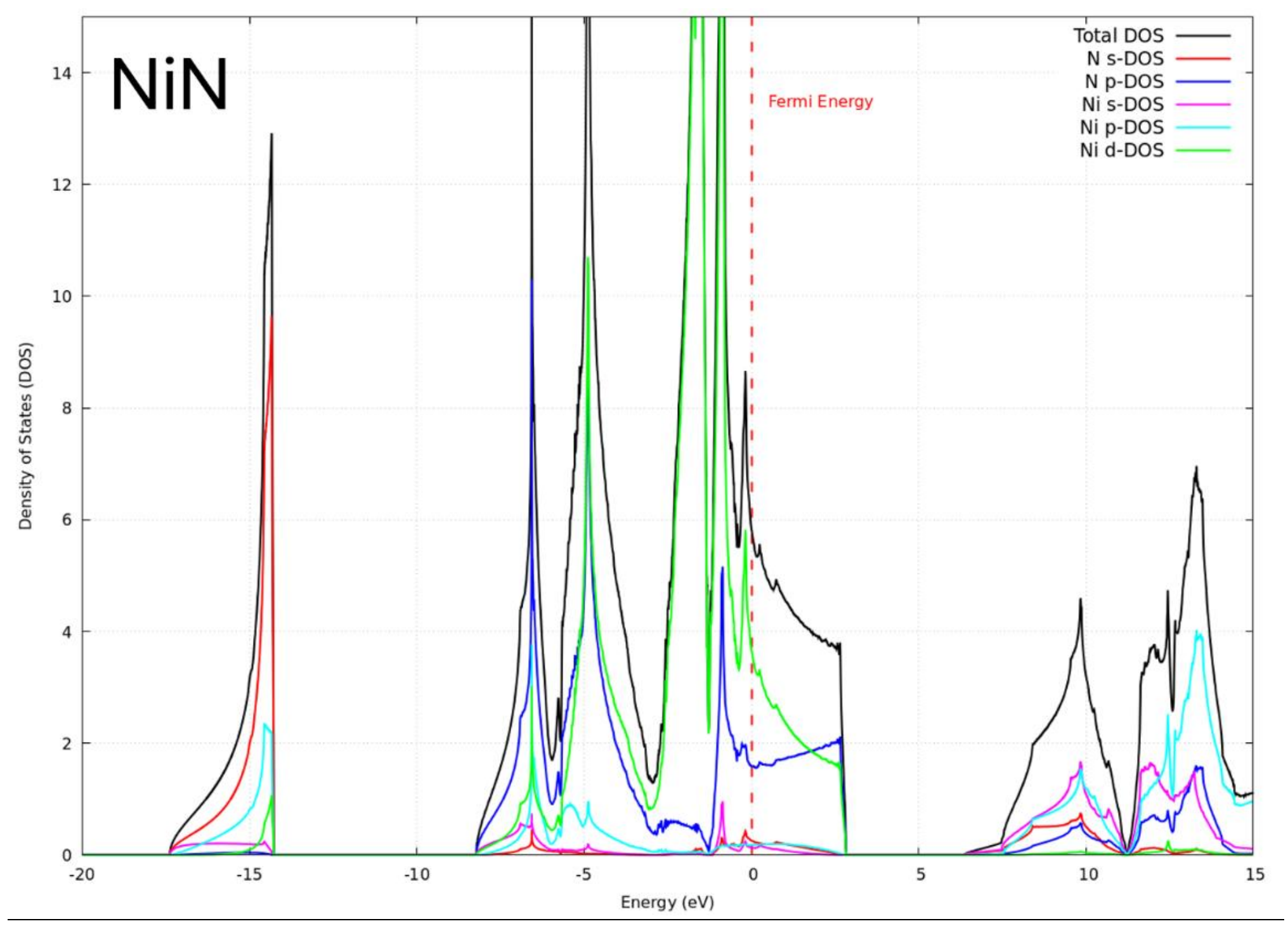

Figure 10: Projected local density of states for rocksalt NiN.

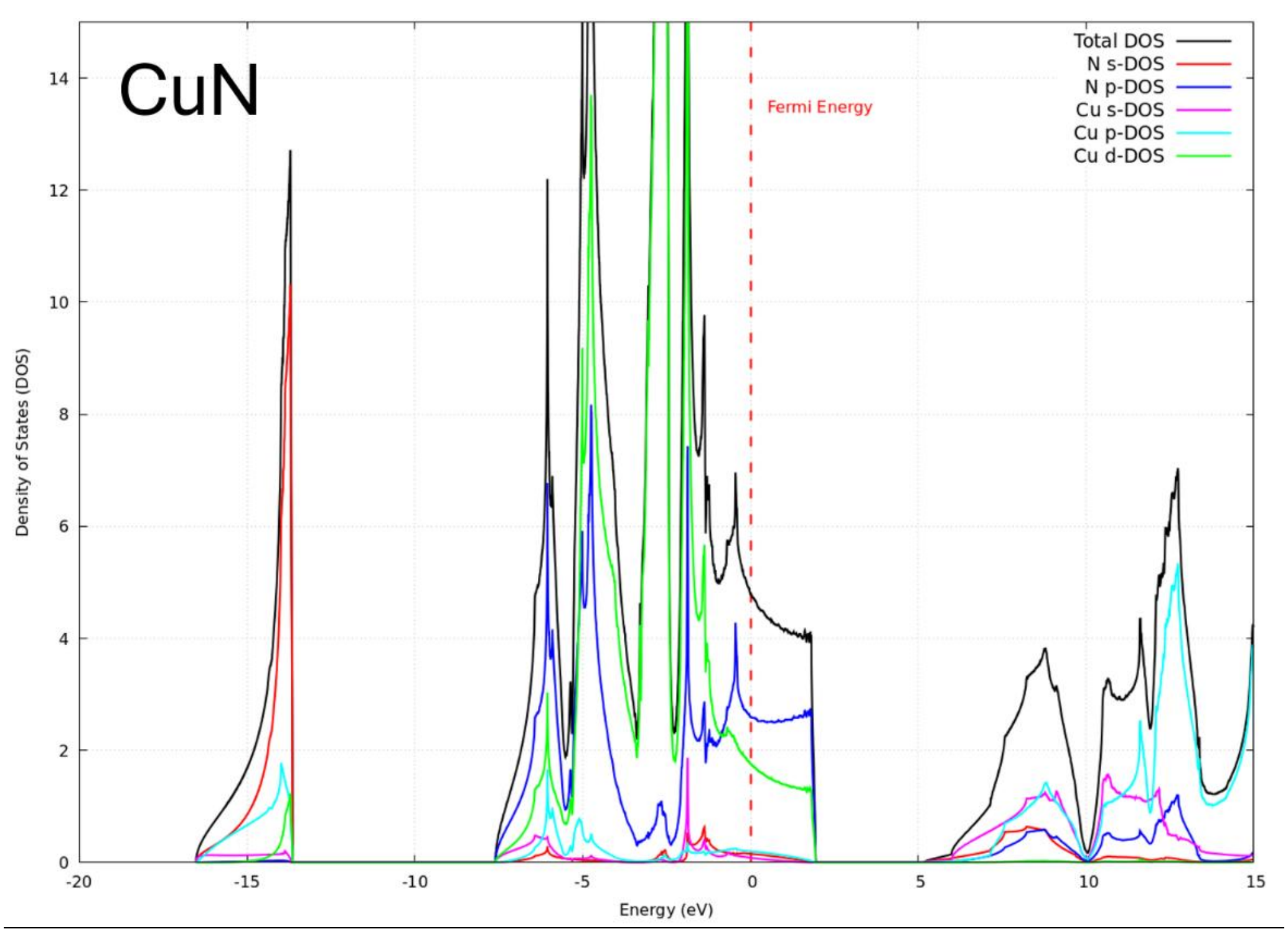

Figure 11: Projected local density of states for rocksalt CoN.

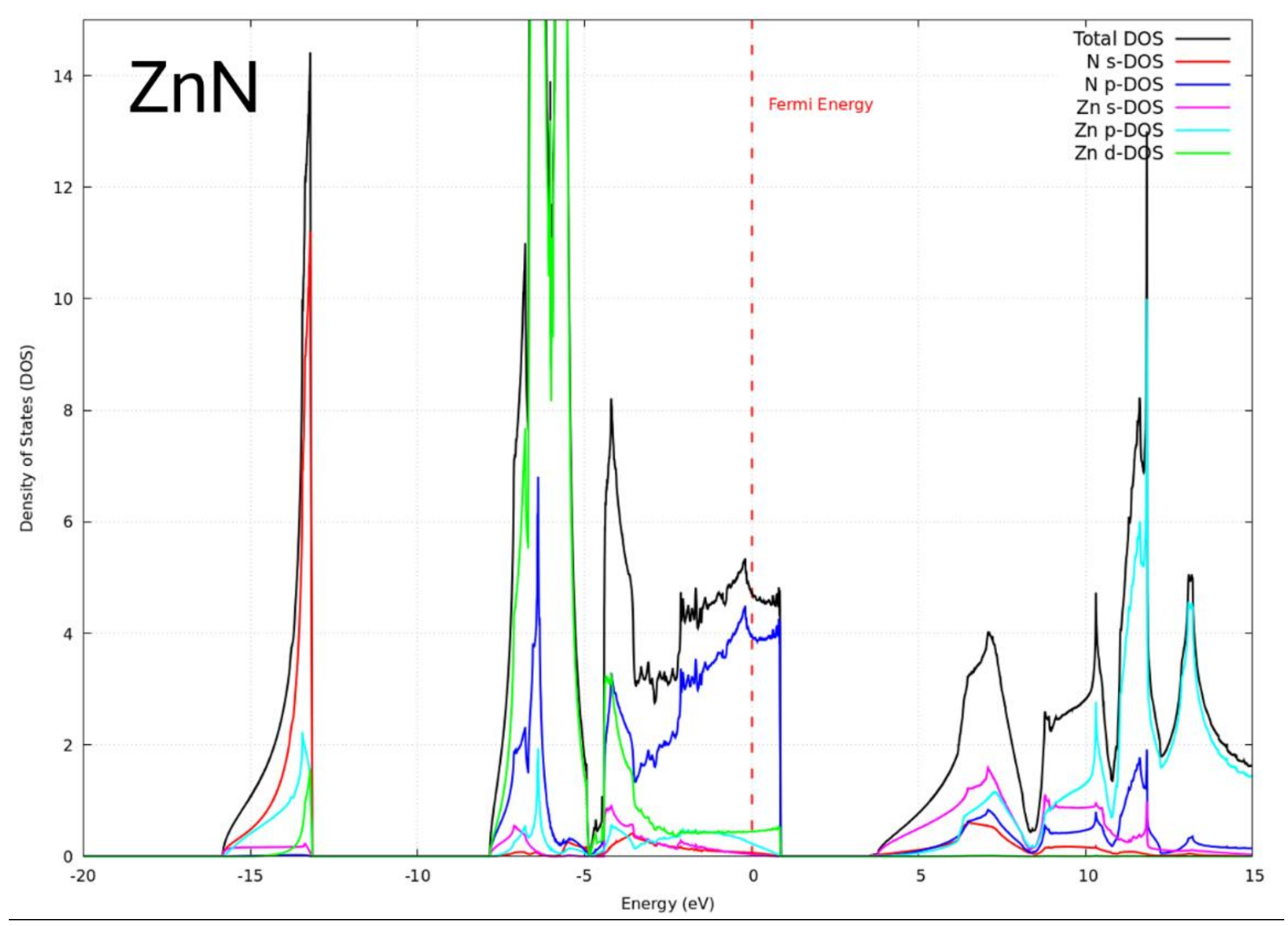

Figure 12: Projected local density of states for rocksalt ZnN.

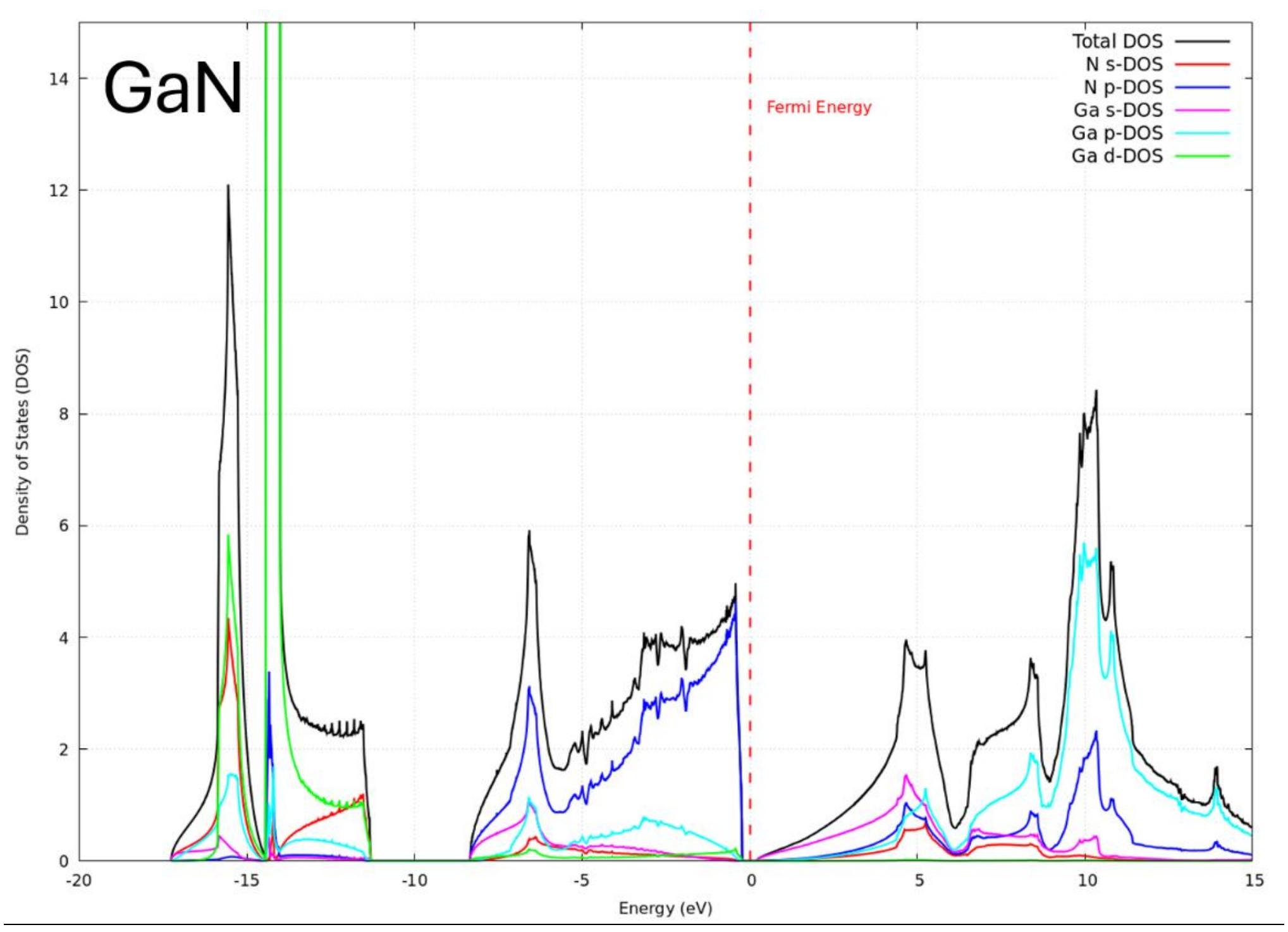

Figure 13: Projected local density of states for rocksalt GaN.

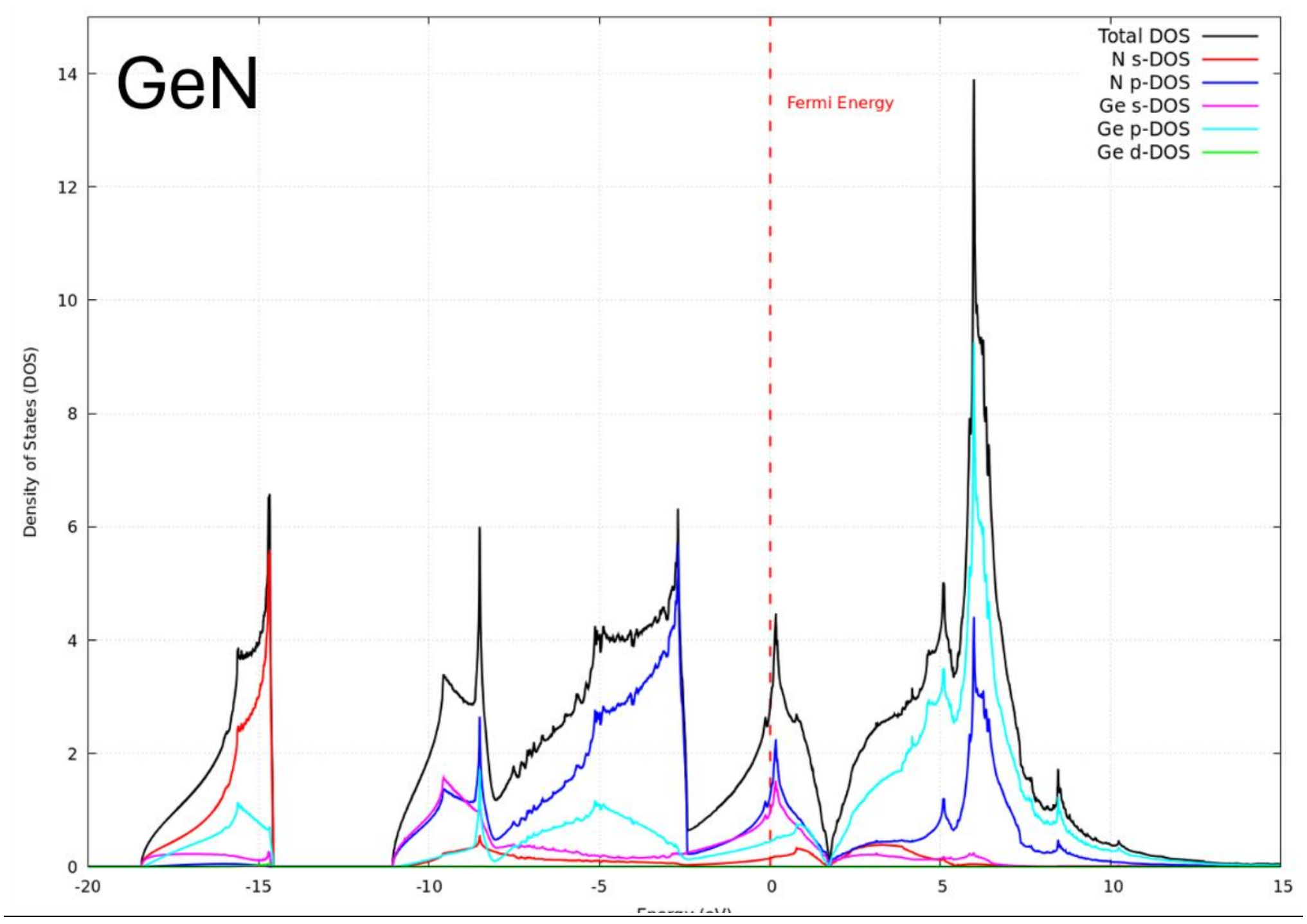

Figure 14: Projected local density of states for rocksalt GeN.

## Projected Density of States for Zincblende 4th Period Nitrides

For the zincblende compounds, the electronic structure of the transition-metal nitrides also evolves systematically across the series, with the progressive filling of the electronic states and the corresponding shift of the Fermi level as the atomic number increases. The projected densities of states for the zincblende binary metal nitrides are presented in Figures 15–28. The different crystal symmetry and coordination environment of the zincblende structure modify the orbital interactions compared with the rocksalt phase, particularly the interactions between the transition-metal $d$ states. Nevertheless, the overall evolution of the electronic structure remains governed by the progressive filling of the available states across the transition-metal series.

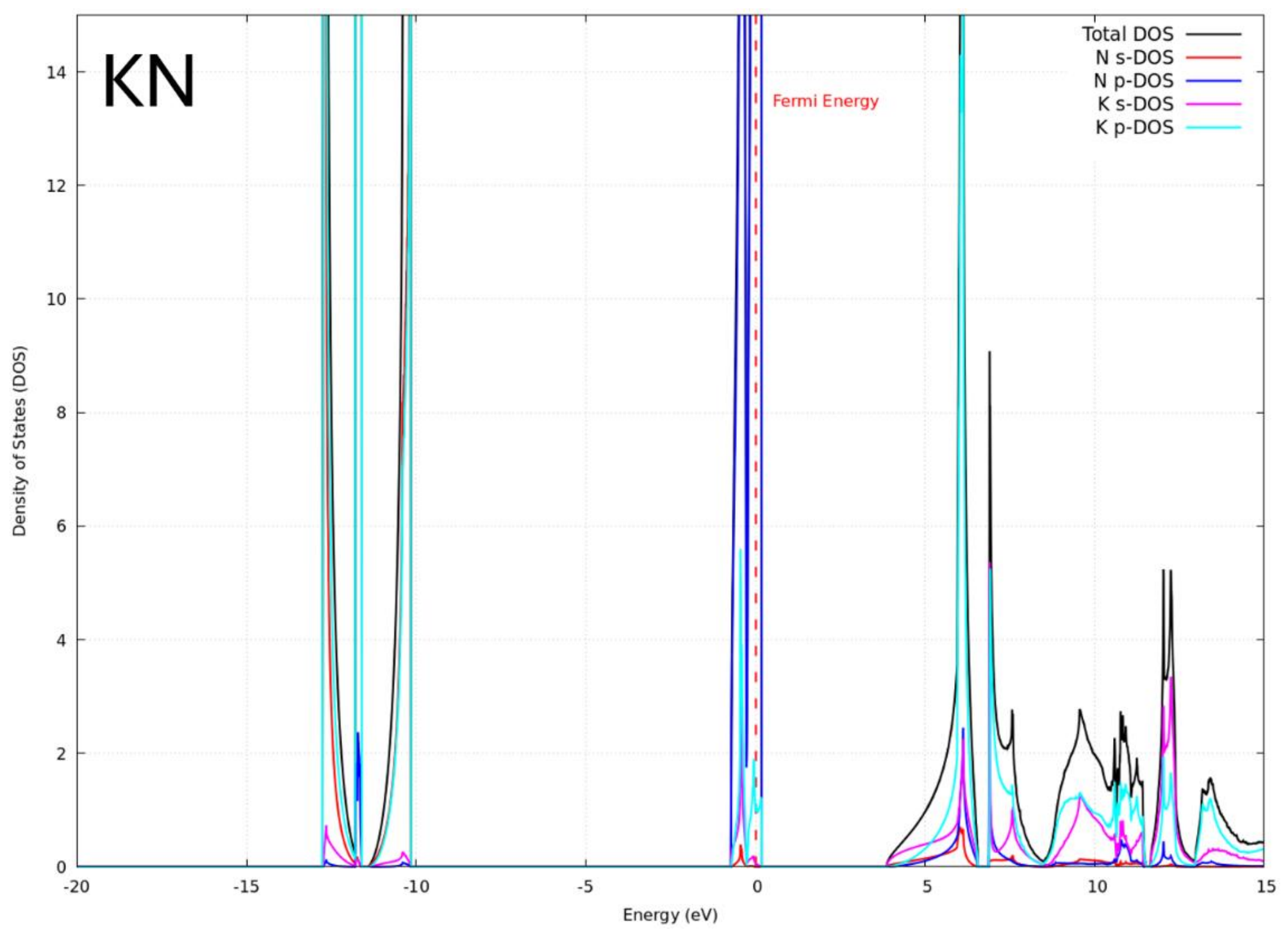

Figure 15: Projected local density of states for zincblende KN.

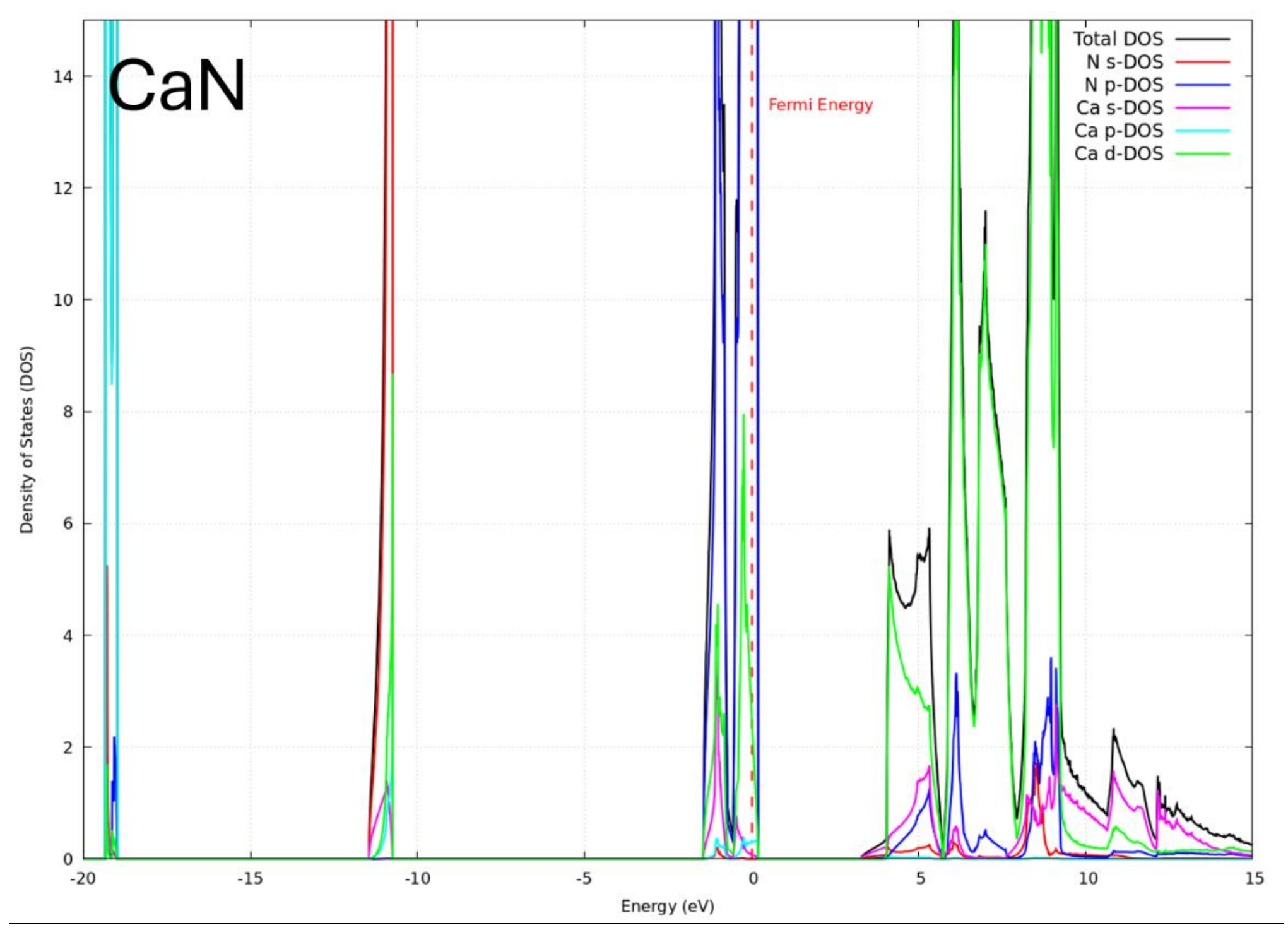

Figure 16: Projected local density of states for zincblende CaN.

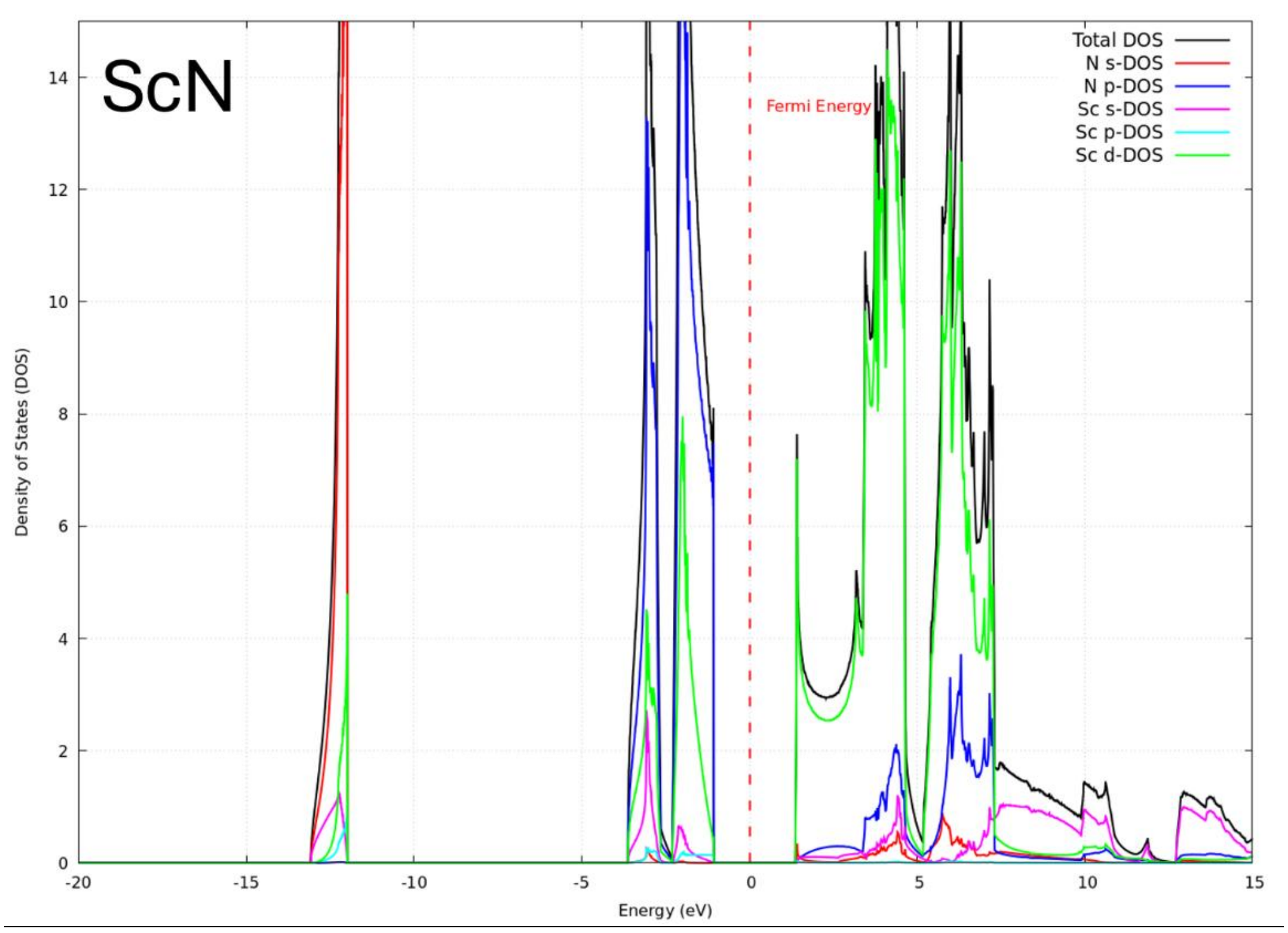

Figure 17: Projected local density of states for zincblende ScN.

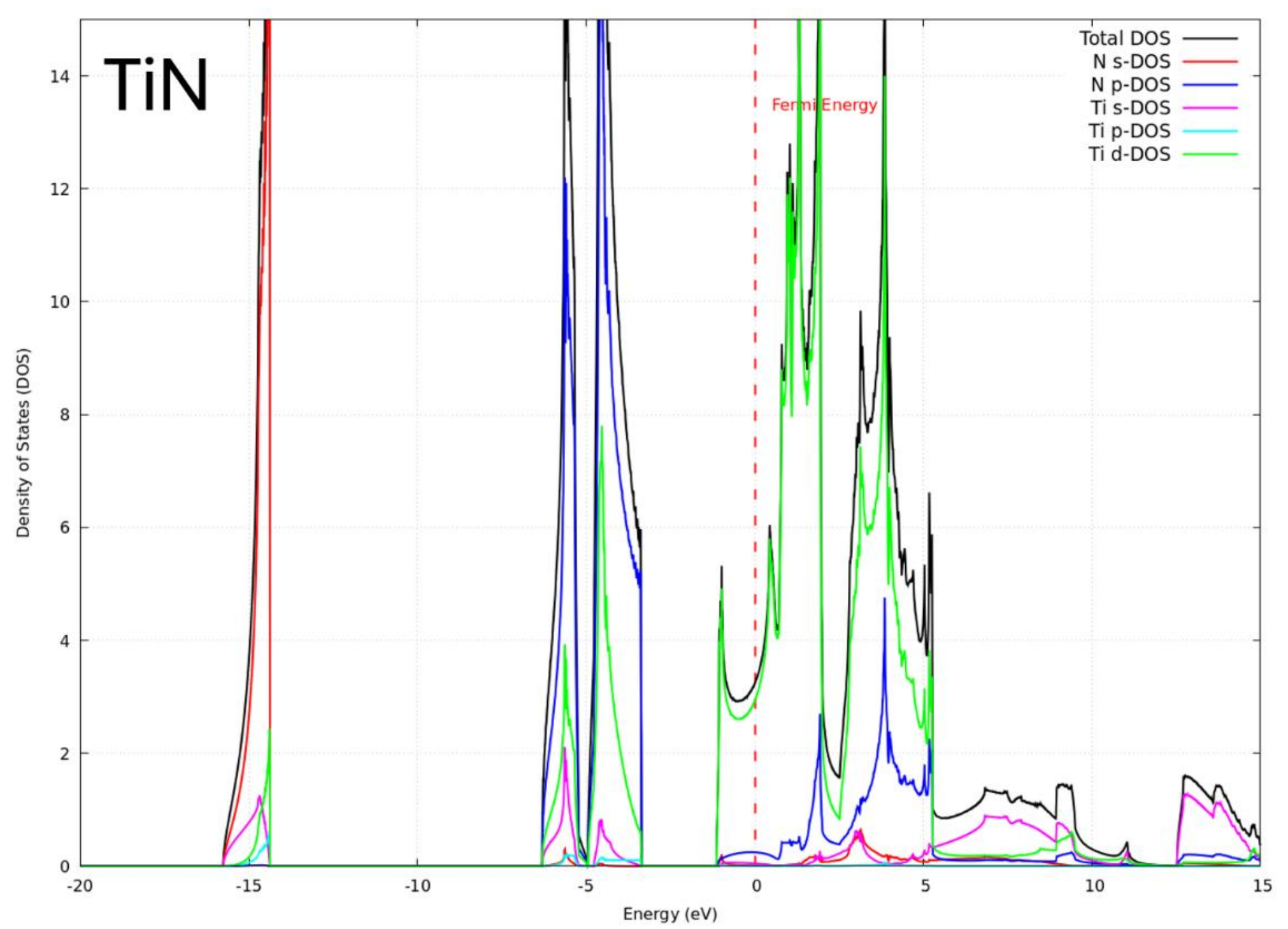

Figure 18: Projected local density of states for zincblende TiN.

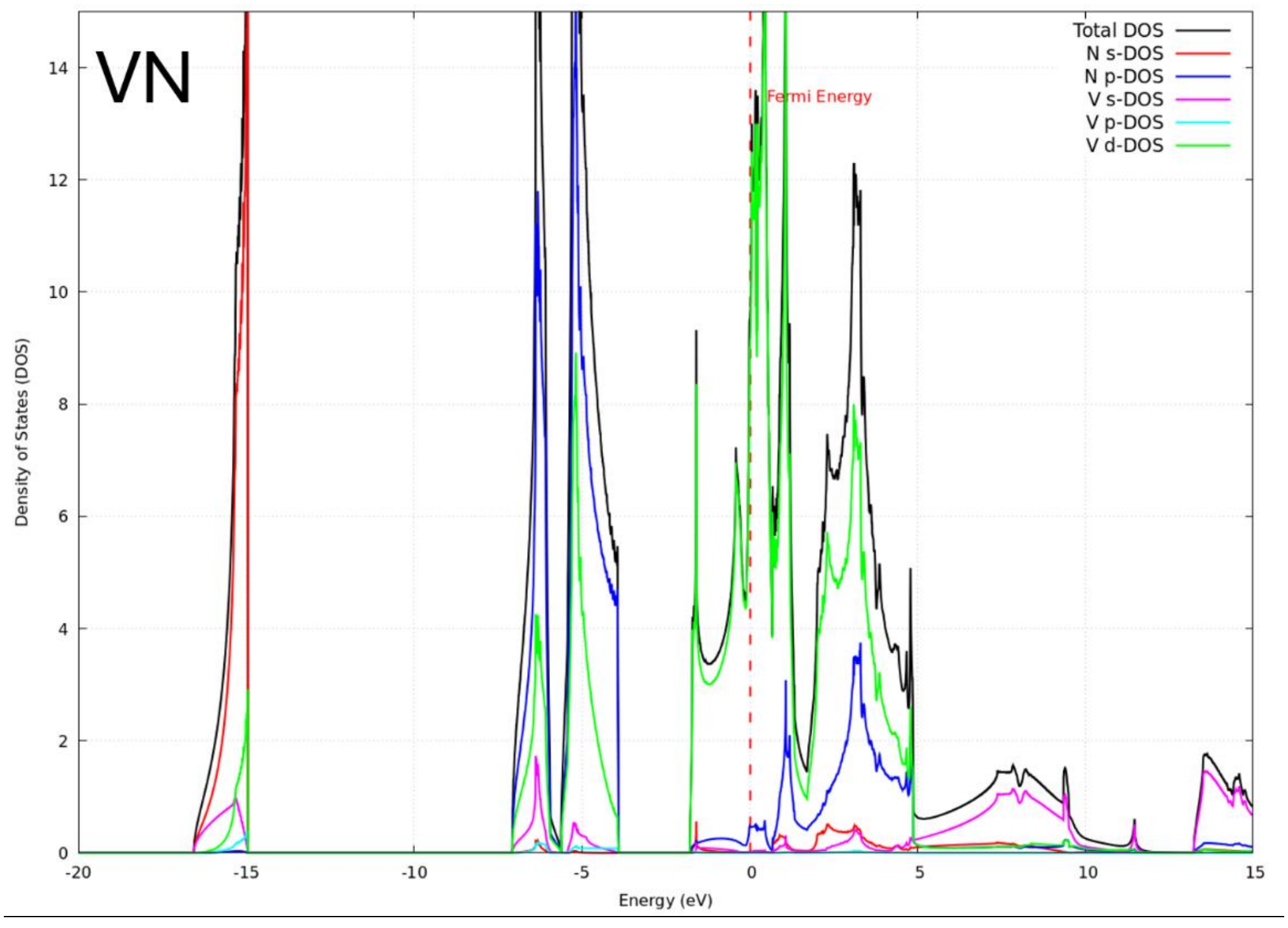

Figure 19: Projected local density of states for zincblende VN.

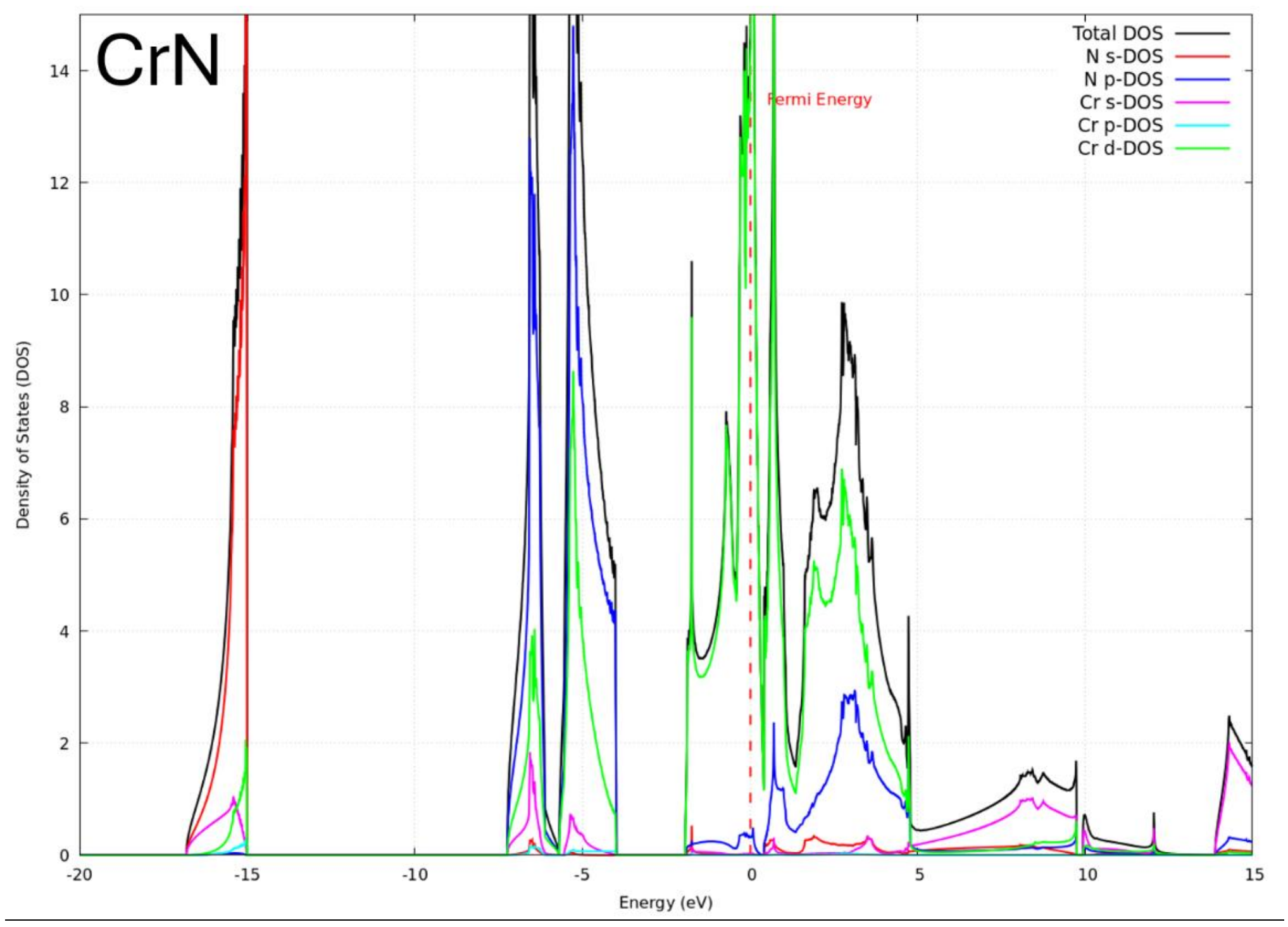

Figure 20: Projected local density of states for zincblende CrN.

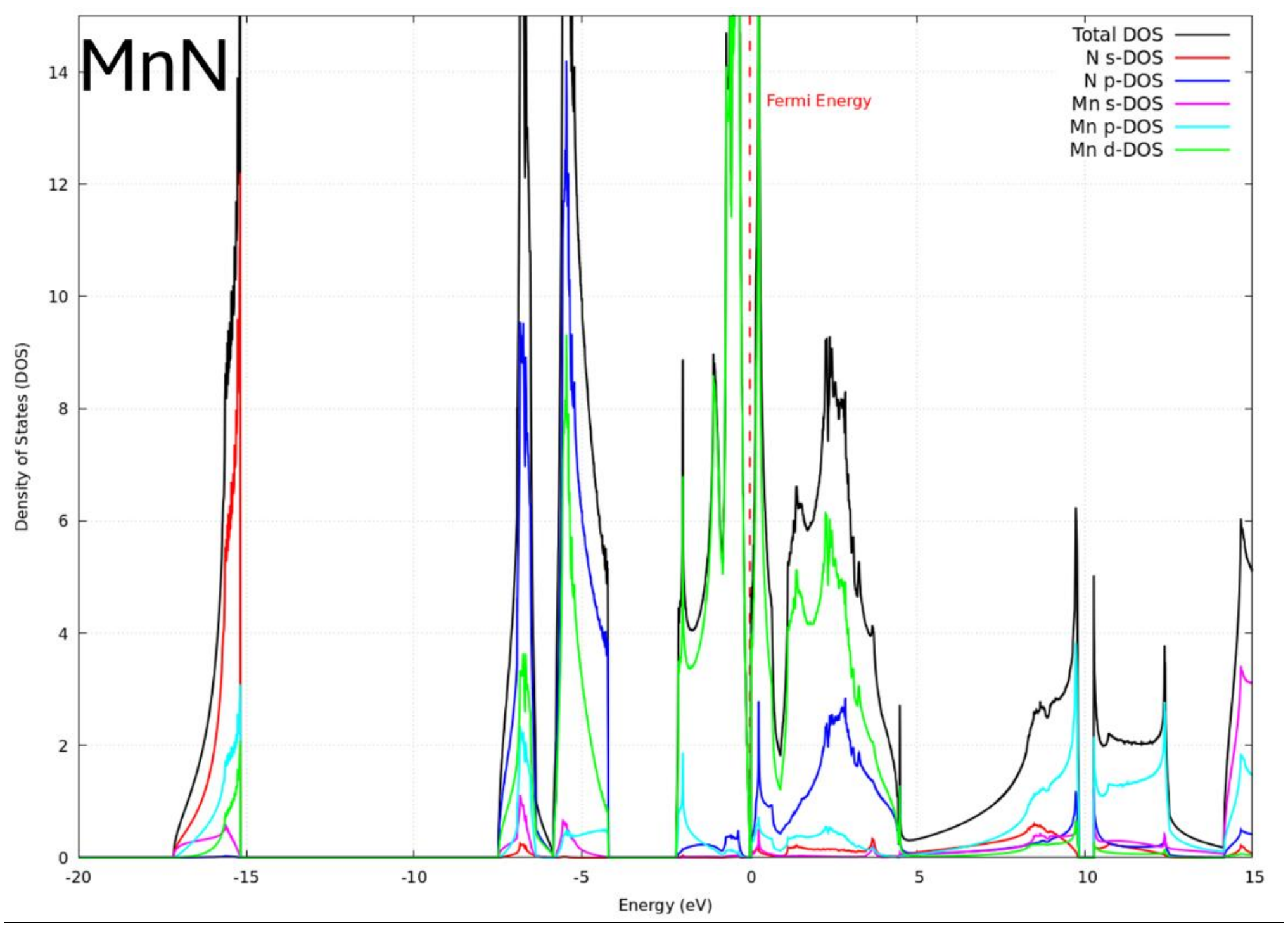

Figure 21: Projected local density of states for zincblende MnN.

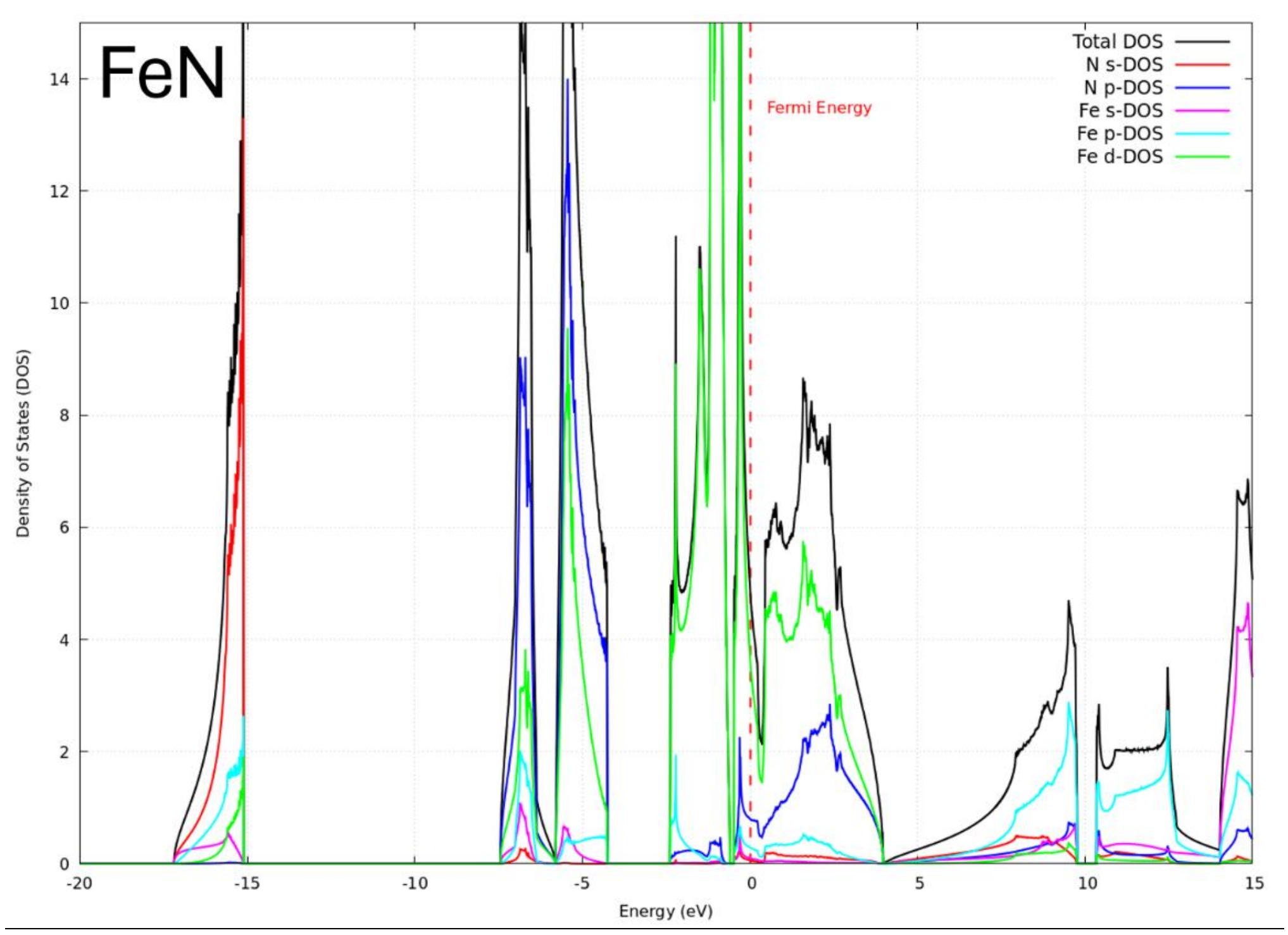

Figure 22: Projected local density of states for zincblende FeN.

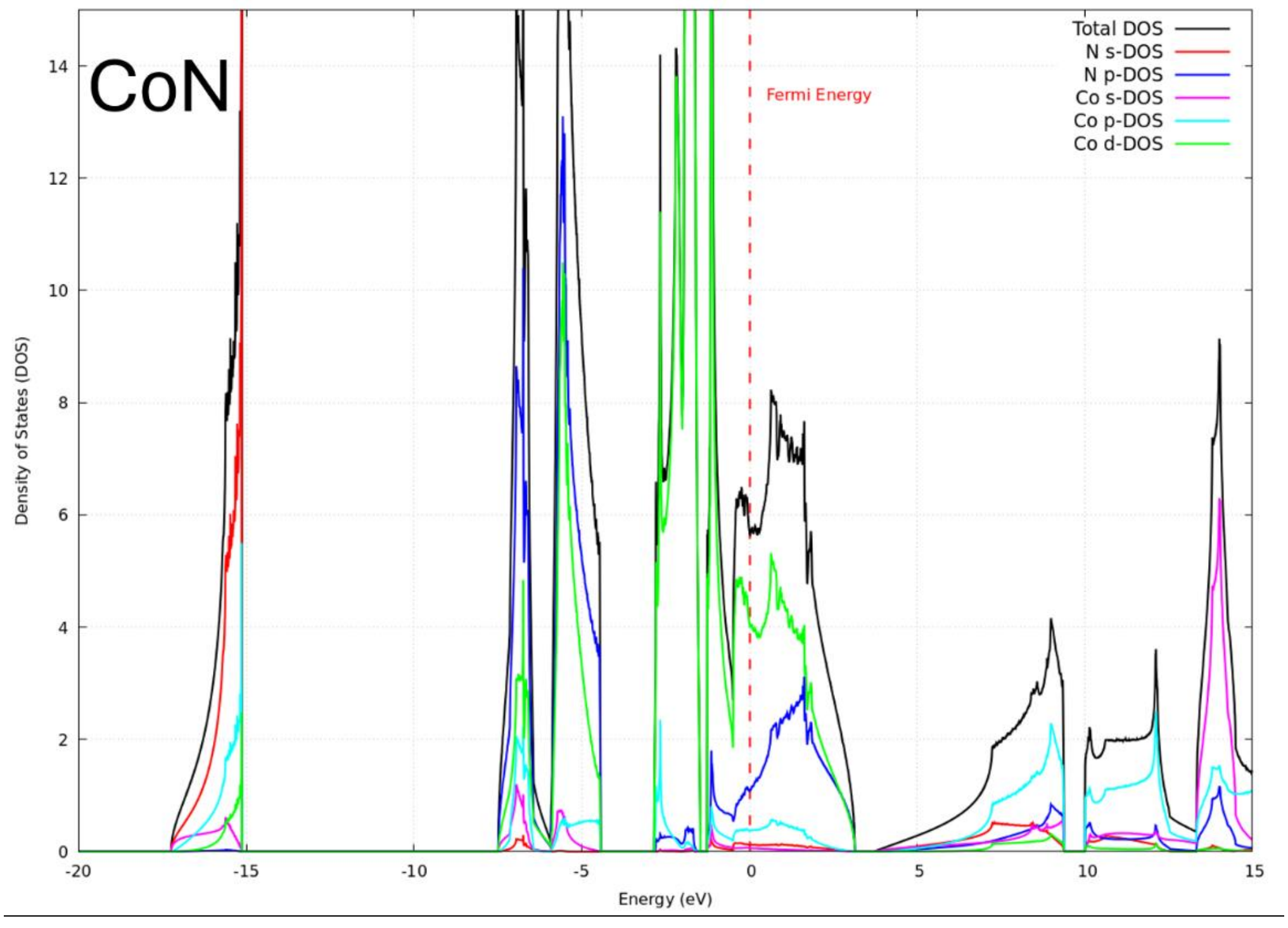

Figure 23: Projected local density of states for zincblende CoN.

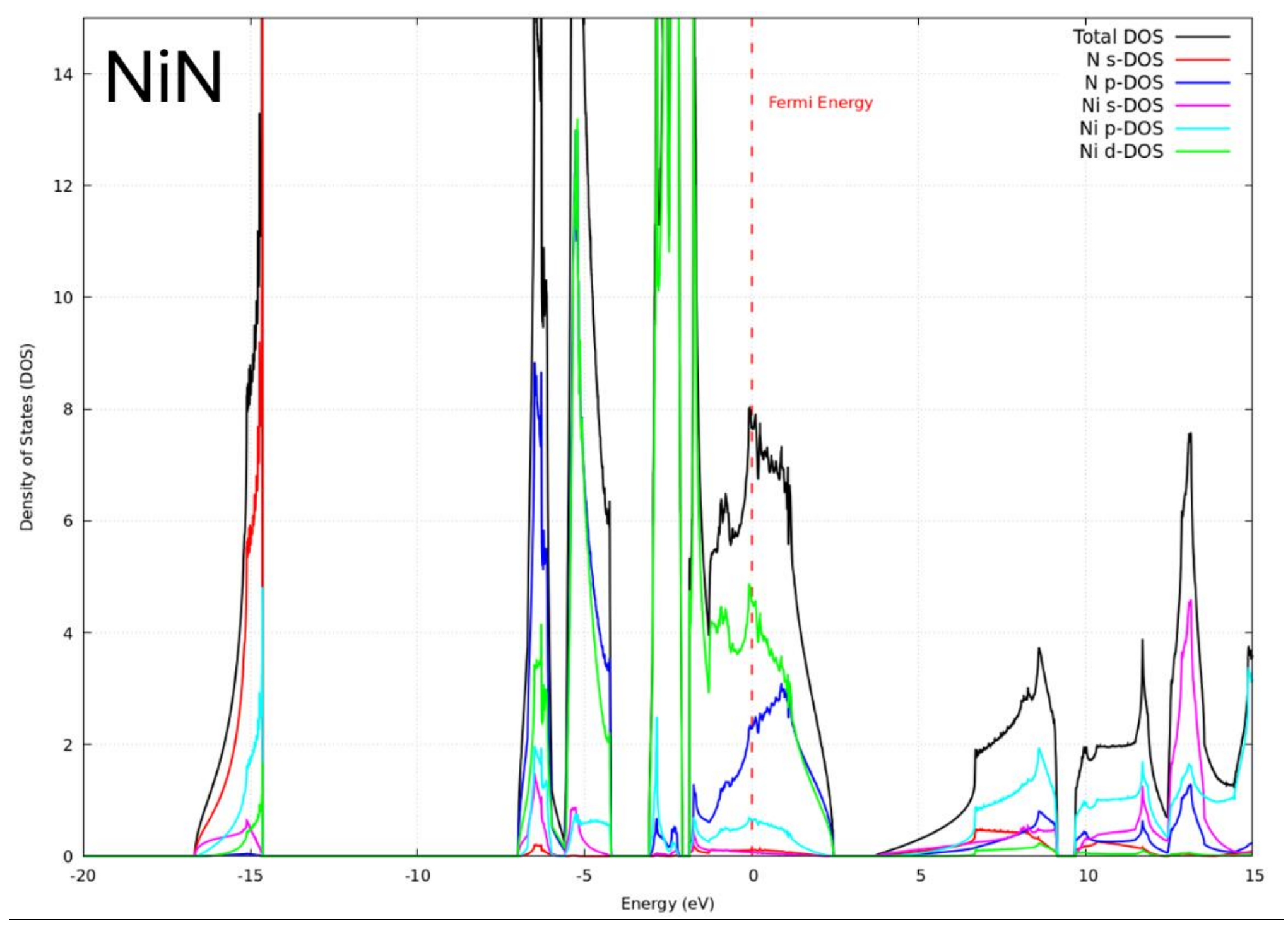

Figure 24: Projected local density of states for zincblende NiN.

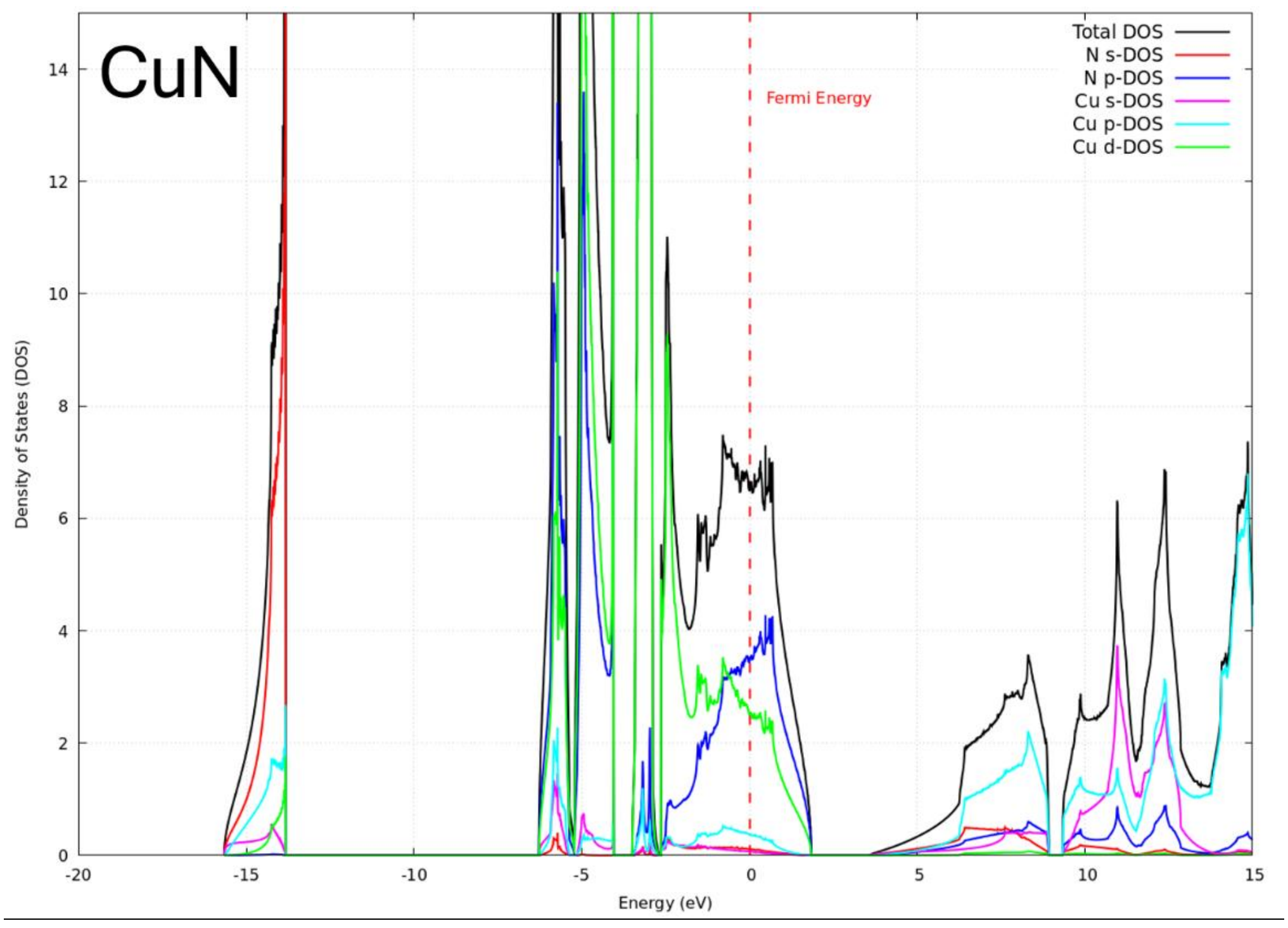

Figure 25: Projected local density of states for zincblende CuN.

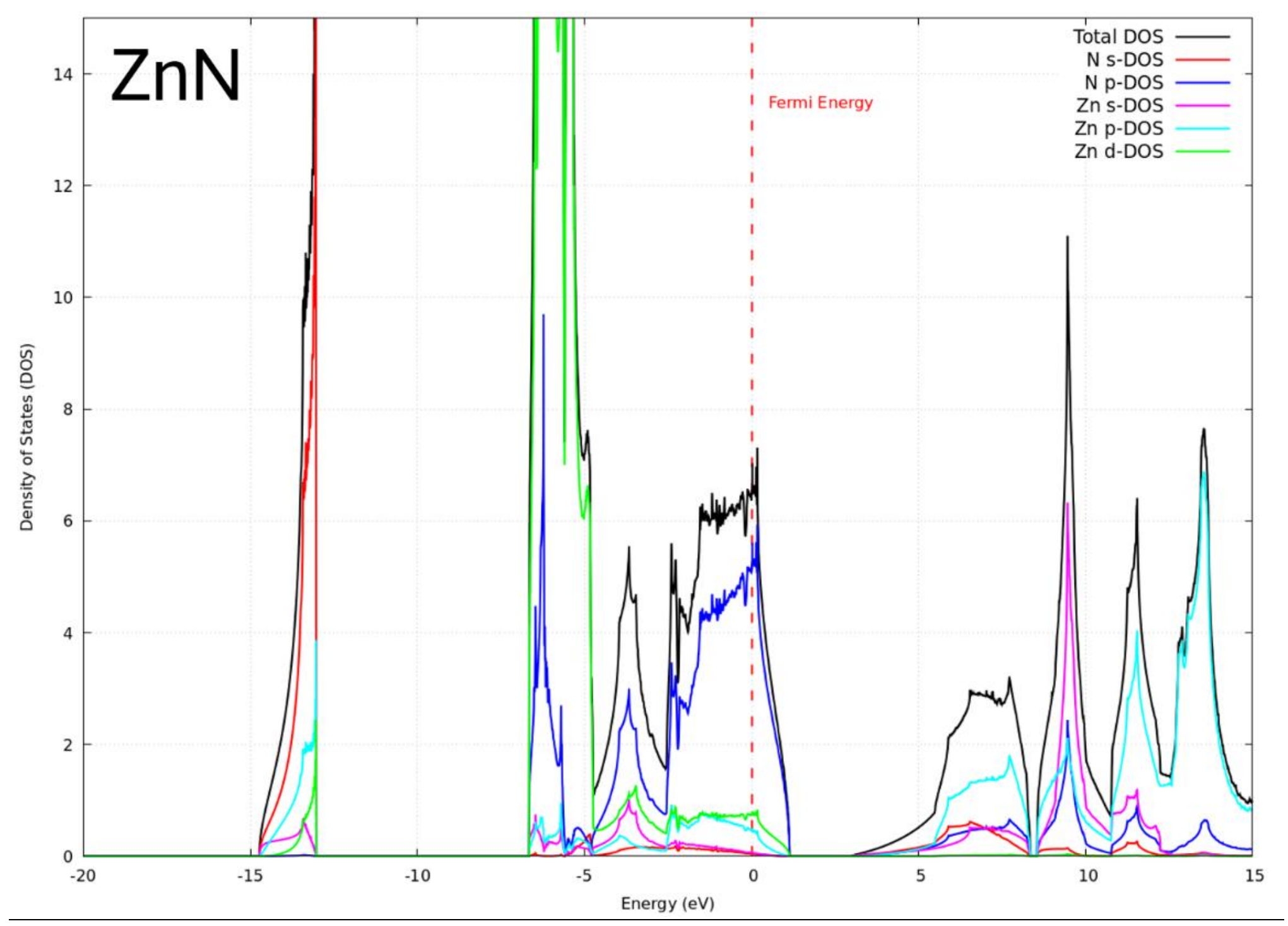

Figure 26: Projected local density of states for zincblende ZnN.

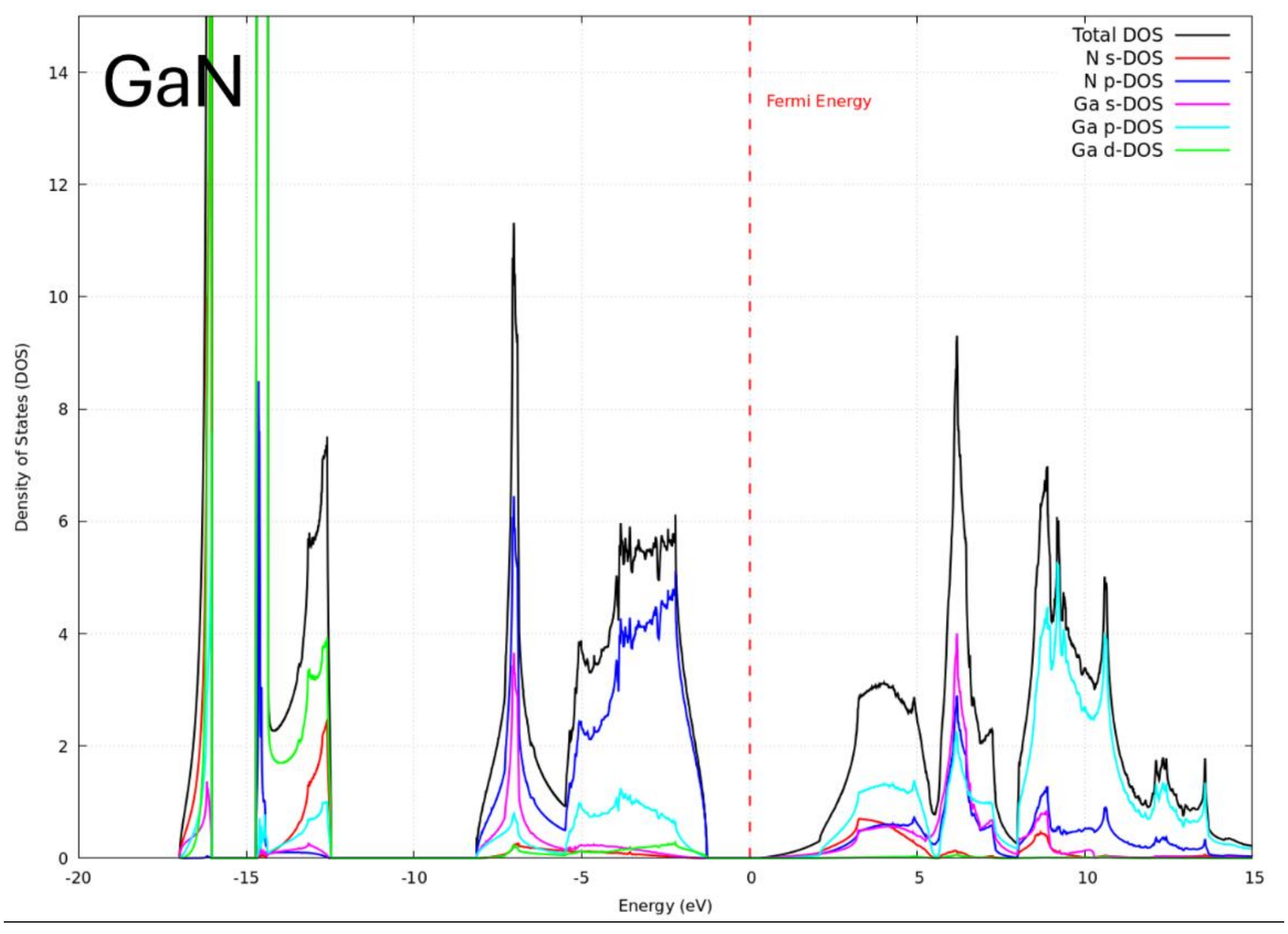

Figure 27: Projected local density of states for zincblende GaN.

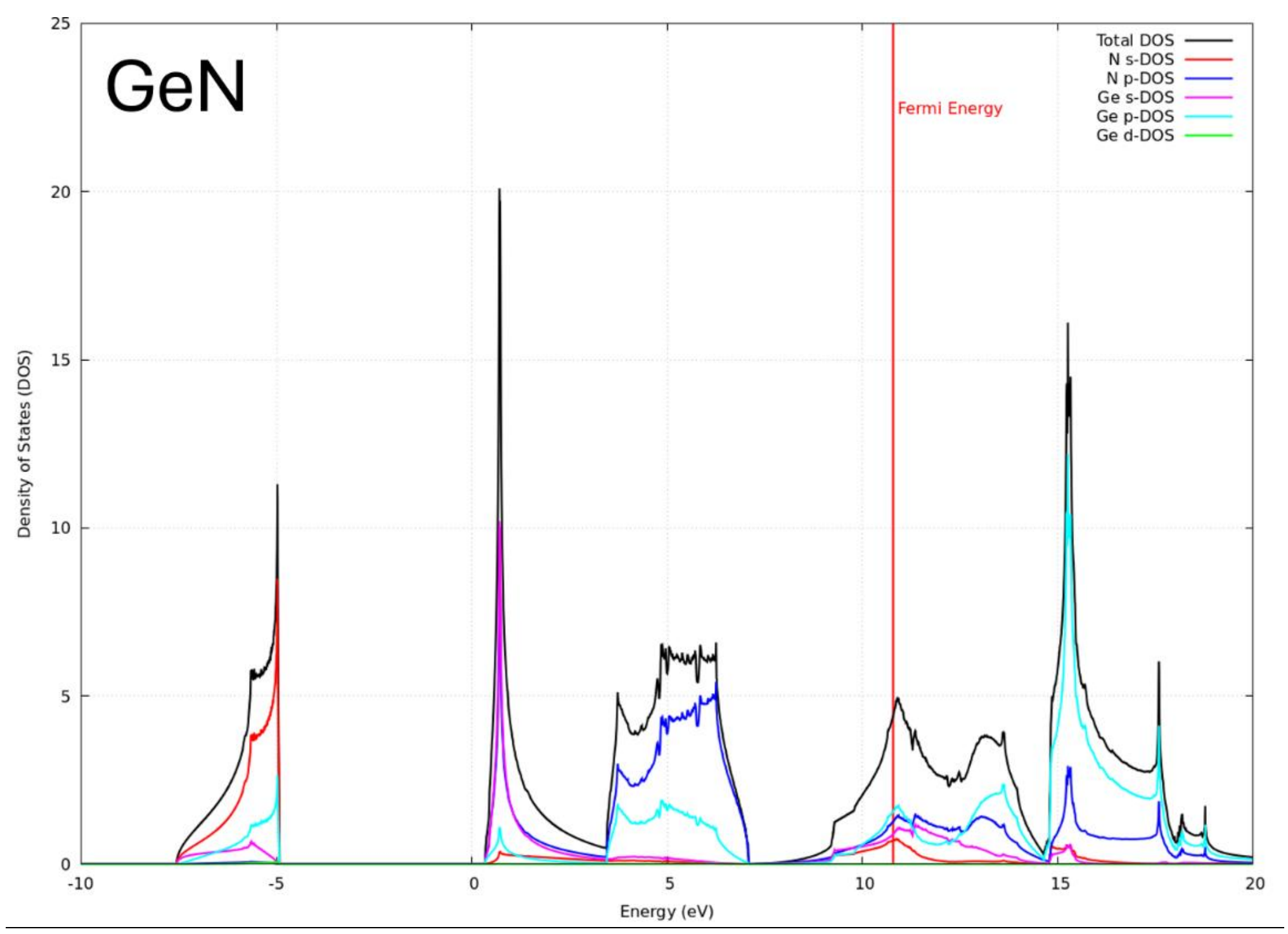

Figure 28: Projected local density of states for zincblende GeN.

### Projected Density of States for Hexagonal 4th Period Nitrides

For the hexagonal compounds, the electronic structure exhibits an evolution closely related to that observed for the zinc-blende phases, with the progressive filling of the electronic states and the corresponding shift of the Fermi level across the transition-metal series. The projected densities of states for the hexagonal binary metal nitrides are presented in Figures 29–42. The internal structural parameter $u$ introduce additional variations in the orbital interactions and hybridization compared with the zincblende structure. In particular, changes in $u$ modify the local coordination environment, influencing the detailed distribution of the electronic states and the resulting bonding characteristics. Interestingly, for the $u = 0.5$ layered hexagonal phases, the electronic structure often exhibits characteristics intermediate between those of the rocksalt and zinc-blende phases. This intermediate character is, for example, reflected in the ScN calculated bandgap, which lies between the corresponding value obtained for the rocksalt and zinc-blende structures.

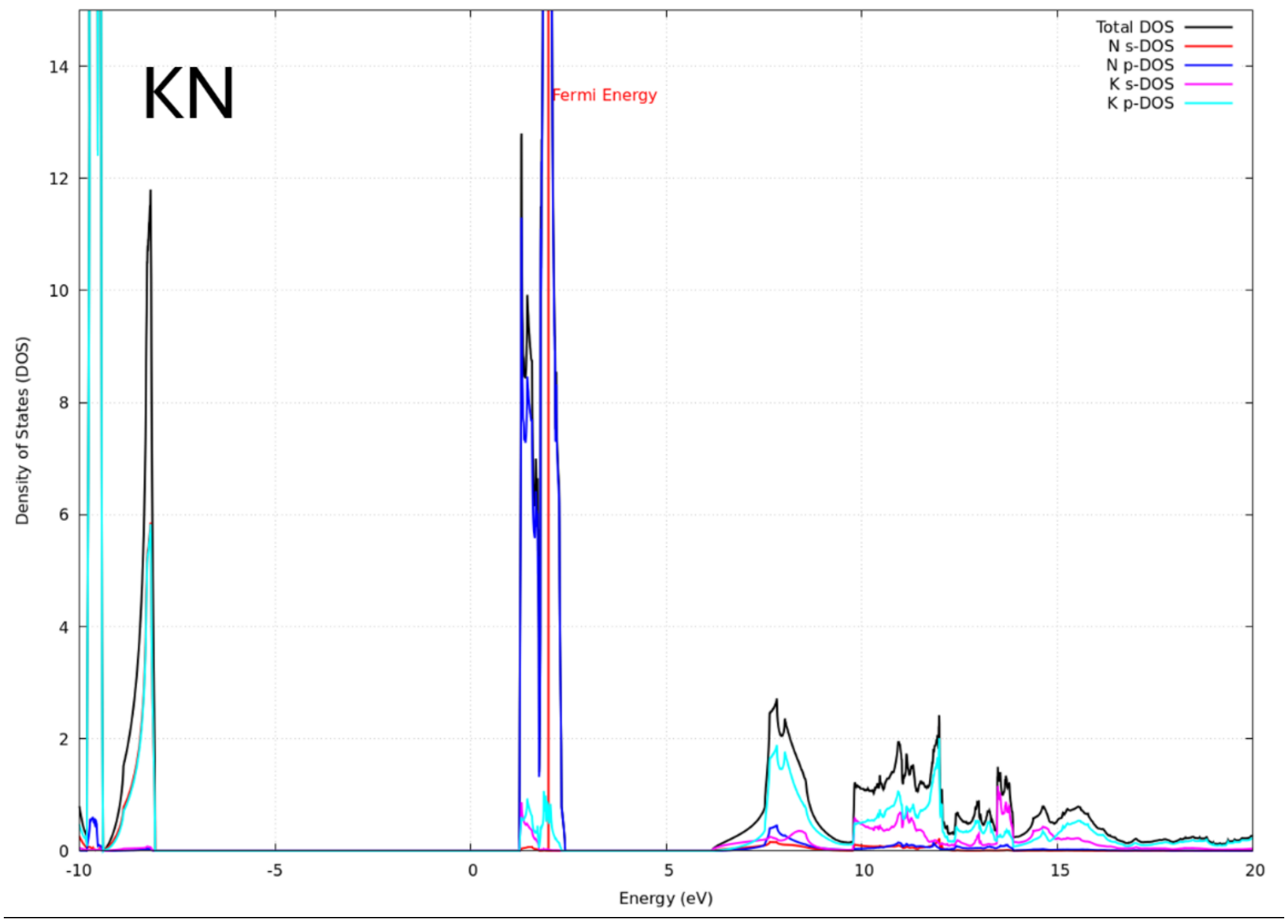

Figure 29: Projected local density of states for hexagonal KN.

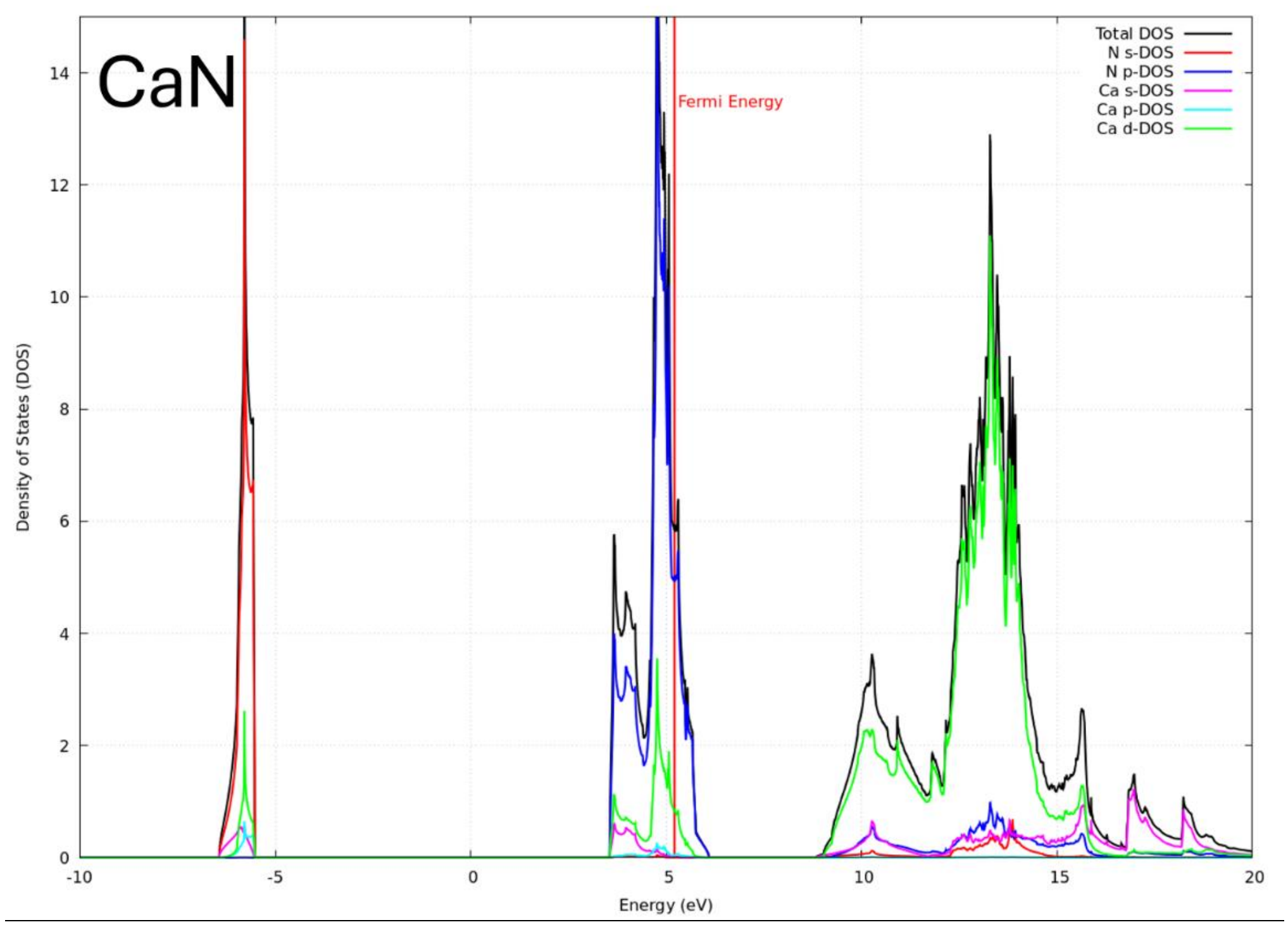

Figure 30: Projected local density of states for hexagonal CaN.

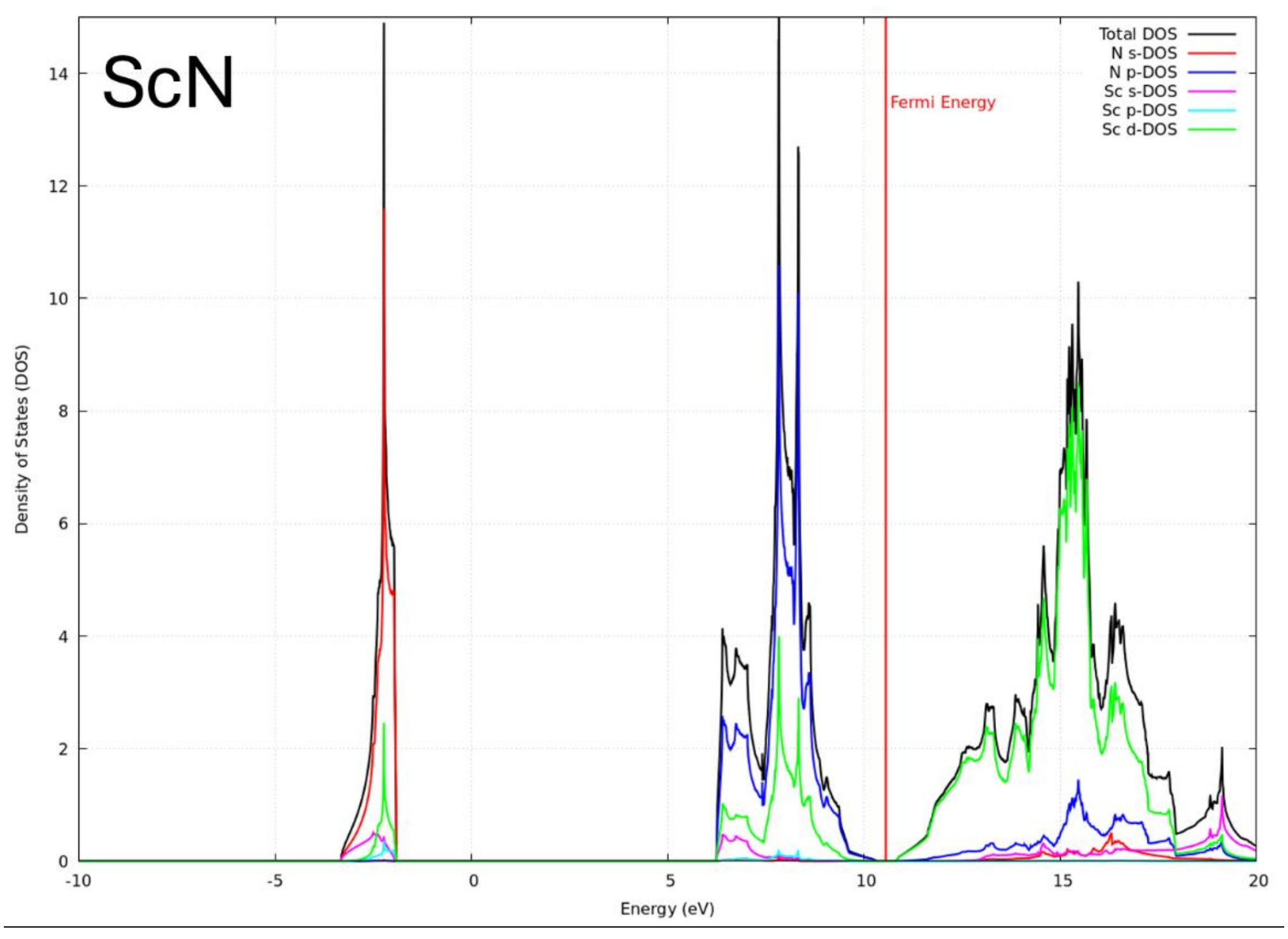

Figure 30: Projected local density of states for hexagonal ScN.

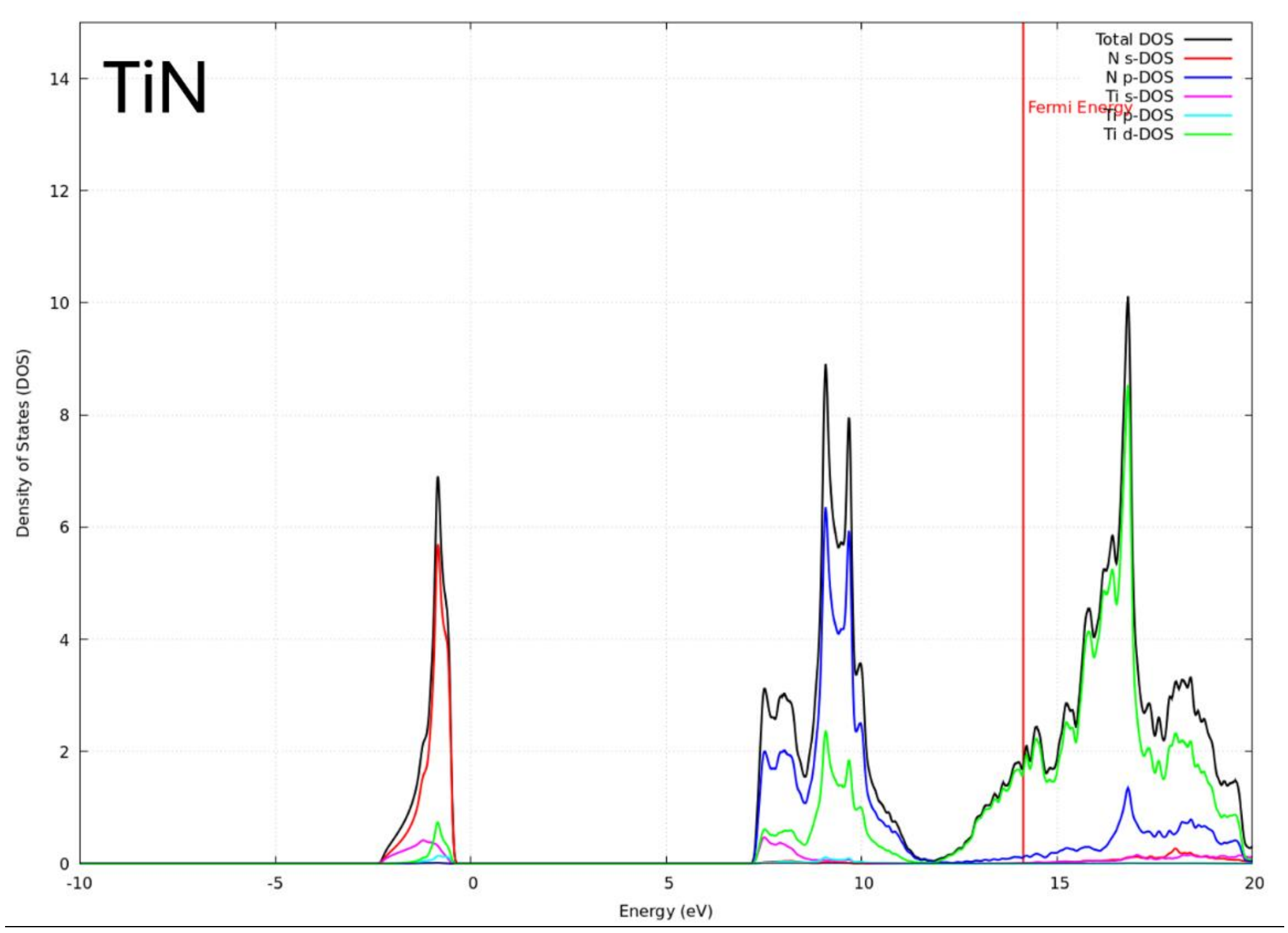

Figure 32: Projected local density of states for hexagonal TiN.

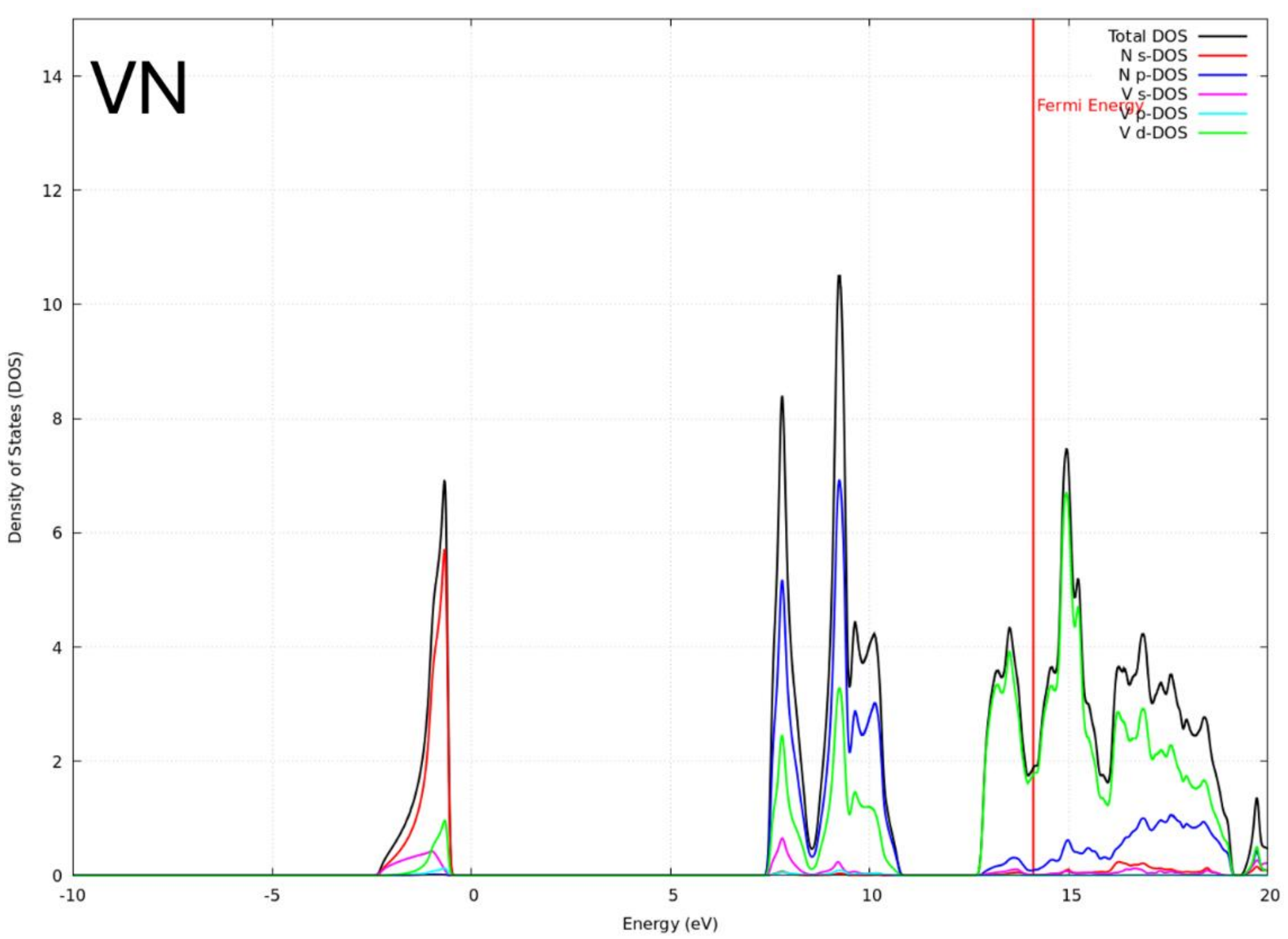

Figure 33: Projected local density of states for hexagonal VN.

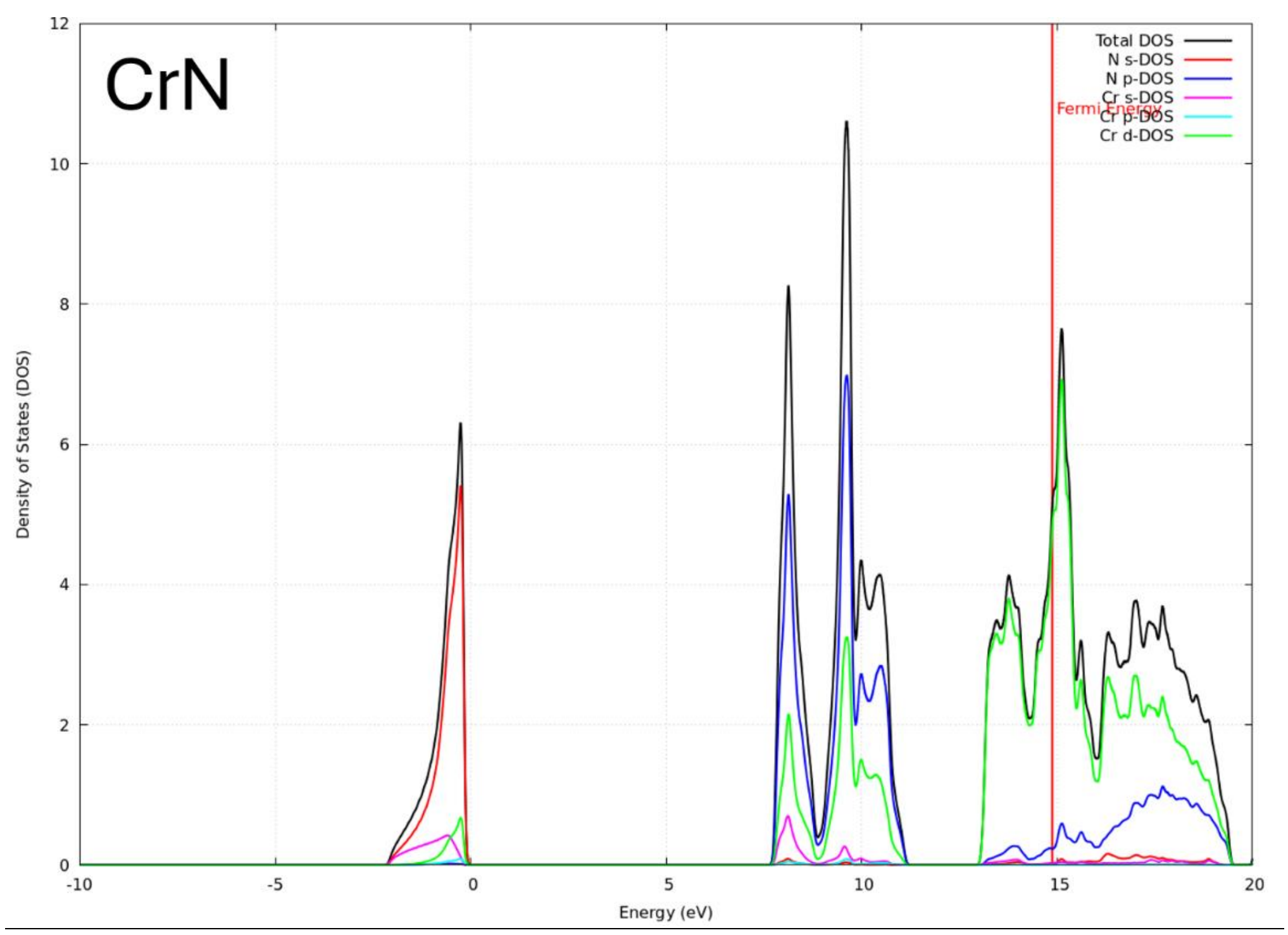

Figure 34: Projected local density of states for hexagonal CrN.

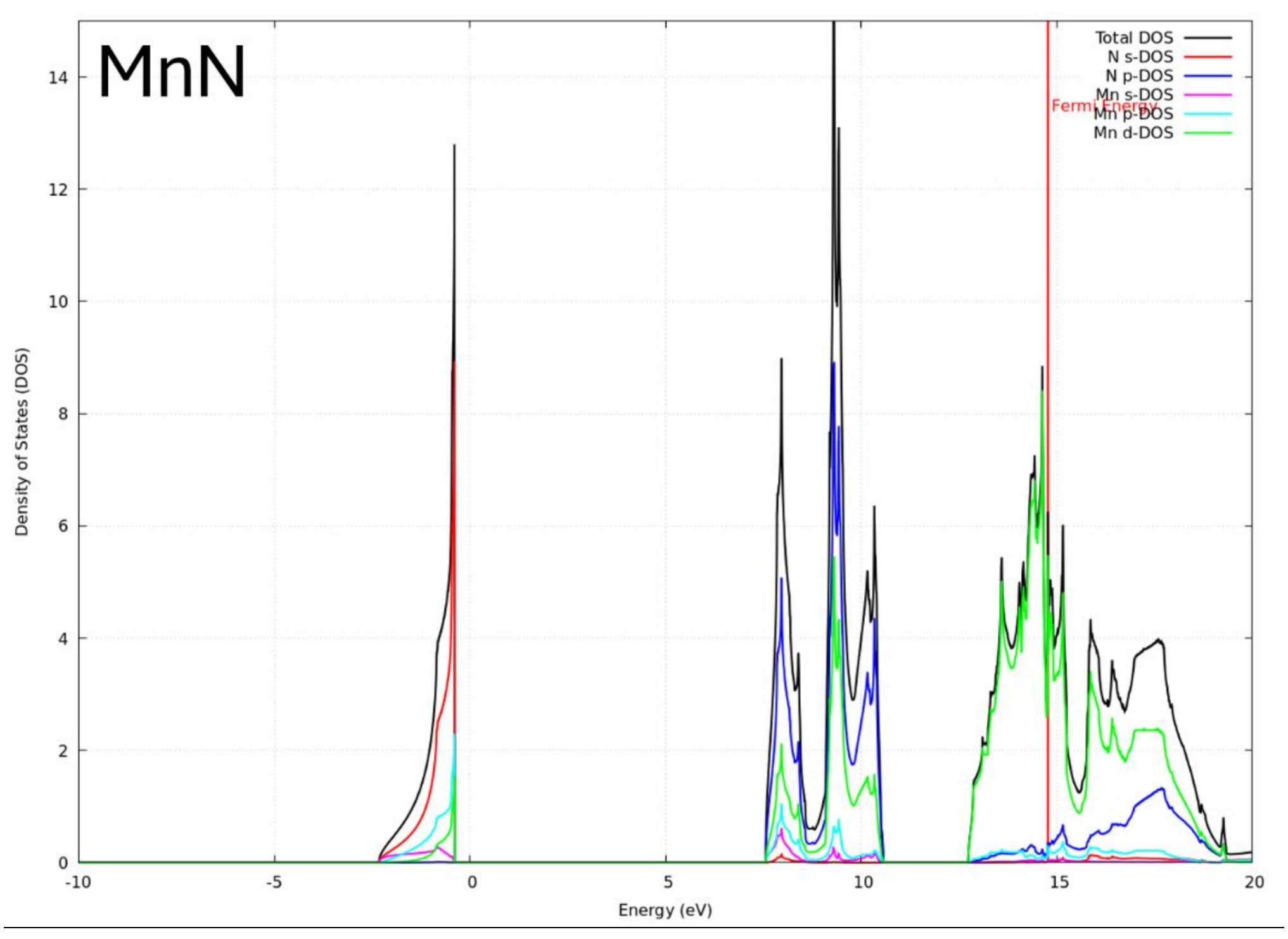

Figure 35: Projected local density of states for hexagonal MnN.

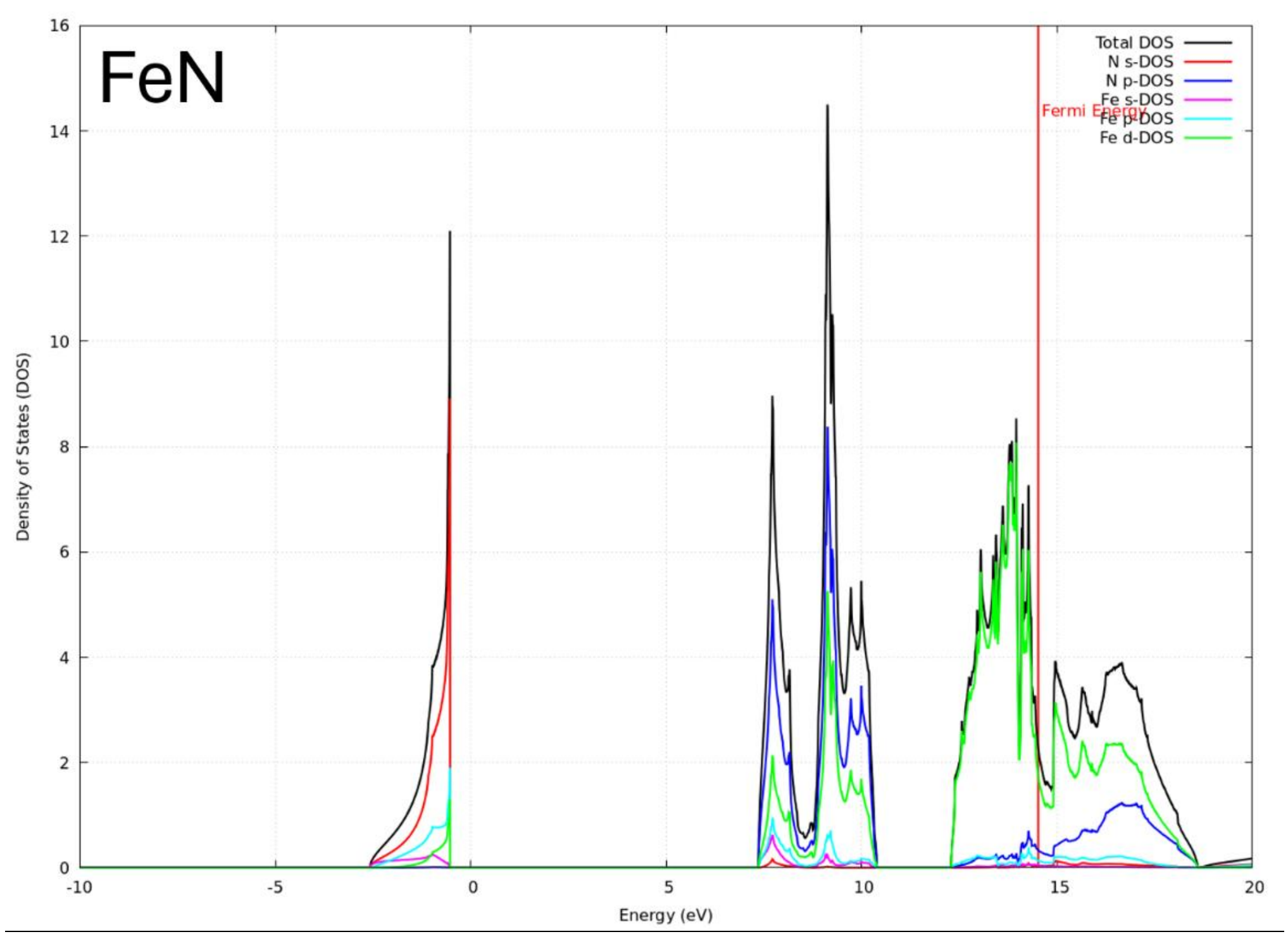

Figure 36: Projected local density of states for hexagonal FeN.

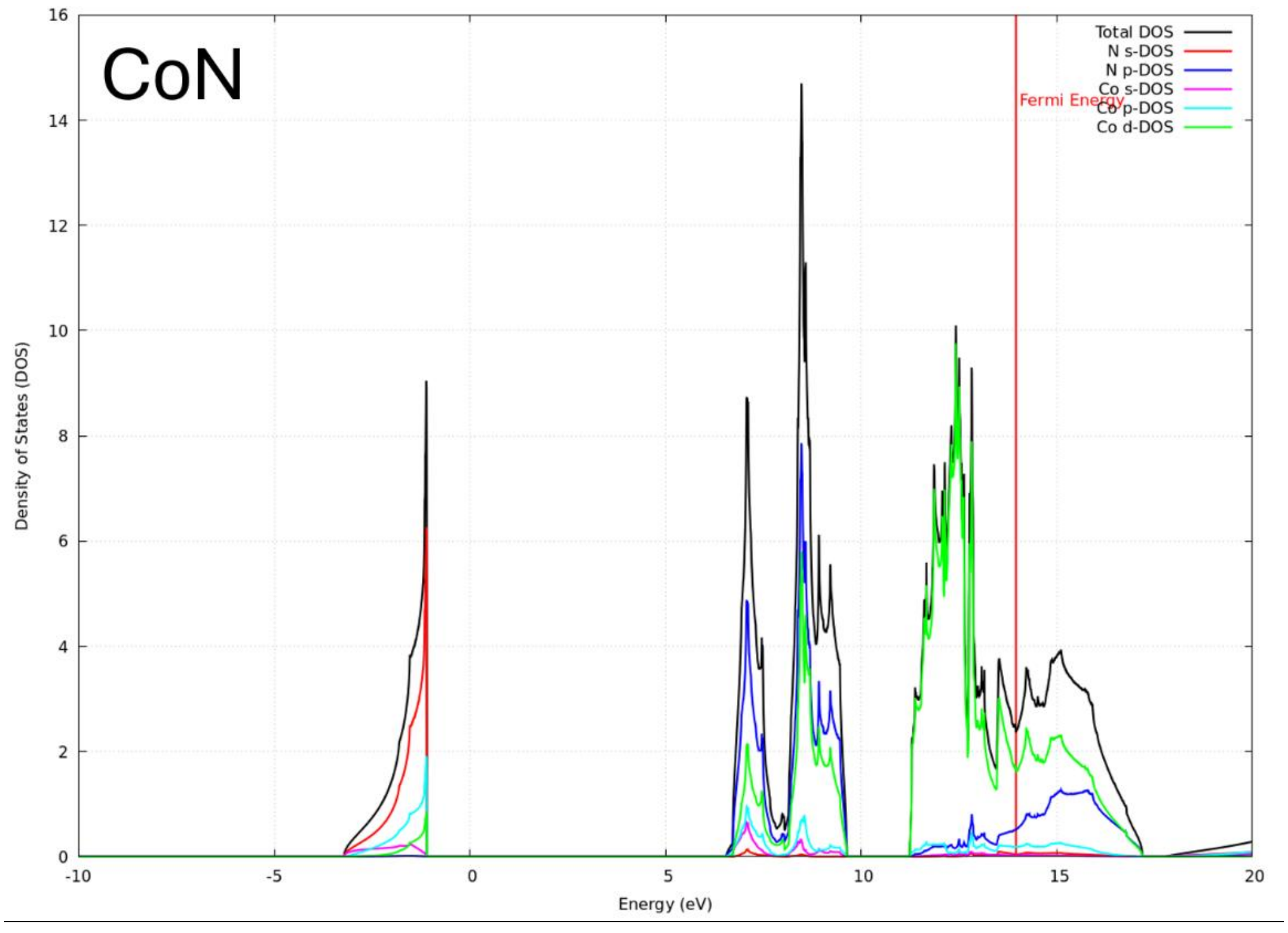

Figure 37: Projected local density of states for hexagonal CoN.

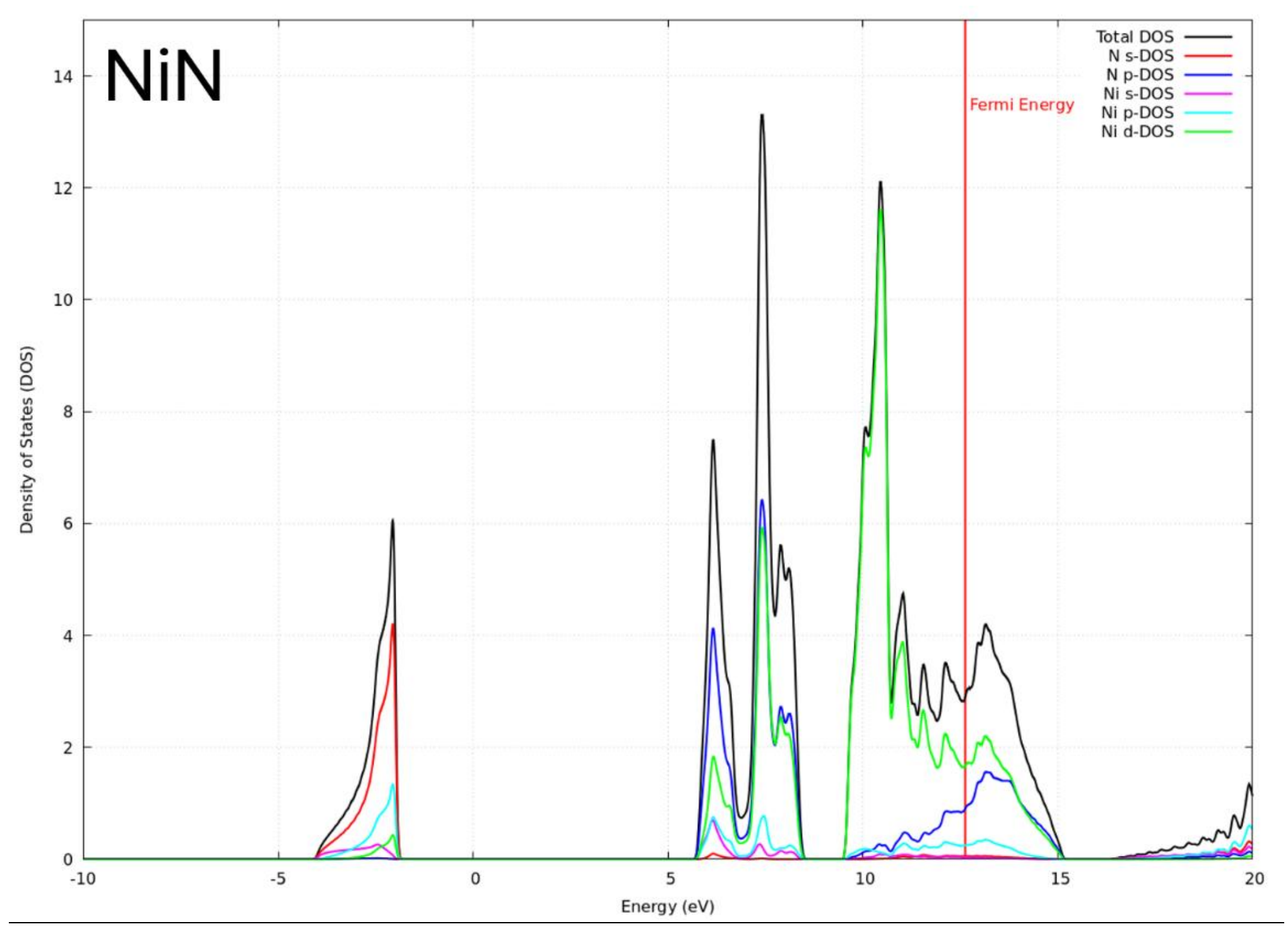

Figure 38: Projected local density of states for hexagonal NiN.

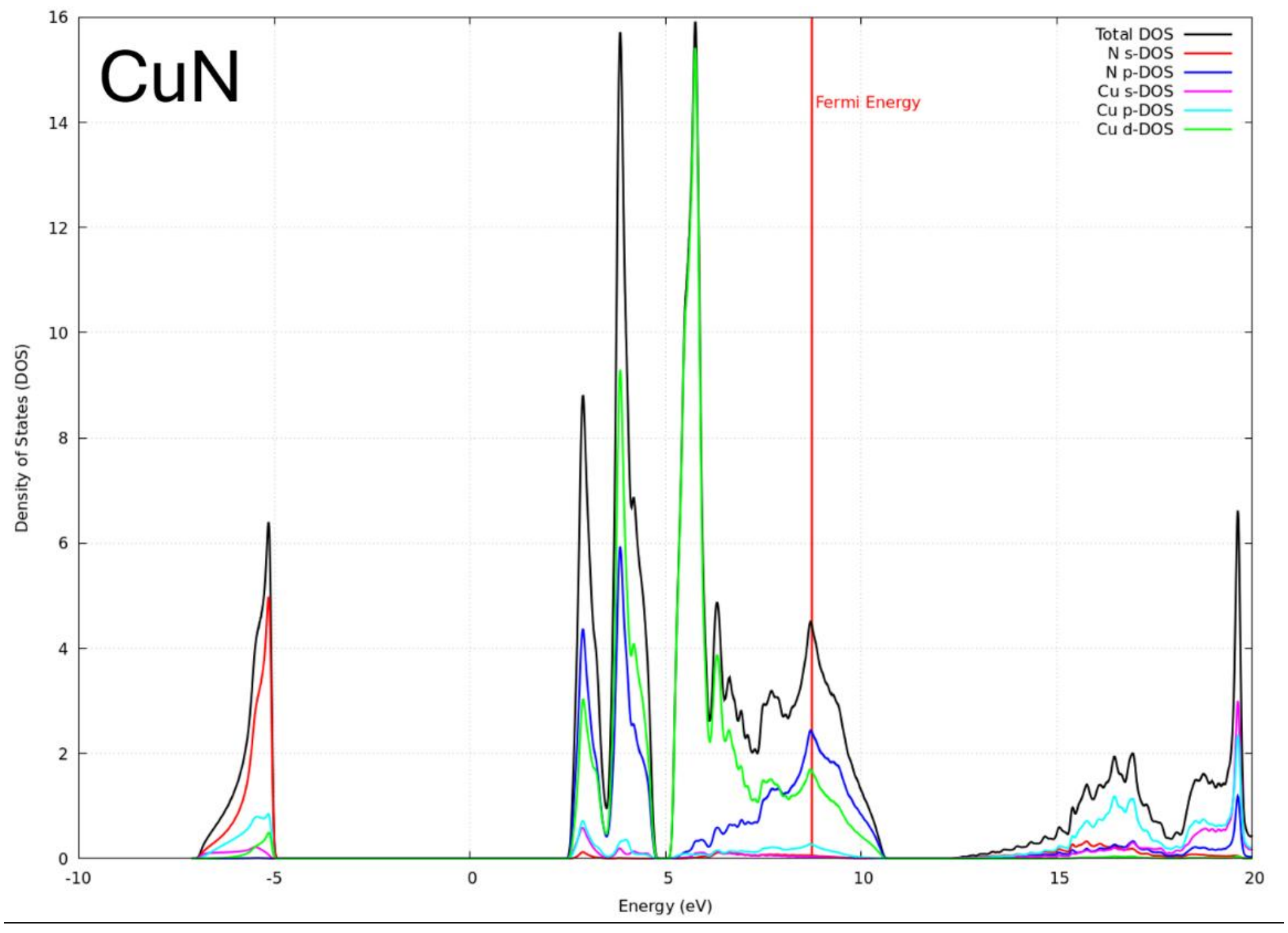

Figure 39: Projected local density of states for hexagonal CuN.

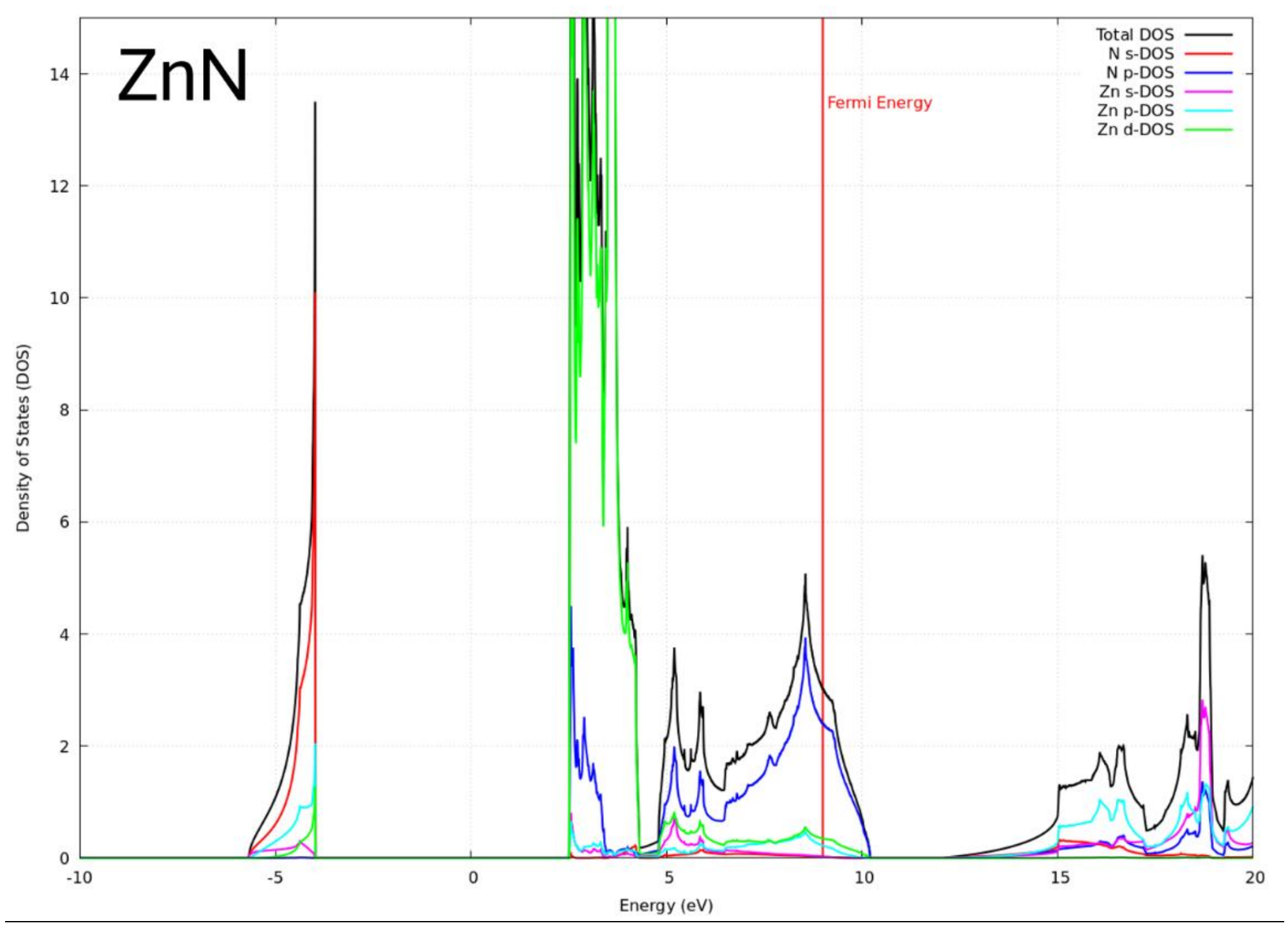

Figure 40: Projected local density of states for hexagonal ZnN.

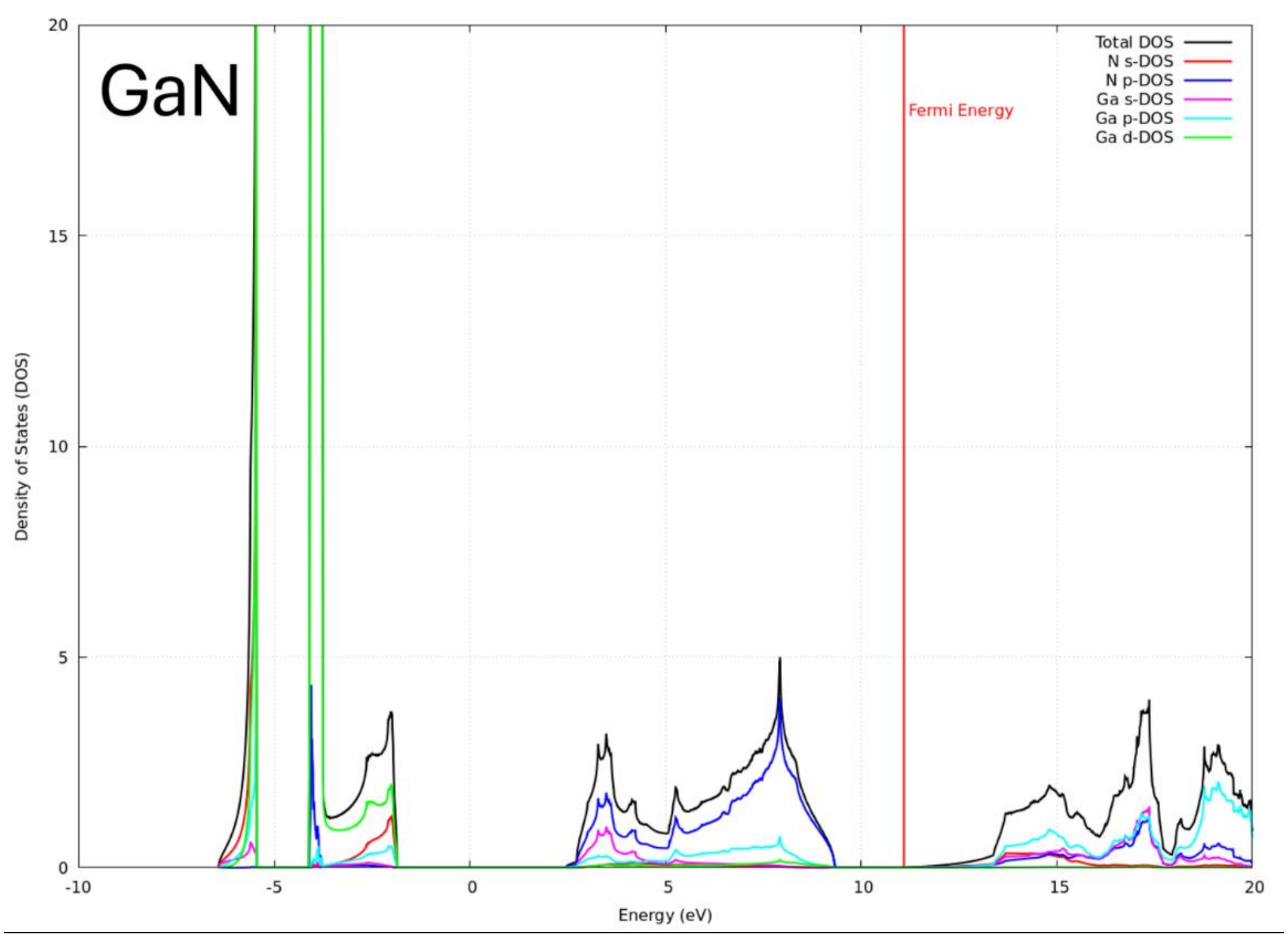

Figure 41: Projected local density of states for hexagonal GaN.

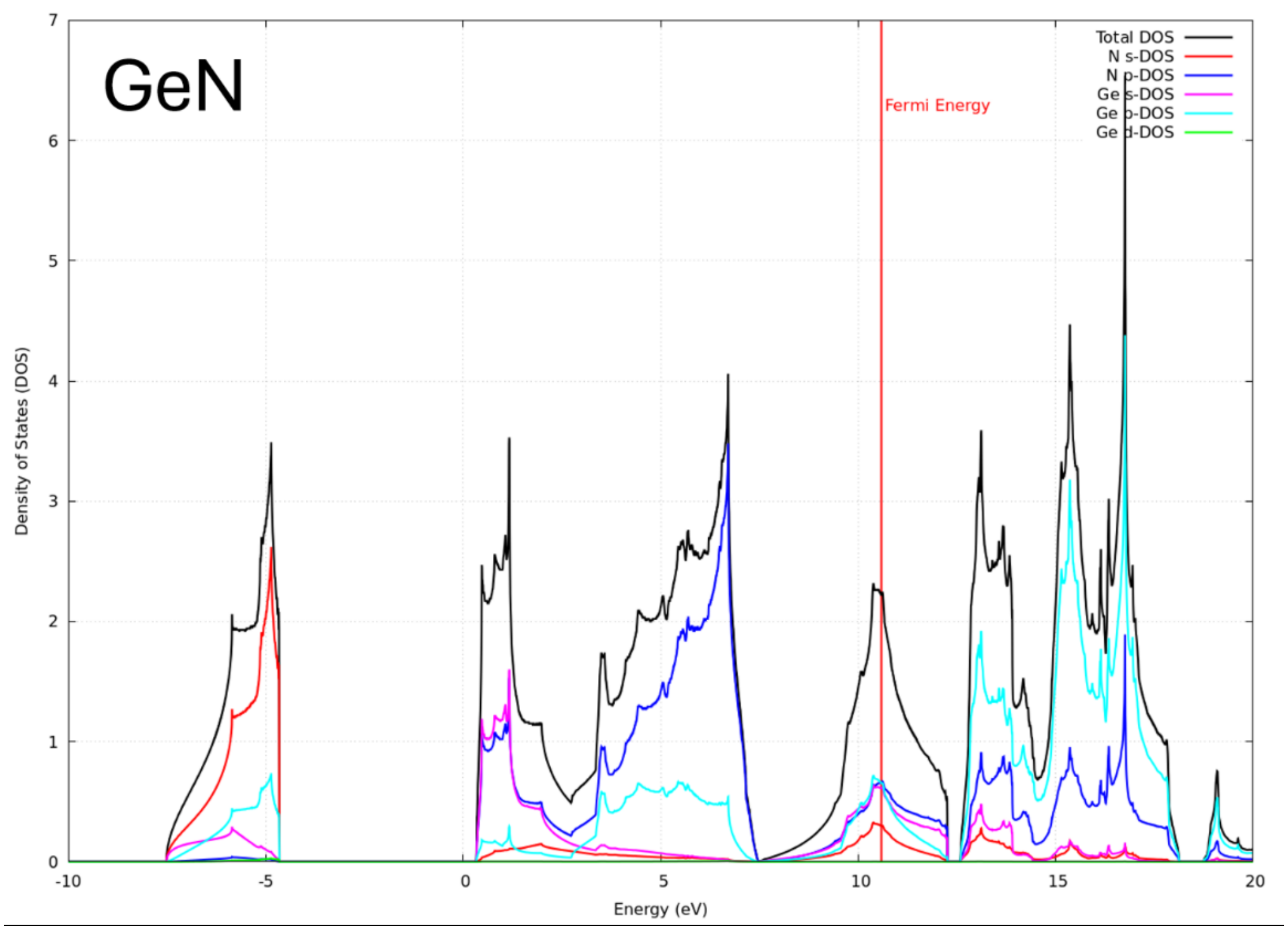

Figure 42: Projected local density of states for hexagonal GeN.

## ILDOS 3D Isosurface representations for Rocksalt, Zincblende and Layered-Hexagonal ScN

Three-dimensional representations of the ILDOS isosurfaces for the rocksalt, zinc-blende, and hexagonal phases are provided respectively in Figures 43 to 45 to facilitate the visualization and interpretation of the spatial distribution of the electronic states. These 3D representations complement the two-dimensional ILDOS projections presented in the main text, providing an alternative view of the shapes and localization of the electronic states and enabling a clearer assessment of the corresponding bonding characteristics.

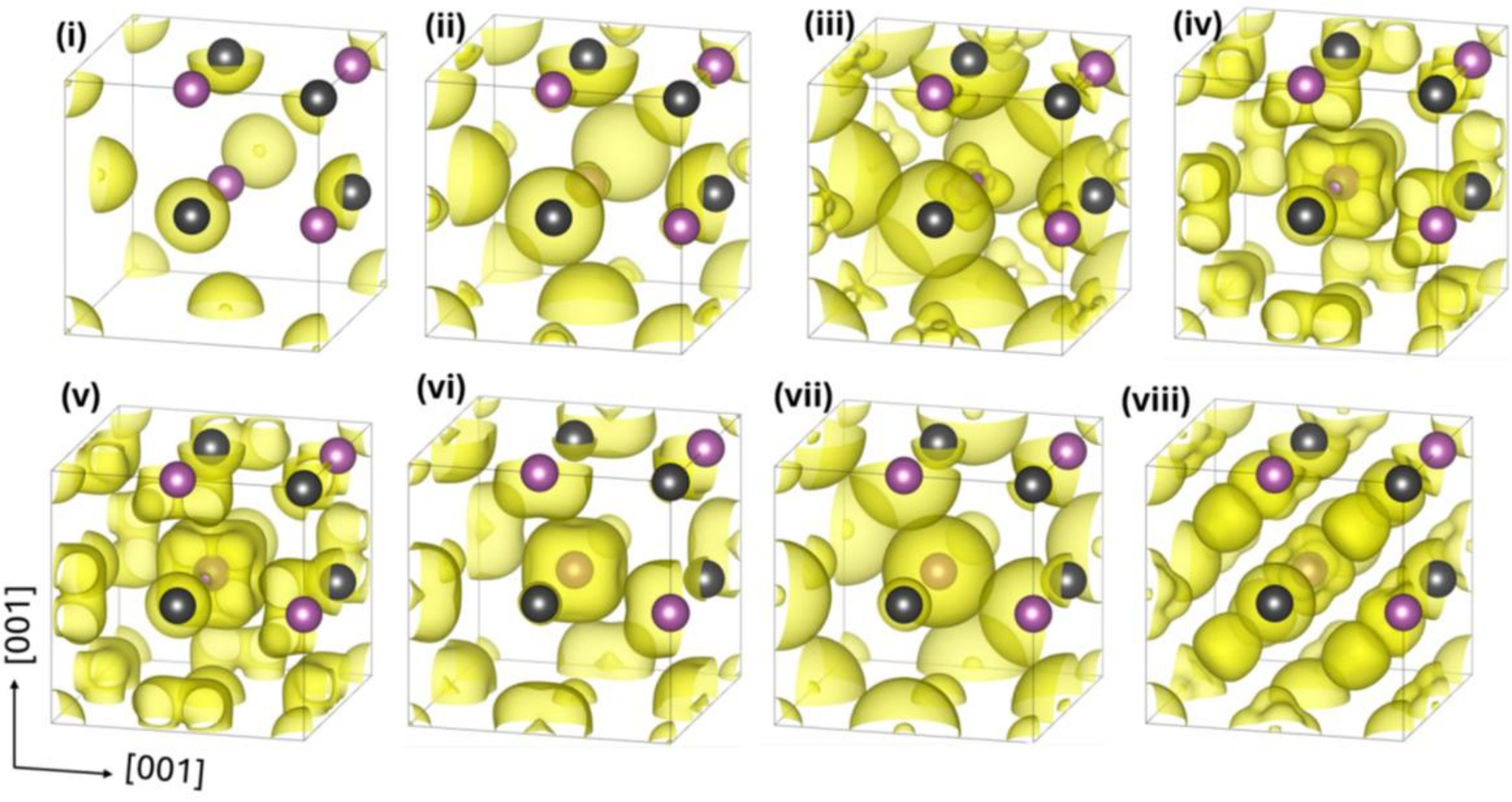

Figure 43: Three-dimensional ILDOS isosurfaces for rocksalt ScN, corresponding to the gray-shaded energy regions indicated in the pDOS plot in the main text. Each panel shows the spatial distribution of the electronic states integrated over a specific energy window, illustrating the three-dimensional character and spatial localization of the states and providing insight into the nature of the associated bonding interactions.

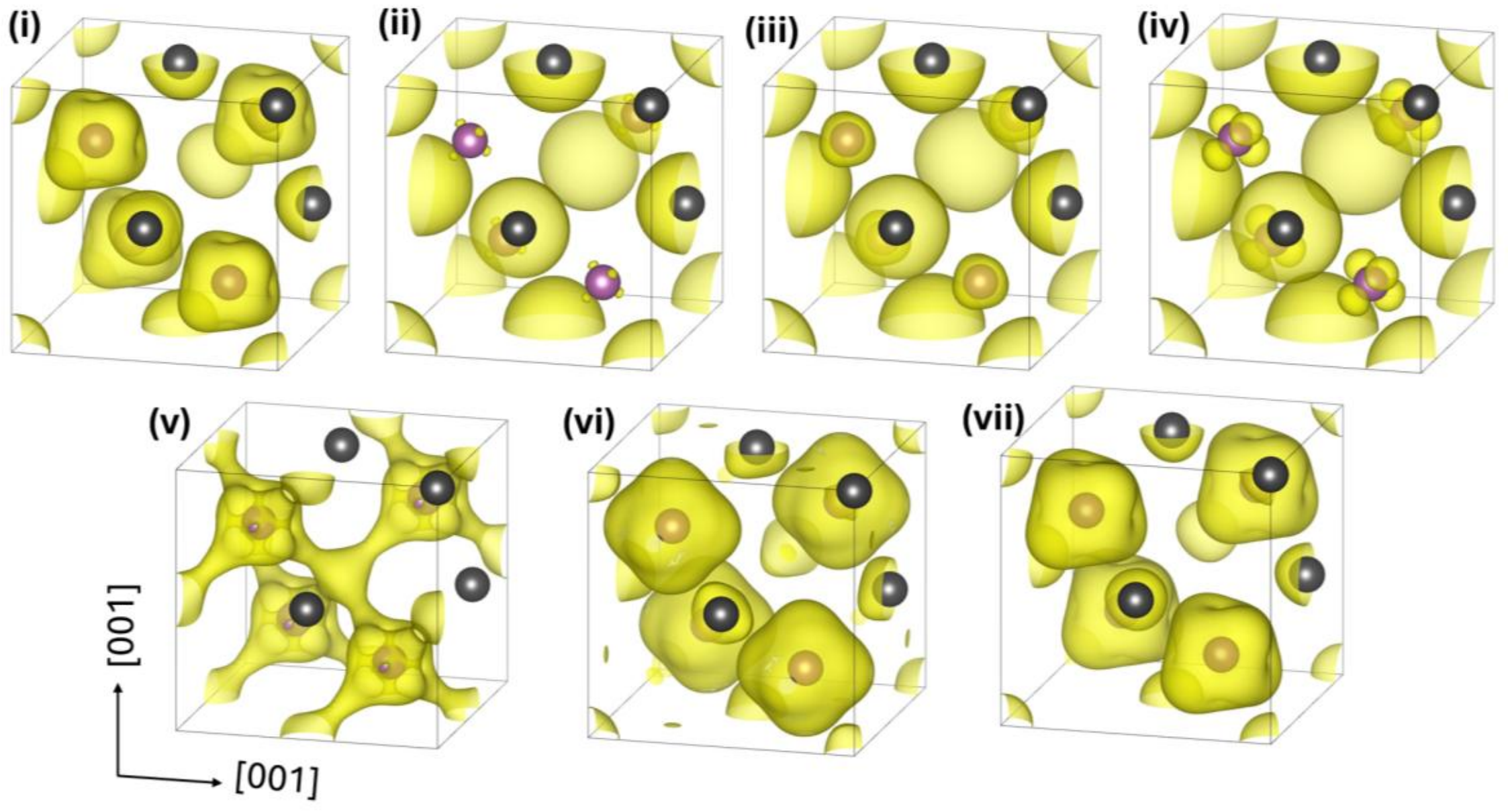

Figure 44: Three-dimensional ILDOS isosurfaces for zincblende ScN, corresponding to the gray-shaded energy regions indicated in the pDOS plot in the main text. Each panel shows the spatial distribution of the electronic states integrated over a specific energy window, illustrating the three-dimensional character and spatial localization of the states and providing insight into the nature of the associated bonding interactions.

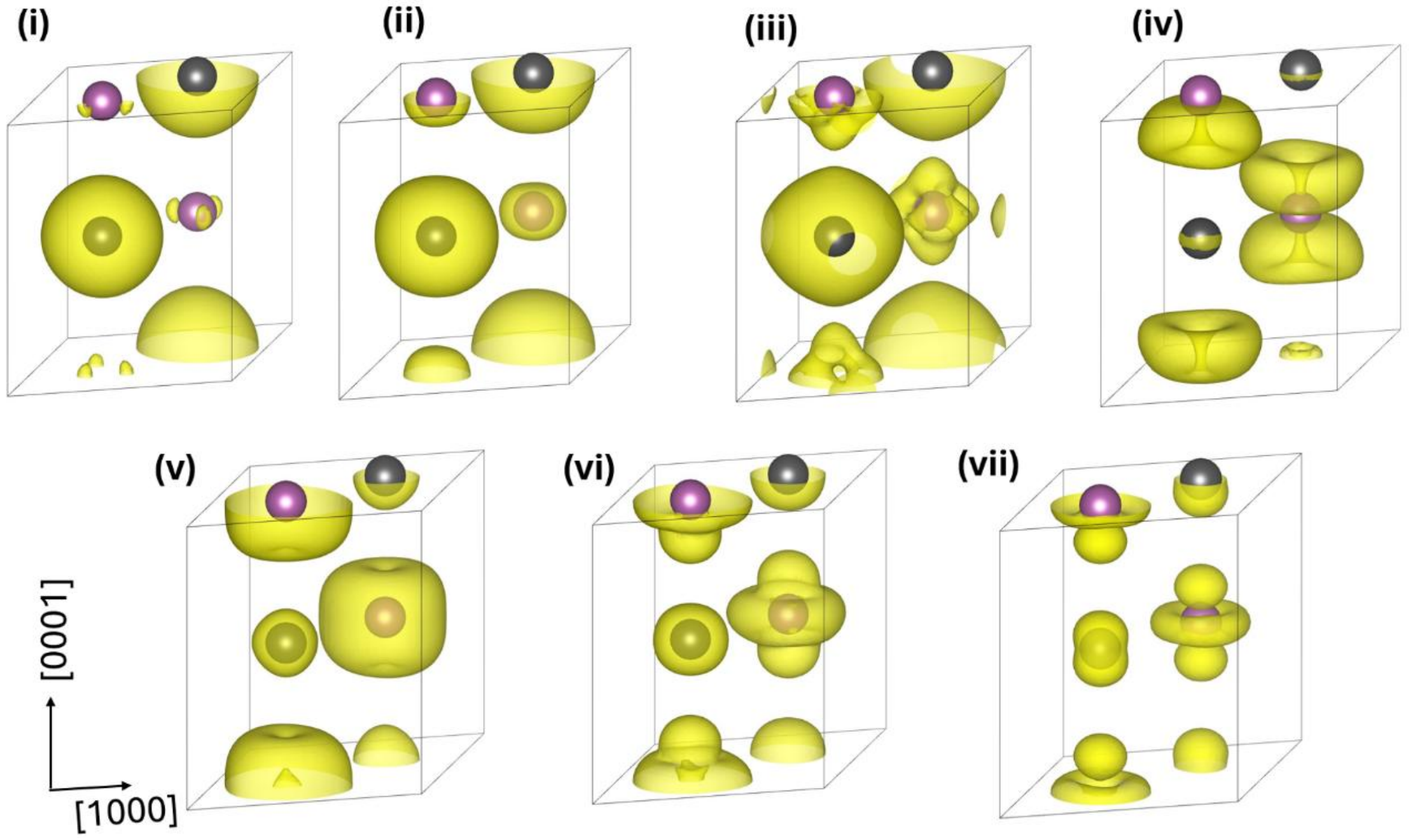

Figure 45: Three-dimensional ILDOS isosurfaces for layered-hexagonal ScN, corresponding to the gray-shaded energy regions indicated in the pDOS plot in the main text. Each panel shows the spatial distribution of the electronic states integrated over a specific energy window, illustrating the three-dimensional character and spatial localization of the states and providing insight into the nature of the associated bonding interactions.